\documentclass[11pt]{article}
\usepackage{graphics}
\usepackage{psfrag}
\usepackage{epsfig}
\usepackage{psfig}
\usepackage{epsf}
\usepackage{float}
\newcommand{\beq}{\begin{equation}}
\newcommand{\eeq}{\end{equation}}
\newcommand{\beqa}{\begin{eqnarray}}
\newcommand{\eeqa}{\end{eqnarray}}

\newcommand{\vs}{\vspace{-0.2cm}}

\newcommand{\no}{\nonumber}
\newcommand{\Mp}{M_\pi}
\newcommand{\Mpz}{M_\pi^2}

\begin{document}


\hfill FZJ-IKP(TH)-1999-19



\vspace{1in}

\begin{center}

{{\Large\bf Nuclear forces from chiral Lagrangians using the method\\[0.3em]
of unitary transformation II~: 
The two--nucleon system}}\footnote{Work supported in part by DFG
under contract number Me 864-16/1.}

\end{center}

\vspace{.3in}

\begin{center}

{\large 
E. Epelbaum,$^\ddagger$$^\dagger$\footnote{email: 
                           evgeni.epelbaum@hadron.tp2.ruhr-uni-bochum.de}
W. Gl\"ockle,$^\dagger$\footnote{email:
                           walter.gloeckle@hadron.tp2.ruhr-uni-bochum.de}
Ulf-G. Mei{\ss}ner$^\ddagger$\footnote{email: 
                           Ulf-G.Meissner@fz-juelich.de}}

\bigskip

$^\ddagger${\it Forschungszentrum J\"ulich, Institut f\"ur Kernphysik 
(Theorie)\\ D-52425 J\"ulich, Germany}

\bigskip

$^\dagger${\it Ruhr-Universit\"at Bochum, Institut f{\"u}r
  Theoretische Physik II\\ D-44870 Bochum, Germany}\\

\end{center}

\vspace{.6in}

\thispagestyle{empty}

\begin{abstract}
\noindent We employ the chiral nucleon--nucleon potential derived in ref.\cite{egm}
to study bound and scattering states in the two--nucleon system. At next--to--leading
order, this potential is the sum of renormalized one--pion and two--pion exchange and
contact interactions. At next--to--next--to-leading order, we have
additional chiral two--pion exchange with low--energy constants
determined from pion--nucleon scattering. Alternatively, we consider
the $\Delta (1232)$ as an explicit degree of freedom in the effective
field theory. The nine parameters related to the contact
interactions can be determined by a fit to the $np$ S-- and P--waves 
and the mixing parameter $\epsilon_1$ for laboratory
energies below 100~MeV. The predicted phase shifts and mixing
parameters for higher energies and higher
angular momenta are mostly well described for energies below 300~MeV. 
The S--waves are described as precisely as in modern
phenomenological potentials.
We find a good description of the deuteron properties. 
\end{abstract}

\vfill

\pagebreak

\section{Introduction}
\def\theequation{\arabic{section}.\arabic{equation}}
\setcounter{equation}{0}
\label{sec:intro}
 
Over the last years, effective field theory methods have been used
to gain a better understanding of the two--nucleon interaction at low
and intermediate energies. While at present these studies do not aim  at
substituting the highly successful ``realistic'' potentials build from
meson exchanges (like the e.g. Bonn--J\"ulich, Nijmegen, Argonne or RuhrPot potential),
effective field theory (EFT) allows for a {\it systematic} and {\it controlled}
expansion of observables in  systems with two or more
nucleons. Apart from 
dealing with the various scales appearing in nuclear systems, it is straightforward
to implement the spontaneously and explicitely broken chiral symmetry of QCD 
as well as external probes in the EFT.
The appearance of shallow bound states (or, equivalently, large scattering
lengths) requires some method of resummation. There are essentially
two ways of tackling this problem.  One approach is
to build the potential from EFT and employ it in a properly
regularized Lippmann--Schwinger (or Schr\"odinger)
equation. Alternatively, one can also do the expansion directly on the
level of the scattering amplitude and resum the leading order, momentum--independent
four--nucleon interaction. While at extremely low energies, much below the
scale set by the pion mass, it is sufficient to consider
four--nucleon interactions only (otherwise the rather successfull effective range
expansion would not work), for typical nuclear momenta of the size
of the pion mass, pions have to be included explicitely. While there has been much 
debate about the way how to treat the pions (perturbative versus non--perturbative
pions), for the range of momenta to be considered here we believe that
it is mandatory to include them at leading order and treat them nonperturbatively. 
The results discussed below give an a posteriori justification of this conjecture.
In fact, our chiral potential at next--to--next--to--leading order in the
power counting leads to results which are only slightly worse than the
ones based on the so--called high accuracy modern potentials.

\medskip
\noindent
In ref.\cite{egm} (referred to as I from here on)
we constructed the two-- and three--nucleon potential based
on the most general chiral effective pion--nucleon Lagrangian using
the method of unitary transformations. For that, we developed a power
counting scheme consistent with this projection formalism. In contrast
to previous results obtained in old--fashioned time--ordered perturbation 
theory, the method employed leads to energy--independent potentials.
This extends the power counting scheme originally proposed by Weinberg~\cite{wein}
in a natural fashion. To leading order (LO), the potential consists of the
undisputed one--pion exchange  and two short--range four--fermion interactions.
Corrections at next--to--leading order (NLO) stem from chiral two--pion exchange
and additional contact terms with two derivatives. We extend the potential 
to next--to--next--to--leading order (NNLO) in two different ways. 
Apart from coupling constant and mass renormalization, one has only to
consider the chiral two--pion--exchange potential (TPEP) with
dimension two insertions from the 
pion--nucleon interaction. In contrast to ref.\cite{ubi}, we take the
novel low--energy constants (LECs) from systematic studies of
pion--nucleon scattering in
CHPT~\cite{bkmlec,moj,fms,paul}.\footnote{We already remark  here that
the determination of these parameters from fitting the invariant
amplitudes inside the Mandelstam triangle
is favored by our fits.} Alternatively, since most of these LECs
are saturated by the $\Delta$--resonance, one can also include the
$\Delta$ explicitely.\footnote{Strictly speaking, one should keep also
the dimension two $\pi N$ operators and subtract the $\Delta$
contribution. Since an explicit $\Delta$ is, however, dynamically
different from integrating it out, we refrain from doing this. The
uncertainty due to this procedure is small in most partial waves.}
That only introduces the $\pi N \Delta$ coupling as a new parameter.
It can be determined  e.g. from an (spin--isospin) SU(4) relation to the
pion--nucleon coupling. In our approach
the exchanges of heavy mesons (like in the CD-Bonn or Nijm93 potential) or
parametrizations thereof (like in  the AV18 or NijmI,II potential) show up as contact
terms, but we refrain from a comparison of similarities and differences to the
various potential model approaches at this point. 
A more detailed introduction of our method is given in I. 
Here, we will be concerned with numerical results obtained on the basis of that 
two--nucleon potential at NLO and NNLO. For that, we first have to
renormalize the potential
and discuss the appropriate cut--off regularization. Having done that, it
is straightforward to solve the corresponding Lippmann--Schwinger equation
for bound and scattering states. In total, we have nine parameters related to the
four--fermion contact terms. These parameters
can be uniquely fixed from fitting $np$ partial waves at low energies, more
precisely the two S--, four P--waves and the mixing parameter $\epsilon_1$.
Predictions for higher energies and the higher partial waves and deuteron 
properties arise (we refrain from fine tuning some of the parameters
in the mixed $^3S_1-^3D_1$ waves to get the exact deuteron  binding
energy as it is done in conventional boson--exchange models).

\medskip
\noindent
To put our calculations in a better perspective, we briefly  review the status
of previous works (directly related to the results presented) without going 
into details. The idea of constructing the potential from effective field theory
was put forward by Weinberg~\cite{wein} in the early nineties. This was taken up 
by van Kolck and collaborators and culminated in ref.\cite{ubi}. In that paper, 
the two--nucleon potential  was constructed at next--next--to--leading 
order\footnote{The leading effects of the  $\Delta (1232)$
resonance were also incorporated in that work.} 
and analyzed. It contained one-- and two--pion
exchange graphs (based on time--ordered perturbation theory diagrams)
and a host of contact interactions. An exponential cut--off at each vertex
was introduced to tame the high energy behaviour.
Fierz reordering was not used so that global fits with 26 parameters
had to be performed and first numerical results for phase shifts and deuteron
properties  were obtained. While promising, these results were not as precise
as the ones usually obtained in boson--exchange models.
Park et al.~\cite{korea} considered a series of interesting applications
mostly related to the deuteron properties. They restricted themselves to 
one--pion--exchange and contact interactions for the relevant phases
$^1S_0, ^3S_1, ^3D_1$ and the mixing parameter $\epsilon_1$. Peripheral
nucleon--nucleon scattering was considered by the Munich
group~\cite{norb}, including two--pion exchange graphs with insertions from the
pion--nucleon Lagrangian of dimension two based on dimensionally regularized Feynman 
graphs. This lead to an improved description of some higher partial waves (for 
similar results, see the work of~\cite{bras}). Since the potential was only 
considered perturbatively, the bound--state problem could not be addressed. 
A different counting scheme was proposed by Kaplan, Savage and Wise~\cite{KSW}. 
In that framework, the low partial waves and mixing parameters were considered to 
NLO and NNLO as well as many deuteron properties. A NNLO calculation of the $^3S_1 
- ^3D_1$ transition potential matrix element was recently
presented~\cite{calt}. Further NNLO investigations for other partial
waves in the framework of the KSW approach are under way by the
Seattle group, see ref.\cite{RS}.

\medskip
\noindent Our manuscript is organized as follows. In sec.\ref{sec:pot}, we
explicitely give the renormalized potential at NLO and NNLO. This follows directly from
the potential derived in~I.\footnote{Note that the NNLO potential was
not given in~I.} We also discuss the regularization procedure
necessecary to render the (iterated) potential finite. The Lippmann--Schwinger
equation underlying the calculation of the scattering and bound states is
briefly discussed in sec.\ref{sec:LS}. The fitting procedure to determine
the LECs and the accuracy of the fits are detailed in sec.\ref{sec:fit}. This
involves the low partial waves at (kinetic) energies (in the lab frame) below 100~MeV.
Results for the low partial waves at higher energies, for the higher partial waves
and the deuteron (bound state) properties are displayed and discussed in 
sec.\ref{sec:res}. Our findings are summarized in sec.\ref{sec:summ}. The appendices
contain details on the renormalization procedure, the inclusion of 
the $\Delta$--resonance and a collection of other useful formulas.
\bigskip
\noindent

\section{The renormalized potential}
\def\theequation{\arabic{section}.\arabic{equation}}
\setcounter{equation}{0}
\label{sec:pot}

In~I, we derived the two--nucleon potential at next--to--leading order.
To leading order, it consists of the one--pion--exchange potential
(which is called the OPEP) and two S--wave 
(non--derivative) contact interactions. The latter are parametrized by
the coupling (low--energy) constants $C_S$ and $C_T$. The OPEP constitutes the
longest range part of the nucleon--nucleon interaction and it completely
dominates the high partial waves. However, the tensor potential related
to one--pion--exchange becomes unrealistic at short distances. Chiral symmetry
enforces a (pseudovector) derivative coupling in harmony with the Goldstone
theorem.  Note also that
while the four--fermion interactions are pointlike on the level of the
Lagrangian, they get smeared out by the regularization procedure to be discussed 
below. At next--to--leading order,
we have three distinct contributions. The first ones are the one--loop
self--energy and vertex corrections to the OPEP, cf. figs.3,4~in~I.
Similarly, it is mandatory to also work out the one--loop corrections to the contact 
terms $\sim C_S, C_T$, cf. fig.6~in~I.
Thirdly, there are the genuine two--pion exchange diagrams, leading to the so--called
TPEP, cf. fig.2~in~I. This TPEP is the second longest range component of the $NN$
interaction. Here, chiral symmetry leads to the $\bar NN\pi\pi$ vertex which in turn
leads to the so--called triangle and football diagrams. These are usually not
accounted for in boson--exchange models. One notices immediately 
that some of the loop corrections contain UV divergences.
Since we are dealing with an effective 
field theory, we can use the standard order--by--order renormalization
machinery for the Lagrangian, 
which was first developed in the context of chiral perturbation theory. All these 
divergent contributions can be renormalized in terms of three divergent (momentum 
space) loop functions,
\beq\label{lfcts}
J_0 = \int_\varepsilon^\infty \frac{dl}{l}~, \quad J_2 = \int_0^\infty
l\, dl~,  \quad J_4 = \int_0^\infty l^3\, dl~.
\eeq
Clearly $J_2\, (J_4)$ is quadratically (quartically) divergent 
while $J_0$ diverges logarithmically
for large momenta and $\varepsilon$ is an IR regulator. Note that we
could introduce other forms of the divergent integrals, which depend
on one dimensionful scale (denoted here by $\varepsilon$). Another
choice of defining this divergent loop integrals would give different 
finite subtractions but  leads to the same non--polynomial terms in the potential.
The precise renormalization procedure is discussed in app.\ref{app:ren}. 
Notice also that different
to standard use, we perform the renormalization of the potential before regularizing
the LS--equation as discussed below. 
Such an additional regularization is needed since loop as well as contact 
term contributions grow quadratically in momenta. More precisely, at
NNLO some of the TPEP contributions even grow with the third power of
momentum. Such an UV 
behaviour of the potential is, of course, unacceptable and requires additional 
regularization.
The solutions of the LS--equation
are adjusted to observables. Therefore, the renormalization constants
are fixed and related to the regularization procedure in the LS--equation.
This procedure is justified since the potential by
itself is {\it not} an observable but rather the bound and scattering states
obtained from the LS--equation. This should always be kept in mind.

\medskip
\noindent
We  consider first the contact terms of the two--nucleon potential. To the accuracy
we are working, the matrix--element of the potential in the center--of--mass
system (cms) for initial and final nucleon momenta $\vec{p}$ and $\vec{p}~'$, 
respectively, takes the form (note that due to the choice of the cms, a reduction 
in the number of terms follows as has been shown in~I)
\beqa\label{Vcon}
V^{{(\rm contact)}} &=& V^{(0)} + V^{(2)}~, \no\\
V^{(0)} &=& C_S  + C_T \, \vec{\sigma}_1 \cdot  \vec{\sigma}_2~,\no\\
V^{(2)} &=& C_1 \, \vec{q}\,^2 + C_2 \, \vec{k}^2 +
( C_3 \, \vec{q}\,^2 + C_4 \, \vec{k}^2 ) ( \vec{\sigma}_1 \cdot \vec{\sigma}_2)
+ iC_5\, \frac{1}{2} \, ( \vec{\sigma}_1 + \vec{\sigma}_2) \cdot ( \vec{q} \times
\vec{k})\no\\
&+& C_6 \, (\vec{q}\cdot \vec{\sigma}_1 )(\vec{q}\cdot \vec{\sigma}_2 ) 
+ C_7 \, (\vec{k}\cdot \vec{\sigma}_1 )(\vec{k}\cdot \vec{\sigma}_2 )~,
\eeqa
with $\vec{q} = \vec{p}~'-\vec{p}$ and $\vec{k} = (\vec{p}+\vec{p}~')/2$. These
terms feed into the matrix--elements of the two S--waves
($^1S_0,^3S_1$), the four P--waves 
($^1P_1,^3P_1,^3P_2,^3P_0)$ and the lowest transition potential
($^3D_1 - ^3S_1$) in the following way:
\beqa\label{VC}
V(^1S_0) &=& {4\pi} \, (C_S-3C_T) + \pi \, ( 4C_1 + C_2 -12C_3
-3C_4 -4C_6 -C_7) ({p}^2+{p}'^2)~,\no\\
&=& \tilde{C}_{^1S_0} + {C}_{^1S_0} ({p}^2+{p}'^2)~,\\
V(^3S_1) &=& {4\pi} \, (C_S+C_T) + \frac{\pi}{3} \, ( 12C_1 + 3C_2 +12C_3
+3C_4 +4C_6 +C_7) ({p}^2+{p}'^2)~,\no\\
&=& \tilde{C}_{^3S_1} + {C}_{^3S_1} ({p}^2+{p}'^2)~,\\
V(^1P_1) &=& \frac{2\pi}{3} \, ( -4C_1 + C_2 +12C_3
-3C_4 +4C_6 -C_7) \, ({p}\, {p}') 
= {C}_{^1P_1}\, ({p}\, {p}')~,\\
V(^3P_1) &=& \frac{2\pi}{3} \, ( -4C_1 + C_2 - 4C_3
+C_4 + 2C_5 +4C_6 + C_7) \, ({p}\, {p}')
= {C}_{^3P_1}\, ({p}\, {p}')~,\no \\ &&\\
V(^3P_2) &=& \frac{2\pi}{3} \, ( -4C_1 + C_2 - 4C_3
+C_4 + 2C_5 ) \, ({p}\, {p}')
= {C}_{^3P_2}\, ( {p}\, {p}')~,\\
V(^3P_0) &=& \frac{2\pi}{3} \, ( -4C_1 + C_2 - 4C_3
+C_4 + 4C_5 +12C_6 - 3C_7) \, ({p}\, {p}')
= {C}_{^1P_0}\, ({p}\, {p}')~,\no \\ &&\\
V(^3D_1 - ^3S_1) &=& \frac{2\sqrt{2}\pi}{3} \, ( 4C_6 + C_7)\,
{p'}^2 = {C}_{^3D_1 - ^3S_1}\, {p'}^2~,\\  
V(^3S_1 - ^3D_1) &=& \frac{2\sqrt{2}\pi}{3} \, ( 4C_6 + C_7)\,
{p}^2 = {C}_{^3D_1 - ^3S_1}\, {p}^2~, \label{VCend}  
\eeqa
with $p = |\vec{p}\,|$ and ${p}' = |\vec{p}\,'|$.
These nine constants are not fixed by chiral symmetry but can be determined
by a fit to these lowest partial waves and one mixing parameter as detailed below.
We have already given the appropriate partial wave decomposition for
the  low--energy constants (LECs) $C_S, C_T, C_1,\ldots,C_7$ here.
From each of the two S--waves, we can determine two parameters, whereas the four
P--waves and the mixing parameter $\epsilon_1$ contain one free
parameter each. Of course, we have to account for the channel coupling
in the mixed triplet partial waves.
It is also important to note that once the $C_{^{2s+1}l_j}$ have been
determined, the original $C_S, C_T, C_1,\ldots,C_7$ are fixed uniquely.
We remark that the values for the $C_{^{2s+1}l_j}$ are renormalized
quantities, see app.\ref{app:ren}.

\medskip
\noindent
Consider now OPEP and TPEP. After vertex and coupling constant renormalization,
as detailed in app.\ref{app:ren}, we find the following expressions in terms of the 
renormalized quantities at NLO:
\beqa\label{OPEP}
V^{\rm OPEP} &=& -\biggl(\frac{g_A}{2f_\pi}\biggr)^2 \, \vec{\tau}_1 \cdot
\vec{\tau}_2 \, \frac{\vec{\sigma}_1 \cdot\vec{q}\,\vec{\sigma}_2\cdot\vec{q}}
{q^2 + M_\pi^2}~,\\ \label{TPEP}
V^{\rm TPEP}_{\rm NLO} 
&=& - \frac{ \vec{\tau}_1 \cdot \vec{\tau}_2 }{384 \pi^2 f_\pi^4}\,
L(q) \, \biggl\{4M_\pi^2 (5g_A^4 - 4g_A^2 -1) + q^2(23g_A^4 - 10g_A^2 -1)
+ \frac{48 g_A^4 M_\pi^4}{4 M_\pi^2 + q^2} \biggr\}\no \\
&& - \frac{3 g_A^4}{64 \pi^2 f_\pi^4} \,L(q)  \, \biggl\{
\vec{\sigma}_1 \cdot\vec{q}\,\vec{\sigma}_2\cdot\vec{q} - q^2 \, 
\vec{\sigma}_1 \cdot\vec{\sigma}_2 \biggr\} + P(\vec{k}, \vec{q}\,)~,
\eeqa
with
\beq\label{Lq}
L(q) = \frac{1}{q}\sqrt{4 M_\pi^2 + q^2}\, 
\ln\frac{\sqrt{4 M_\pi^2 + q^2}+q}{2M_\pi}~,
\eeq
and we have set $q \equiv |\vec{q}\,|$. Furthermore, $f_\pi = 93\,$MeV
is the pion decay constant, $M_\pi = 138.03\,$MeV the pion
mass and $g_A = 1.26$ for the axial--vector coupling.\footnote{Note
  that we have changed our conventions as compared
  to~I for the pion decay constant, the isospin generators and the
  relative momentum $\vec{q}$.}
Furthermore, $P(\vec{k}, \vec{q}\,)$ is a polynom in momenta of at most second
degree and has the same structure as the expressions in eq.(\ref{Vcon}). More
precisely, after performing the partial decomposition,  $P(\vec{k}, \vec{q}\,)$ 
leads in each partial wave to polynoms in $\vec{p}$ and $\vec{p}\,'$ of at most
second degree. Thus, its explicit form is of
no relevance here, since it only contributes to the renormalization of the
couplings $C_S, C_T, C_i$, as detailed in app.\ref{app:ren}.
The TPEP agrees with the one given 
by the Munich group. It is important to stress the differences to the calculation
of ref.\cite{norb}. While there the potential was treated perturbatively, we iterate it
to all orders in a Lippmann--Schwinger (LS) equation. Second, we use a cut--off
regularization within the LS--equation and not dimensional regularization on the
level of the diagrams. Of course, for the peripheral partial waves, the iteration
is not of importance. We are, however, more ambitious in that we want to get
a description of all partial waves as well as of the bound state (deuteron) properties.
The TPEP at NNLO has also been given in ref.\cite{norb} using
dimensional regularization. Within our  renormalization  scheme, 
it reads:\footnote{Note that in~I we have used the power counting such that
$Q/m \sim Q^2/\Lambda^2_\chi$, where $Q$ corresponds to the low momentum scale.
Accordingly, the $1/m$ corrections in eq.(\ref{TPEP2}) are smaller than those
given by the $c_1$, $c_3$, and $c_4$ terms and therefore contribute to the
N$^3$LO potential. In fact, this follows also if one compares the numerical
vaues of these two types of corrections ($1/m$ versus $c_i$). We have decided
here to keep these $1/m$ corrections, since otherwise one cannot directly use 
the values of the LECs $c_{1,3,4}$ as determined from the $\pi N$
sector in the presence of the  $1/m$ terms.}
\beqa\label{TPEP2}
V^{\rm TPEP}_{\rm NNLO} &=& -\frac{3g_A^2}{16\pi f_\pi^4} \biggl\{ -\frac{g_A^2
  M_\pi^5}{16 m (4M_\pi^2+q^2)} + \biggl(2M_\pi^2(2c_1 -c_3) -q^2 \,
\bigl( c_3 + \frac{3g_A^2}{16m} \bigr) \biggr) (2M_\pi^2+q^2) A(q) \biggr\}
\no\\
&-& \frac{g_A^2}{128\pi m f_\pi^4} (\vec \tau_1 \cdot \vec \tau_2 ) \,
\biggl\{ -\frac{3g_A^2  M_\pi^5}{4M_\pi^2+q^2} + \bigl( 4M_\pi^2 +
2q^2 -g_A^2 (4M_\pi^2 + 3q^2) \bigr)  (2M_\pi^2+q^2) A(q) \biggr\}
\no\\
&+&  \frac{9g_A^4}{512\pi m f_\pi^4} \biggl( (\vec \sigma_1 \cdot \vec
q\,)(\vec \sigma_2 \cdot \vec q\,) -q^2 (\vec \sigma_1 \cdot\vec \sigma_2
)\biggr) \, (2M_\pi^2+q^2) A(q)  \no\\
&-& \frac{g_A^2}{32\pi f_\pi^4} (\vec \tau_1 \cdot \vec \tau_2 ) \,
\biggl( (\vec \sigma_1 \cdot \vec q\,)(\vec \sigma_2 \cdot \vec q\,) 
-q^2 (\vec \sigma_1 \cdot\vec \sigma_2 )\biggr) \no \\
&& \qquad   \qquad \qquad \qquad   \qquad \qquad \times
\biggl\{ \bigl( c_4 + \frac{1}{4m} \bigr) (4M_\pi^2 + q^2) 
-\frac{g_A^2}{8m} (10M_\pi^2 + 3q^2) \biggr\} \, A(q) \no \\
&-& \frac{3g_A^4}{64\pi m f_\pi^4} \, i \, (\vec \sigma_1 +  \vec
\sigma_2 ) \cdot (\vec{p}~' \times \vec{p} ) \, (2M_\pi^2+q^2) A(q) \no \\
&-& \frac{g_A^2(1-g_A^2)}{64\pi m f_\pi^4} (\vec \tau_1 \cdot \vec \tau_2 )
\, i \, (\vec \sigma_1 +  \vec \sigma_2 ) \cdot (\vec{p}~' \times
\vec{p} ) \, (4M_\pi^2+q^2) A(q) + \tilde{P} (\vec{k}, \vec{q}\,)~,
\eeqa
with
\beq
A(q) = \frac{1}{2q} \arctan \frac{q}{2M_\pi}~.
\eeq
The polynom $\tilde{P}(\vec{k}, \vec{q}\,)$  
plays a similar role as $P(\vec{k}, \vec{q}\,)$  in eq.(\ref{TPEP}).
Note that we have used the standard dimension two $\pi N$ Lagrangian
as in ref.\cite{norb}.
For the LECs $c_{1,3,4}$ we should take the values obtained from
fitting $\pi N$ phases in the threshold region, see
e.g. ref.\cite{fms} or, alternatively, from fitting the invariant amplitudes
inside the Mandelstam triangle, i.e. in the unphysical region~\cite{paul}.
The so determined parameters are only slightly different, but these small differences
will play a role later on. For example, the LECs $c_{1,3,4}$ from fit~1 of
ref.\cite{fms} are $c_1 = -1.23\,$GeV$^{-1}$, $c_3 = -5.94\,$GeV$^{-1}$ 
and $c_4 = 3.47\,$GeV$^{-1}$. A recent investigation of the subthreshold
amplitudes~\cite{paul} leads to slightly different values,  $c_1 =
-0.81\,$GeV$^{-1}$, $c_3 = -4.70\,$GeV$^{-1}$  and $c_4 = 3.40\,$GeV$^{-1}$. 
It is this latter set we will use in the following. These values are also
consistent with the recent determination from the proton--proton interaction
based on the chiral two--pion exchange potential~\cite{nijmci}.

\medskip
\noindent
The total renormalized potential at NLO is now given as the sum of the OPEP, TPEP
and contact potentials as given in
eqs.(\ref{OPEP},\ref{TPEP},\ref{Vcon}). At NNLO, we have to add the
additional TPEP from eq.(\ref{TPEP2}). Alternatively, we have also included the leading
contribution from the $\Delta (1232)$ resonance. This is formally a NNLO contribution
as demanded by the decoupling theorem. The pertinent
equations are collected in app.~\ref{sec:delta}. We only remark that there is one
new parameter, namely the $\Delta N\pi$ coupling constant. 
It can be determined from the SU(4) relation
\beq\label{gpiND}
g_{\Delta N\pi} = \frac{3}{\sqrt{2}} \, \frac{g_A \, m}{F_\pi}~,
\eeq
making use of the Goldberger--Treiman relation (GTR) (higher order
corrections to the GTR are not considered here). This
leads to $g_{\Delta N\pi} = 27.2$, as detailed in app.~\ref{sec:delta}. 
The corresponding theory is called NNLO--$\Delta$.
The chiral TPEP has a different high momentum behaviour as in the
strict NNLO approach because of the explicit $\Delta$--propagator
(assuming that we only include the $\Delta$ to generate the NNLO TPEP
and no additional $\pi N$ contact interactions).
Only for small momenta NNLO and NNLO--$\Delta$ are essentially the same (resonance
saturation). At this point, we already mention that the NNLO TPEP is
too strong at short distances (large momenta). To the order we are
working, we have counterterms in the S-- and P--waves to balance this, but
e.g. not in the D--waves. This will be discussed in more details below. 

\medskip
\noindent
 Since the potential is only meaningful for 
momenta below a certain scale, it needs regularization. Stated
differently, the large momentum behaviour of the potential is not correct.
To the order we are working, the contact interactions diverge
quadratically for large momenta and some components of the TPEP grow
even with the third power of momentum. As it is appropriate in
effective field theory, we regularize the potential. That is done in the following way:
\beq\label{Vreg}
V( \vec{p},\vec{p}~'\,) \to f_R ( \vec{p}\,) \, V( \vec{p},\vec{p}~'\,) \,
f_R (\vec{p}~'\,)~,
\eeq
where $f_R ( \vec{p}\,)$ is a regulator function chosen in harmony with the
underlying symmetries. In what follows, we work with two different regulator functions,
\beqa\label{reg1}
 f_R^{\rm sharp} ( \vec{p}\,) &=& \theta (\Lambda^2 -p^2)~,\\\label{reg2}
 f_R^{\rm expon} ( \vec{p}\,) &=& \exp(-p^{2n} / \Lambda^{2n})~,
\eeqa
with $n = 2,3,\ldots\,$. The sharp cut--off is most appropriate to the
projection formalism. For the calculation of some observables, however,
it cannot be used since at $p,p' = \Lambda$ it leads to discontinous derivatives.
For very large integers $n$ the exponential cut--off approximates the
sharp one. Throughout, we work with $n=2$. To the order we are working, the
choice $n=1$ has to be excluded since the contact terms of order $p^2, {p'}^2$ would
be modified. 
For $n=2$, the error we make  is beyond the accuracy of the order we are calculating. 
We are now in the position to calculate observables with this potential.

\section{Bound and scattering state equation}
\def\theequation{\arabic{section}.\arabic{equation}}
\setcounter{equation}{0}
\label{sec:LS}

In ref.\cite{ubi}, the Schr\"odinger equation was solved after the
chiral NN potential had been transformed from momentum into
co-ordinate space. We
consider it more natural to work directly in momentum space. The
corresponding equation describing the bound and scattering states is
the Lippmann--Schwinger equation. Here, we briefly discuss it to
keep the manuscript self--contained. For a more detailed exposition
concerning also methods of solving the LS--equation, we refer to the
monograph~\cite{WGbook}.

\medskip \noindent
The LS--equation (for the T--matrix) projected into states with orbital angular momentum
$l$, total spin $s$ and total angular momentum $j$ is 
\beq\label{LSeq}
T^{sj}_{l',l} (p',p) = V^{sj}_{l',l} (p',p) + \sum_{l''} \,
\int_0^\infty \frac{dp'' \, {p''}^2}{(2 \pi )^3} \,  V^{sj}_{l',l''} (p',p'')
\frac{1}{p^2-{p''}^2 -i\eta} T^{sj}_{l'',l} (p'',p)~,
\eeq
with $\eta \to 0^+$.
In the uncoupled case, $l$ is conserved. The partial wave projected
potential $V^{sj}_{l',l} (p',p)$ is obtained as follows.
We first rewrite the potential $V$  in the form~\cite{erk}
\begin{eqnarray}
\label{pot_dec}
V &=& V_C + V_\sigma \; \vec{\sigma}_1 \cdot \vec{\sigma}_2 + V_{SL} \; i \; \frac{1}{2} 
(\vec{\sigma}_1 + \vec{\sigma}_2 )  \cdot
( \vec{k} \times \vec{q} )  + V_{\sigma L} \; \vec{\sigma}_1 \cdot ( 
\vec{q} \times
\vec{k} ) \; \vec{\sigma}_2 \cdot ( \vec{q} \times \vec{k} ) \nonumber \\
&& {} + V_{\sigma q} \; ( \vec{\sigma}_1 \cdot \vec{q} ) \; (\vec{\sigma}_2 \cdot
\vec{q} ) + V_{\sigma k} \; (\vec{\sigma}_1 \cdot \vec{k}) \; (\vec{\sigma}_2 \cdot \vec{k})
\end{eqnarray}
with six functions $V_C (p, p', z), \; \ldots , \; V_{\sigma k} (p, p', z)$
depending on $p \equiv | \vec{p}\,|$, $p' \equiv | \vec{p} \, '|$ and
the cosine of the angle between the two momenta $z$.
These functions may depend on the isospin matrices $\vec{\tau}$ as well.      
For $j>0$,  one obtains in the usual $lsj$ representation
\begin{eqnarray}
\langle j0j | V | j0j \rangle &=& 2 \pi \int_{-1}^{1} \, dz \, \left\{ V_C - 3 V_\sigma
+ {p '}^2 {p}^2 (z^2 -1) V_{\sigma L} - q^2 V_{\sigma q} - k^2 V_{\sigma k} 
\right\} P_j (z)~, \nonumber \\
\langle j1j | V | j1j \rangle 
&=& 2 \pi \int_{-1}^{1} \, dz \, \left\{ \left[ V_C + V_\sigma + 2 
p' p z V_{SL} - {p'}^2 p^2 (1 + 3 z^2) V_{\sigma L} + 4 k^2 V_{\sigma q} + 
\frac{1}{4} q^2 V_{\sigma k} \right]   \right. \nonumber \\
&&   {} \times P_j (z) + \left. \left[- p' p \, V_{SL} + 
2 {p'}^2 p^2 z V_{\sigma L} -2 p' p \, (V_{\sigma q} - \frac{1}{4} V_{\sigma k} )\right] \right. \nonumber \\
&& {} \times \left( P_{j-1} (z) +
P_{j+1} (z) \right) \bigg\}~,  \nonumber 
\end{eqnarray}
\begin{eqnarray}
\langle j\pm 1,1j | V | j \pm 1, 1j \rangle &=& 2 \pi \int_{-1}^{1} \, dz \, \left\{
p' p \left[ -V_{SL}  \pm \frac{2}{2j+1} \left( - p' p z V_{\sigma L} + V_{\sigma q} -
\frac{1}{4} V_{\sigma k} \right) \right] \right.  \nonumber \\
&&  {} \times P_j (z) + \bigg[ V_C + V_\sigma + p' p z V_{SL} + {p'}^2 p^2 (1-z^2) V_{\sigma L} \\
&&  {} \left. \left. \pm\frac{1}{2j+1} \left( 2 {p'}^2 p^2 V_{\sigma L} - 
({p'}^2 + p^2)(V_{\sigma q} +
\frac{1}{4} V_{\sigma k} ) \right) \right] P_{j\pm 1} (z) \right\}  \nonumber \\
\langle j\pm 1,1j | V | j \mp 1, 1j \rangle &=& \frac{\sqrt{j(j+1)}}{2j+1} 2 \pi \int_{-1}^{1} \, dz \, 
\Bigg\{- p' p \, (4 V_{\sigma q} - V_{\sigma k} ) P_j (z)~,  \nonumber \\
&& {} + \left[\mp \frac{2 {p'}^2 p^2}{2j+1} V_{\sigma L} + {p'}^2 \Big(2 V_{\sigma q}+\frac{1}{2}
V_{\sigma k}\Big) \right] P_{j\mp 1} (z)  \nonumber \\
&& {} \left. + \left[\pm \frac{2 {p'}^2 p^2}{2j+1} V_{\sigma L} + {p}^2 \Big(2 V_{\sigma q}+\frac{1}{2}
V_{\sigma k}\Big) \right] P_{j\pm 1} (z) \right\}~.  \nonumber 
\end{eqnarray} 
Here, $P_j (z)$ are the conventional Legendre polynomials.
For $j=0$ the two nonvanishing matrix elements are
\begin{eqnarray} 
\langle 000 | V | 000 \rangle &=& 2 \pi \int_{-1}^{1} \, dz \, \left\{
V_C - 3 V_\sigma
+ {p '}^2 {p}^2 (z^2 -1) V_{\sigma L} - q^2 V_{\sigma q} - k^2 V_{\sigma k} 
\right\}~, \nonumber \\
\langle 110 | V | 110 \rangle &=& 2 \pi \int_{-1}^{1} \, dz \, \bigg\{ z V_C
+ z V_\sigma + p' p (z^2 -1) V_{SL} + {p'}^2 p^2 z (1-z^2) V_{\sigma L}  \\
&& {}\left.  - \left( ( {p'}^2 + p^2 )
z - 2 p' p \right)  V_{\sigma q} - \frac{1}{4} 
\left( ( {p'}^2 + p^2 )
z + 2 p' p \right) V_{\sigma k}  \right\}~.
\nonumber
\end{eqnarray}
Note that sometimes another notation is used in which an additional overall ``-'' 
sign enters the expressions for the 
off--diagonal matrix elements with $l=j+1, \, l'=j-1$ and $l=j-1, \, l'=j+1$.

\medskip
\noindent
The relation between the on--shell $S$-- and $T$--matrices is given by 
\beq
S_{l' l}^{s j} (p) = \delta_{l' l} - \frac{i}{8 \pi^2} 
\, p \, m \,  T_{l' l}^{s j} (p)~,
\eeq
where $p$ denotes the two--nucleon center--of--mass three--momentum. The
phase shifts in the uncoupled cases can be obtained from the
$S$--matrix via
\beq
S_{jj}^{0j} = \exp{ \left( 2 i \delta_{j}^{0j} \right)} \; , \quad 
S_{jj}^{1j} = \exp{ \left( 2 i \delta_{j}^{1j} \right)} \;,
\eeq
where we have used the notation $\delta^{sj}_l$.
Throughout, we use the so--called Stapp parametrization~\cite{stapp}
of the $S$--matrix in the coupled channels ($j>0$):
\beq
S = \left( \begin{array}{cc} S_{j-1 \, j-1}^{1j} &  S_{j-1 \, j+1}^{1j} \\
S_{j+1 \, j-1}^{1j} &  S_{j+1 \, j+1}^{1j} \end{array} \right) = 
\left( \begin{array}{cc} \cos{(2 \epsilon)} \exp{(2 i \delta^{1j}_{j-1})} &
i \sin{(2 \epsilon)} \exp{(i \delta^{1j}_{j-1} +i \delta^{1j}_{j+1})} \\
i \sin{(2 \epsilon)} \exp{(i \delta^{1j}_{j-1} +i \delta^{1j}_{j+1})} &
\cos{(2 \epsilon)} \exp{(2 i \delta^{1j}_{j+1})} \end{array} \right)~.
\eeq
The bound state is obtained from the homogeneous part of eq.(\ref{LSeq})
and obeys
\beq\label{LSb}
\phi_l (p) = \frac{1}{E_d -p^2/m} \,  \sum_{l'} \, \int_0^\infty \frac{dp'
\, {p'}^2}{(2 \pi )^3} \,  V^{sj}_{l,l'} (p,p') \, \phi_{l'} (p')~,
\eeq
with $s=j=1$ and $l=l'=0,2$ and $E_d$ denotes the bound--state energy.
The LS-equations for the scattering and
the bound state(s) are solved by standard Gauss--Legendre quadrature. 
It goes without saying that in this discussion, the potential is to be
understood in its regularized form as detailed in
eqs.(\ref{Vreg}, \ref{reg1}, \ref{reg2}).

\section{The fits}
\def\theequation{\arabic{section}.\arabic{equation}}
\setcounter{equation}{0}
\label{sec:fit}

In this section we discuss the determination of the various coupling constants.
The leading OPEP is constrained by chiral symmetry, its strength is
given by the pion--nucleon coupling constant, which due to the
Goldberger--Treiman relation is proportional to $g_A/f_\pi$.
For the NNLO--$\Delta$ approach, we in addition have 
the strong pion--nucleon--$\Delta$ coupling constant.

\medskip
\noindent To pin down the nine parameters $C_S, C_T, C_1, \ldots,C_7$ we do not
perform global fits as done in ref.\cite{ubi}. Rather we introduce the
independent new parameters as given in eqs.(\ref{VC}--\ref{VCend}). So to leading order,
the two S-- waves are depending on one parameter each. At NLO, we have one
additional parameter for $^1S_0$ and $^3S_1$ as well as one parameters in each of the
four P--waves and in $\epsilon_1$. We thus can fit each partial wave separately,
which makes the fitting procedure not only extremely simple but also
unique. Of course, in case of the triplet coupled waves ($^3S_1 -
^3D_1, ^3P_2 - ^3F_2)$ the fitting is performed for the
corresponding $2\times 2$ S--matrix parametrizations (parametrized by
two partial waves and one mixing parameter). 
At NNLO (NNLO--$\Delta$), we have no new parameters, but must refit
the 
various contact interactions
due to the TPEP contribution in all partial waves. We have used two
different methods to fix the LECs of the contact interactions. First, we
fit to the phase shifts of the Nijmegen partial wave
analysis~\cite{Nij93} for laboratory energies smaller than
(50)~100~MeV at (NLO)~NNLO.\footnote{Note that equally well we could
  fit directly to the data. However, for easier comparison with other
  EFT calculations, we use the Nijmegen PSA to simulate the data.}
 Alternatively, we used the effective range
parameters for the phases $^1S_0$, $^3S_1$ and $\epsilon_1$ to fix
$\tilde{C}_{^1S_0},{C}_{^1S_0},\tilde{C}_{^3S_1}, {C}_{^3S_1},$ and
${C}_{^3D_1 - ^3S_1}$. In what follows we will mark the corresponding 
potentials by a ``$\star$'' if the LECs were fixed from the effective range parameters.
We have found that the observables (phase
shifts) depend rather weekly on what fitting procedure is used. The
values of the LECs do not change much except for some $C$'s at NLO,
where significant variations were observed. Note further that in the
S--waves, especially in $^1S_0$, isospin breaking effects like
e.g. the charged to neutral pion mass difference, are known to be
important. We do not consider such effects in this work and thus take
an average value for the pion mass. The actual values of the S--wave
LECs seem to be rather sensitive to the choice of the pion mass.
For the phases $^1S_0$, $^3S_1$ and $\epsilon_1$ we use the errors as
given in ref.\cite{Nij93}, for all other partial waves we assign an
absolute error of 3\%.
This number is arbitrary, but taking any other
value would not change the fit results, only the total $\chi^2$. 
To perform the fits, we have to specify the value of the
cut--off $\Lambda$ in the regulator functions as defined in 
eqs.(\ref{reg1},\ref{reg2}). 
At NLO, for any choice between 380~MeV and 600~MeV, we get very similar 
fits (a very shallow $\chi^2$ distribution in each partial wave). 
At NNLO, this shallow distribution turns into a plateau,
which  shifts to higher values of the cut--off. We can now use
values between 600~MeV and 1~GeV. The results using the sharp or the exponential 
cut--off are very similar. For illustration,
we consider  the sharp cut--off with $\Lambda = 500\,$MeV at NLO
and $\Lambda = 875\,$MeV at NNLO. 
We show  in fig.~\ref{figfit}  the absolute quadratic deviations
of the fit to the Nijmegen phase shift analysis (PSA), defined by
$(\delta^{\rm fit} - \delta^{\rm PSA})^2$.
Consider first
the S--waves. Both in $^1S_0$ and $^3S_1$, we observe a clear
improvement when going from LO to NLO to NNLO (NNLO--$\Delta$). 
A similar pattern holds for the P--waves and $\epsilon_1$, although
the differences between NLO and NNLO are somewhat less
pronounced. Note that the $^3P_0$ wave is very senstive to the value
of the pion mass, therefore the slightly better NLO fit should not be
considered problematic. The corresponding parameters of the
coupling constants are collected in table~\ref{tab:LECs}. 

\renewcommand{\arraystretch}{1.2}
\begin{table}[htb] 
\begin{center}

\begin{tabular}{||l||c|c|c|c|c|c|c|c|c||}
  \hline 
 & $\tilde{C}_{^1S_0}$ & ${C}_{^1S_0}$ & $\tilde{C}_{^3S_1}$ &
 ${C}_{^3S_1}$ & $C_{^1P_1}$ & $C_{^3P_1}$ & $C_{^3P_0}$ & $C_{^3P_2}$ &
 $C_{^3D_1 -^3S_1}$\\ \hline\hline  
 NLO & $-$0.134 & 1.822 & $-$0.130 & $-$0.393 & 0.344 &
 $-$0.394 & 1.335 & $-$0.1907 & $-$0.0317 \\
 NLO$\star$ & $-$0.0928 & 2.125 & $-$0.102 & 0.0243 & 0.344 &
 $-$0.394 & 1.335 & $-$0.191 & $-$0.0357 \\
 NNLO & $-$4.249 & 11.945 & $-$6.508 & 11.293 & $-$2.045 &
 $-$7.061 & $-$2.832 & $-$8.056 & $-$3.424 \\
 NNLO$\star$ & $-$4.246 & 11.943 & $-$6.655 & 11.279 & $-$2.045 &
 $-$7.061 & $-$2.832 & $-$8.056 & $-$3.516 \\
 NNLO-$\Delta$ & $-$5.731 & 13.823 &  $-$0.977 & 3.446 &
 $-$2.494 & $-$8.188 & $-$2.993 & $-$8.431 & $-$0.562 \\
\hline
  \end{tabular}
\caption{The values of the LECs as determined from the low partial
 waves. We use a sharp cut--off with $\Lambda= 500$~MeV and 875~MeV at NLO and NNLO,
 respectively. The $\tilde{C}_i$ are in $10^4$ GeV$^{-2}$ while the others are in
 $10^4$ GeV$^{-4}$. The parameters of the NLO, NNLO and NNLO--$\Delta$ potentials
 are obtained from fitting to the Nijmegen PSA. The LECs of the NLO$\star$, NNLO$\star$ 
 potential in the $^1 S_0$ and $^3 S_1 - ^3 D_1$ channels  are fixed to reproduce exactly the 
 effective range parameters. 
\label{tab:LECs}} 
\end{center}
\end{table}

\noindent To illustrate the
dependence on the cut--off, we show in fig.\ref{figCrun} the running of the
two (three) couplings in the $^1S_0$ ($^3S_1$) channels at NNLO (note that the
third parameter in $^3S_1$ comes in via the mixing with the $^3D_1$ wave).
We notice that the variation of these LECs over a wide range of cut--offs
is rather modest. We also mention that using the $\pi N$ parameters from
refs.{\cite{bkmlec,moj,fms} leads to a considerably worse $\chi^2$ in the fits.
We take that as an indication that the determination of the $c_i$ based on
the method employed in ref.\cite{paul} is more reliable than fitting to
$\pi N$ phase shifts (as long as one works to third order in the chiral
expansion). We remark that using the parameters of ref.\cite{paul},
the deuteron binding energy $E_d$ comes out as
\beqa
{\rm NLO}  &:& E_d = -2.175\,{\rm MeV}~, \no\\  
{\rm NNLO} &:& E_d = -2.208\,{\rm MeV}~,
\eeqa
i.e. the NNLO result is already within 7.5 permille of the experimental number. 
Fine tuning in the parameters in the deuteron channel would
allow to get the binding energy at the exact value of $-2.224575(9)\,$MeV
without leading to any noticeable change in the phase shifts.
We later consider the deuteron channel separately with an exponential
regulator. This will lead to an improved binding energy but no attempt
is made to match the exact value in all digits.
 
\medskip
\noindent At this point we would like to comment on the  increase in the 
cut--off values when going from NLO to NNLO (NNLO--$\Delta$).
Consider first the leading order result. Lepage \cite{Lep} has pointed
out that inclusion of the one--pion exchange does not lead to a remarkable increase 
in the cut--off values compared to a pionless theory. 
In order to fit the phase shifts in the S--waves one should choose the  cut--off  
below 500--600 MeV even if the contact interactions with two derivatives are 
taken into account (within our power counting scheme such contact interactions
contribute first at next--to--leading order and are of the same size as the 
lading two--pion exchange terms). Lepage assumed that such a low value 
of the cut--off is due to the missed physics assosiated with the two--pion exchange.
So, naively, one would expect that the inclusion of the leading two--pion exchange
contributions at NLO would  allow to take larger cut--off values. However,
that does not happen. This was also pointed out in ref.\cite{Park}.
According to our analysis only at NNLO, after the subleading two--pion exchange 
contributions are taken into account, one can increase the cut--off up to 
800~to~1000 MeV. The inclusion of the dimension two operators
of the pion--nucleon interaction at NNLO encodes some information about heavy
meson exchange as well as virtual isobar excitations, 
as discussed in detail in ref.\cite{bkmlec}.
In this work we were able to separate the leading effects of $\Delta$--resonance
(NNLO--$\Delta$).  The clear increase in the cut--off values 
when going from NLO to NNLO--$\Delta$ indicates the importance of physics assosiated
with heavier mass states like e.g. the $\Delta$-resonance. Our
NNLO (NNLO--$\Delta$) TPEP is sensitive to momentum scales sizeably larger than twice the
pion mass (as it would be the case for uncorrelated TPE) and delta--nucleon 
mass splitting. Consequently,
the cut--off has to be chosen safely above these scales, say above
500~MeV (with the sharp regulator). The upper limit of about 1~GeV
is related to the cancellations in the S--waves (fine--tuning), for
too large values of $\Lambda$ it is not longer possible to keep this
intricate balance. It is, however, comforting to see that including
more physics in the potential leads indeed to a wider range of applicability
of the EFT.
 
\medskip
\noindent Finally, we need to discuss one further topic. Performing the
fits, we have found two minima in both the $^1S_0$ and the $^3S_1$
channel. This is not unexpected and can easily be understood in the
case of a pionless theory. In that case, exact analytical calculations
are possible. As was shown in ref.\cite{BCP}, the requirement of
reproducing exactly the scattering length and the effective range
leads to the following conditions for the LECs in the $^1S_0$ partial
wave,
\beqa\label{2sol}
\frac{m}{16\pi^2 a}  &=& \frac{(C_{^1S_0} I_3 -1)^2}{\tilde{C}_{^1S_0} +
C_{^1S_0}^2 I_5} - I_1 \quad, \no\\
\frac{mr}{32\pi^2}  &=& \biggl( \frac{m}{16\pi^2 a} + I_1 \biggr)^2 \,
\frac{C_{^1S_0} (2 - C_{^1S_0} I_3)}{(C_{^1S_0} I_3 -1)^2} + 
\frac{m}{(2\pi)^3 \Lambda}~,
\eeqa
where
\beq
I_n = -m \int_0^\Lambda \frac{d q}{(2\pi)^3} q^{n-1} =
-\frac{m}{(2\pi)^3} \frac{\Lambda^n}{n}~.
\eeq
The second equation in (\ref{2sol}) can be further simplified
if $a \ll 1/\Lambda$. Then, one obtains the following quadratic
equation for $C_{^1S_0}$,
\beq
\frac{m}{32\pi^2} \biggl( r -\frac{4}{\pi \Lambda} \biggr) \simeq
I_1^2 \, \frac{C_{^1S_0} (2 - C_{^1S_0} I_3)}{(C_{^1S_0} I_3 -1)^2}~,
\eeq
which leads to
\beq
C_{^1S_0} = -\frac{24\pi^3}{m\Lambda^3} \biggl( 1\pm
\frac{2\sqrt{3}}{\sqrt{16 -\pi r \Lambda}} \biggr)~.
\eeq
Note that the existence of real solutions for $C_{^1S_0}$ requires
that $\Lambda \le 16/(\pi r) \sim 376\,$MeV. Such a situation with two
solutions appears also in the NLO and NNLO theory with pions. At NLO, we find
very similar predictions for the phase shifts and observables as
well as a very similar quality of the fits in the $^1S_0$ and $^3S_1 -
^3D_1$ channels for both solutions, see the upper panel in
fig.\ref{fig2sol}. So there is no real criterion to prefer one of these
solutions. However, at NNLO, the behaviour of the phase shifts at
higher energies differs quite remarkbaly as it is illustrated in the
lower panel of fig.\ref{fig2sol}. Also, the $\chi^2$ for these two
solutions differs typically by factors of $2...10$. In what follows,
we will only discuss the best solution at NNLO.

\section{Results and discussion}
\def\theequation{\arabic{section}.\arabic{equation}}
\setcounter{equation}{0}
\label{sec:res}

Having determined the parameters, we can now predict the S-- and P--waves
for energies larger than 100~MeV and all other partial waves are parameter
free predictions for all energies considered. 
Since at the laboratory energy of 280~MeV the first inelastic
channel opens, we only plot the  phase shifts up to $E_{\rm lab} = 300\,$MeV. 
We show the results using the sharp regulator function with $\Lambda =
500$ and $875\,$MeV for NLO and NNLO, as discussed in
sec.\ref{sec:fit}. This is a best global fit for {\it all} partial
waves. Using the exponential cut--off, the curves come out very
similar. In fact, for the calculation of some deuteron observables, 
we have to use the exponential regulator. Obviously, the corresponding
cut--offs are somewhat larger than in case of the sharp regulator (to
preserve the area to be integrated). For the NNLO-$\Delta$ approach, 
we only show some selected partial waves in a separate paragraph. 
\subsection{Phase shifts}

\subsubsection{S--waves}

\noindent
In fig.\ref{SW} we show the two S--waves at LO, NLO and NNLO for the
cut--off values given above.  Clearly, the lowest order OPEP plus non--derivative 
contact terms is insufficient to describe the $^1S_0$ phase (as it is well--known from
effective range theory and previous studies in EFT approaches).
The much more smooth $^3S_1$ phase is already fairly well described at leading order. 
For energies above 100~MeV, the improvement by going from NLO to NNLO
is clearly visible. The corresponding values of the S--wave phase shifts at
certain energies are given
in tables~\ref{tab:3S1},\ref{tab:1S0}. For comparison, we also give the
results of the Nijmegen and VPI PSA~\cite{vpipwa} and of three modern
potentials (Nijmegen~93~\cite{Nij93}, Argonne V18~\cite{V18} and 
CD-Bonn~\cite{cdbonn}). Our NNLO result for $^1S_0$ is visibly better
than the one obtained in ref.\cite{RS}.

\medskip\noindent
It is also of interest to consider the scattering lengths and
effective range parameters. The effective range expansion takes
the form (written here for a genuine partial wave)
\beq\label{ERE}
p\, \cot(\delta) = -\frac{1}{a} + \frac{1}{2}\, r \, p^2
+ v_2 \, p^4 + v_3 \,p^6 + v_4 \, p^8 + {\cal O}(p^{10})~,
\eeq
with $p$ the nucleon cms momentum, $a$ the scattering length and
$r$ the effective range. It has been stressed in ref.\cite{CH}
that the shape parameters are a good testing ground for the range
of applicability of the underlying EFT since a fit to say the
scattering length and the effective range at NLO leads to predictions
for the $v_i$.\footnote{We remark in passing that the so--called
  ``low--energy theorems'' discussed in that paper do not qualify
  under this title. For a pedagogical discussion on this point, see
  e.g. ref.\cite{EM}.} In table~\ref{tab:range}, we present our
results for the S--waves in comparison to the ones obtained from the
Nijmegen PSA. Note that in the coupled channel we have used the so--called
Blatt and Biedenharn parametrization of the S--matrix in order to be able
to compare our findings with those of the ref.\cite{swart}.
To show that our results are stable and do not depend 
on the fitting procedure we present the NNLO predictions obtained with 
two different methods of fixing the LECs as discussed above. 
Overall, the agreement is quite satisfactory. The same
holds for the NLO results, but as stated before, these are more
sensitive to the fitting procedure.

\begin{table}[H] 
\begin{center}
\begin{tabular}{|l||c|c|c|c|c|c|c|}
    \hline
 $E_{\rm lab}$ [MeV] & $\,\,$NNLO$\star \,\,$ & $\,\,$NNLO$\,\,$ & Nijm PSA & 
VPI PSA &  Nijm93 & AV18 & CD-Bonn   \\ 
    \hline   \hline  
1$\star$   & 147.735 & 147.727& 147.747 & 147.781 & 147.768 & 147.749 & 147.748  \\
2   & 136.447 & 136.450& 136.463 & 136.488 & 136.495 & 136.465 & 136.463  \\
3   & 128.763 & 128.781& 128.784 & 128.788 & 128.826 & 128.786 & 128.783  \\
5$\star$   & 118.150 & 118.196& 118.178 & 118.129 & 118.240 & 118.182 & 118.175  \\
10$\star$  & 102.56 & 102.67& 102.61  & 102.41  & 102.72    & 102.62  & 102.60   \\
20  & 85.99  & 86.21& 86.12  & 85.67  & 86.35        &  86.16  & 86.09   \\
30  & 75.84  & 76.14& 76.06  & 75.46  & 76.40        & 76.12 & 75.99   \\
50$\star$  & 62.35  & 62.79& 62.77  & 62.12  & 63.36        & 62.89 & 62.63   \\
100$\star$ & 42.33  & 43.06& 43.23  & 42.98  & 44.33        & 43.18 & 42.93   \\
200 & 19.54  & 20.68& 21.22  & 20.88  & 22.82        & 21.31 & 20.88    \\
300 & 4.15 & 5.58 & 6.60  & 5.08   &  8.44        &  7.55 &  6.70   \\
    \hline
  \end{tabular}
\caption{$^3S_1$ $np$ phase shift for the global fit at NNLO (sharp  cut--off, 
$\Lambda = 875\,$MeV) compared
to phase shift analyses and modern potentials. The parameters of the 
NNLO potential are fixed by fitting the Nijmegen PSA at six energies
($E_{\rm lab} =1,5,10,25,50(,100)\,$MeV). These energies are marked by
the star. The parameters of the NNLO$\star$ potential are choosen to 
reproduce exactly the scattering length and the effective range as described 
in the text.
\label{tab:3S1}}
\end{center}
\end{table}

\begin{table}[H] 
\begin{center}
\begin{tabular}{|l||c|c|c|c|c|c|c|}
    \hline
 $E_{\rm lab}$ [MeV] & $\,\,$NNLO$\star \,\,$ & $\,\,$NNLO$\,\,$ & Nijm PSA & 
VPI PSA &  Nijm93 & AV18 & CD-Bonn   \\ 
    \hline   \hline  
1$\star$   & 62.071 & 62.063 & 62.069 & 62.156 & 62.065 & 62.015 & 62.078  \\
2   & 64.472 & 64.469 & 64.573 & 64.573 & 64.460 & 64.388 & 64.478  \\
3   & 64.671 & 64.671 & 64.762 & 64.762 & 64.650 & 64.560 & 64.671  \\
5$\star$   & 63.659 & 63.663 & 63.708 & 63.708 & 63.619 & 63.503 & 63.645  \\
10$\star$  & 60.02 & 60.03  & 59.96  & 60.00  & 59.94  & 59.78  & 59.97   \\
20  & 53.66 & 53.68  & 53.57  & 53.77  & 53.54  & 53.31  & 53.56   \\
30  & 48.55 & 48.58  & 48.49  & 49.00  & 48.42  & 48.16  & 48.43   \\
50$\star$  & 40.49 & 40.54  & 40.54  & 41.66  & 40.38  & 40.09  & 40.37   \\
100$\star$ & 26.30 & 26.38  & 26.78  & 27.86  & 26.17  & 26.02  & 26.26   \\
200 & 7.63 & 7.76   & 8.94   & 7.86   & 7.07   & 8.00   & 8.14    \\
300 & $-$6.41 & $-$6.24& $-$4.46& $-$5.55& $-$7.18& $-$4.54& $-$4.45 \\
    \hline
  \end{tabular}
\caption{$^1S_0$ $np$ phase shift for the best fit at NNLO 
(sharp  cut--off, $\Lambda = 875\,$MeV) compared
to phase shift analyses and modern potentials. The parameters of the 
NNLO potential are fixed by fitting the Nijmegen PSA at six energies
($E_{\rm lab} =1,5,10,25,50,100\,$MeV). These energies are marked by
the star. The parameters of the NNLO$\star$ potential are choosen to 
reproduce exactly the scattering length and the effective range as described 
in the text.
\label{tab:1S0}}
\end{center}
\end{table}

\noindent
We also remark that the momentum
expansion for $\epsilon_1$ discussed in ref.\cite{CH} depends
sensitively on the parametrization one uses for the coupled triplet
waves and we thus refrain from discussing this issue any 
further.\footnote{Note, however, that in order to fix the LECs in the $^3 S_1 - ^3 D_1$
channel we have used the leading coefficient $g_1=1.66$ in the momentum expansion 
of the $\epsilon_1$. This value agrees with the one given in ref.\cite{CH}.
In principle, we could take apart from the $^3 S_1$ scattering length and 
effective range the deuteron binding energy as the third quantity to fix
the three free parameters in the potential. However we refrain from doing that
because of a strong correlation between the deuteron binding energy and
the $^3 S_1$ effective range parameters.} 
\begin{table}[H] 
\begin{center}
\begin{tabular}{|l||c|c|c|c|c|}
    \hline
 & a [fm] & r [fm] & $v_2$ [fm$^3$] & $v_3$ [fm$^5$] & $v_4$ [fm$^7$] \\
    \hline   \hline  
$^1S_0$ NNLO$\star$ & $-$23.739 & 2.68 & $-$0.61 & 5.1 & $-$29.7  \\
$^1S_0$ NNLO        & $-$23.722 & 2.68 & $-$0.61 & 5.1 & $-$29.8  \\
$^1S_0$ NPSA        & $-$23.739 & 2.68 & $-$0.48 & 4.0 & $-$20.0  \\
\hline
$^3S_1$ NNLO$\star$ &   5.420 & 1.753 & 0.06 & 0.66  & $-$3.8  \\
$^3S_1$ NNLO        &   5.424 & 1.741 & 0.05 & 0.67  & $-$3.9  \\
$^3S_1$ NPSA        &   5.420 & 1.753 & 0.04 & 0.67  & $-$4.0  \\
    \hline
  \end{tabular}
\caption{Scattering lengths and range parameters for the S--waves at
 NNLO  (global fit) compared to the Nijmegen PSA (NPSA). The values
 for $v_{2,3,4}$ in the $^1S_0$ channel are based on the $np$ Nijm~II 
 potential and the values of the scattering length and the effective range 
 are taken from the ref.\cite{rentm}. The effective 
 range parameters for the $^3 S_1 - ^3 D_1$ channel are discussed in \cite{swart}.
\label{tab:range}}
\end{center}
\end{table}

\noindent
For completeness, we collect the experimental values for the
S--wave scattering lengths and efffective ranges:
\beqa 
^1S_0 &:&  a_s = (-23.758\pm 0.010)~{\rm fm} \, , 
\quad r_s = (2.75\pm 0.05)~{\rm fm}~,
\\
^3S_1 &:&  a_t = (5.424\pm 0.004)~{\rm fm} \, , 
\quad \quad r_t = (1.759\pm 0.005)~{\rm fm}~.
\eeqa

\subsubsection{P--waves} 
In fig.\ref{PW} we show the corresponding partial waves together with 
the mixing parameter $\epsilon_1$ 
for the best global fit. In some cases, the differences between NLO and NNLO
are modest, in $^1P_1$ and $^3P_1$ NLO is even somewhat better. That means 
that the chiral TPEP is too strong in these phases. Note also that in
the $^3P_1$ phase OPEP is dominant. Thus, the inclusion of 
the contact interaction does not lead to a visible change. In $^3P_2$, NNLO
is still too strong but the prediction is considerably better than the NLO one.
The energy dependence of $\epsilon_1$ is fairly precisely described at NLO and
NNLO. These results are visibly better than the ones obtained in ref.\cite{ubi} 
or in ref.\cite{calt}, the latter being a NNLO calculation in the KSW scheme.
This is shown in detail in fig.\ref{eps1}, where $\epsilon_1$ is plotted versus the
cms nucleon momentum ($p < 350$~MeV) in comparison to the Nimegen PSA and the results from
refs.\cite{calt}.\footnote{We are grateful to Iain Stewart for
  supplying us with the KSW NLO and NNLO results for larger momenta than
  given in refs.\cite{KSW,calt}.} In the KSW approach, the pions are treated
perturbatively. From these results we conclude that in the two--nucleon system,
pions have to be treated non--perturbatively (if one intends to describe data
above $p \simeq 150\,$MeV).

\subsubsection{D-- and F--waves} 
\label{sec:DF}
These are the most problematic phases in our approach. It is known that
TPEP alone is too strong in some of these partial waves. To the order
we are working, there are no contact interactions so that apart from
varying the cut--off we have no freedom here. Still, for our best global
fit, the D--waves are quite reasonably described, see fig.\ref{DW}, in particular $^3D_1$
and $^3D_3$. The result for the latter phase is rather astonishing, since e.g.
in the Bonn potential~\cite{bonn} the correlated two--pion exchange is
very important in the description of  $^3D_3$. Such $\pi\pi$ correlations are only
implicitely present to the order we are working. They are hidden in the strengths 
of the LECs $c_1$ and $c_3$ as explained in ref.\cite{bkmlec}. The very accurate
(and parameter-free) description of the $^3D_1$ wave is of course important
for the deuteron channel to be discussed in section~\ref{sec:deut}. Even in $^1D_2$,
we get a fair description at NNLO. This partial
wave is generally believed to be very sensitive 
to the $\Delta$--resonance. We come back
to this point when we discuss the NNLO-$\Delta$ approach. 
Note that the D--waves are rather sensitive to the choice of the
regulator cut--off, see the upper panel in fig.\ref{fig:DL}.
At N$^3$LO we have besides new pion--exchange terms the first contact
interactions that feed into the D--waves,
\beq
\langle ^m{D}_n | V_{\rm cont}^{\rm {N}^3{\rm LO}} | ^m{D}_n \rangle
\sim p^2 \, {p'}^2~.
\eeq
More precisely, we have one independent parameter in each D--wave. Thus
the cut--off independence will be restored by the running of the
corresponding LECs. To illustrate this, we show in the lower panel
of fig.\ref{fig:DL} the partial N$^3$LO results for the $^1D_2$
channel. The corresponding potential  consists of the NNLO terms plus
one  N$^3$LO contact interaction. As expected, the cut--off dependence
of the phase shift is very much reduced compared to the NNLO result. 
Of course, this illustrative example can not substitute for a complete
N$^3$LO calculation, but one should expect very similar results.
Note that according to our findings, the NNLO potential in the D--wave 
channels is not weak enough to be treated perturbatively, as it has been
done in ref.\cite{norb}. The potential has to be iterated to all orders
in the LS equation. Only then one obtains a reasonable description of the 
phase shifts in these partial waves.
Concerning the F--waves, which are shown in fig.\ref{FW},
$^1F_3$ and $\epsilon_3$ are well described, whereas the NNLO
TPEP is visibly too strong in $^3F_2$, $^3F_3$, and $^3F_4$. This can be cured
at higher orders by contact interactions. More precisely, a N$^3$LO calculation
should be sufficient.

\subsubsection{Peripheral waves} 
In figs.\ref{GW},~\ref{HW}~and~\ref{IW} we show the G--, H-- and
I--waves together with  the
mixing parameters $\epsilon_{4,5,6}$. These partial waves were first
discussed in detail by the Munich group~\cite{norb}.
Their calculation was perturbative and based on dimensional regularization
of the TPE graphs. However, for these partial waves the iteration becomes
unimportant and our findings confirm their results. The description of
$^1G_4$, $^3G_3$, $^3G_5$, $^3H_5$, $^3H_6$,  $^3I_5$ and $^3I_7$ is visibly improved
by the NNLO TPEP. Only in $^1I_6$ the NLO result is better than the NNLO one.
Of course, for the peripheral partial waves OPE does already a fairly
good job, but the improvement in some of these phases due to the NNLO
TPEP clearly underlines the importance of chiral symmetry in a precise
description of low--energy nuclear physics.

\subsection{Deuteron properties}
\label{sec:deut}
We now turn to the bound state properties. At NNLO (NLO), we consider an exponential
regulator with $\Lambda =1.05\,(0.60)\,$GeV, which reproduces the deuteron
binding energy within an accuracy of about one third of a permille
(2.5 percent).
We make no attempt to reproduce this number with better precision.\footnote{Note
that the deuteron binding energy is {\it not} used to fit the free
parameters in the potential.}
The results for the phase shifts, which correspond to these values of 
the exponential regulator, are very similar to those obtained with the 
sharp cutoff  $\Lambda =0.875\,(0.50)\,$GeV.
For completeness, we list in table~\ref{coupl:exp} the values of the coupling constants 
in the $^3 S_1 - ^3 D_1$ channel corresponding to the exponential regulator.
\begin{table}[H] 
\begin{center}
\begin{tabular}{||l||c|c|c||}
  \hline 
 & $\tilde{C}_{^3S_1}$ & ${C}_{^3S_1}$ & $C_{^3D_1 -^3S_1}$\\ \hline\hline  
NLO & $-$0.0363 & 0.186 & $-$0.190 \\
NNLO & $-$14.497 & 15.588 & $-$4.358 \\
NNLO--$\Delta$ & $-$8.637 & 7.264 & $-$0.447 \\
\hline
  \end{tabular}
\caption{The values of the LECs as determined from the $^3 S_1 - ^3 D_1$ channel.
 We use an exponential cut--off with $\Lambda= 0.6$~GeV and 1.05~GeV
 at NLO and NNLO (NNLO--$\Delta$),
 respectively. The LEC $\tilde{C}_{^3 S_1}$ is in $10^4$ GeV$^{-2}$,
 while the others are in
 $10^4$ GeV$^{-4}$. The parameters of the NLO, NNLO and NNLO--$\Delta$ potentials
 are obtained from fitting to the Nijmegen PSA. 
\label{coupl:exp}} 
\end{center}
\end{table}

\medskip
\noindent
In table~\ref{tab:D} we collect the deuteron properties in comparison to the
data and two realistic potential model predictions (the pertinent 
formulae are given in app.~\ref{app:deut}). We give the results for NLO  
and NNLO. We note that deviation of our prediction for the quadrupole
moment compared to the empirical value slightly larger than for the
realistic potentials. The asymptotic $D/S$ ratio,
called $\eta$, and the strength of the asymptotic wave function, $A_S$, are well
described. The D--state probability, which is not an observable, is
most sensitive to small variations in the cut--off. At NLO, it is comparable
and at NNLO somewhat larger than obtained in the CD-Bonn or the
Nijmegen-93 potential. This increased value of $P_D$ is related to the
strong NNLO TPEP. At N$^3$LO, we expect this to be compensated by
dimension four counterterms.
Altogether, 
we find a much improved
description of the deuteron as compared to ref.\cite{ubi}. Our results are almost as
precise as the ones obtained in the much more complicated and less systematic
meson--exchange models.
\renewcommand{\arraystretch}{1.2}
\begin{table}[H] 
\begin{center}

\begin{tabular}{||l||c|c|c||c|c||c||}
    \hline
    & NLO  &  NNLO  & NNLO--$\Delta$ & Nijm93 & CD-Bonn & Exp.    \\ 
    \hline   \hline  
$E_d$ [MeV] & $-$2.1650  & $-$2.2238 & $-$2.1849 & $-$2.224575 
            & $-$2.224575 & $-$2.224575(9) \\
    \hline
$Q_d$ [fm$^2$] & 0.266  & 0.262 & 0.268 & 0.271 & 0.270 & 0.2859(3) \\
    \hline
$\eta$ & 0.0248  & 0.0245 & 0.0247 & 0.0252 & 0.0255 & 0.0256(4)\\
    \hline
$r_d$ [fm] & 1.975  & 1.967 & 1.970 & 1.968 & 1.966 & 1.9671(6)\\
    \hline
$A_S$ [fm$^{-1/2}$] & 0.866  & 0.884 & 0.873 & 0.8845 & 0.8845 & 0.8846(16)\\
    \hline
$P_D [\%]$ & 3.62  & 6.11 & 5.00 & 5.76 & 4.83 &  -- \\
    \hline
  \end{tabular}
\caption{Deuteron properties derived from our chiral potential
    compared to two ``realistic'' potentials (Nijmegen--93 and 
    CD--Bonn) and the data. Here, $r_d$ is the
    root--mean--square matter radius. An exponential regulator
    with $\Lambda = 600\,$MeV and $\Lambda = 1.05\,$GeV 
    at NLO and NNLO (NNLO--$\Delta$), in order, is used. \label{tab:D}}
\end{center}
\end{table}

\noindent
The coordinate space S-- and D--state wave functions obtained in our approach are
shown in fig.\ref{figdwf}. At NLO they look quite similar to the 
ones obtained from various potential models. At NNLO one obtains a lot 
of structure in the wave functions below 2 fm. This is because 
two additional spurious (unphysical) very deeply bound states appear 
in the $^3 S_1 - ^3 D_1$ channel. The binding energy of these 
states varies strongly by changing the cut--off. For the exponential regulator with 
$\Lambda = 1.05\,$GeV
we get binding energies of $E_1=47.1\,$GeV and $E_2=2.5 \,$GeV, respectively.
This values correspond to center--of--mass momenta of about a  few GeV
which is clearly out of the applicability range of the low--momentum   
effective theory. Further, because of such huge values of the binding
energy these unphysical states obviously do not influence physics
in the energy region below 350 MeV that we are interested in and 
can, in  principle, be integrated out.\footnote{As soon as one
is dealing with only the two--nucleon system there is no need to 
integrate out such unphysical states, since this does 
not modify the low--energy observables. We refrain here from the 
discussion of the complications which may arise in three-- and 
more--body calculations due to such spurious states. Note, however, that
according to our power counting one has an additional contact 
three--body force at NNLO which possibly can compensate the effects
of the spurious states.} 
A very similar situation with the spurious states happens at NNLO also 
in other S--, P-- and D--waves.
Note that these unphysical states are  purely short range effects:
as one can see from the fig.\ref{figdeep} the wave--functions corresponding
to such states become negligeble for distances above 2 fm. The  corresponding 
root--mean--square matter radii are $\langle r_1^2 \rangle^{1/2} =0.27 \,$fm  and 
$\langle r_2^2 \rangle^{1/2} =0.40 \,$fm.
For separations above 2 fm the NNLO deuteron wave--function is very close
to the one obtained with the CD-Bonn potential.\footnote{We would like to thank 
Hiroyuki Kamada for supplying us with the deuteron wave--function calculated
with the CD-Bonn potential.}
This is shown in fig.\ref{cdbonn}.  For a discussion on such deeply
bound states in effective field theories, see ref.\cite{Lep}.
We end this paragraph with the following remark: According
to Levinson's theorem, the difference between the phase shift at the
origin and at infinity is given by $n\pi$, with $n$ the number of
bound states. Thus, the phase shifts in  the S--, P-- and
D--waves should become unphysical at large energies. This is, however,
of no relevance for the EFT since we do not attempt to correctly
reproduce (or predict) the phase shift behaviour for all energies
(from threshold to infinity).

\subsection{Results for the NNLO-$\Delta$ approach} 

As already stated before, our NNLO-$\Delta$ approach is not complete
since we are omitting the dimension two pion--nucleon interactions not
generated by the isobar. While the LECs $c_3$ and $c_4$ are dominated
by the $\Delta$, see ref.~\cite{bkmlec}, most of the correlated two--pion
exchange is parametrized in $c_1$. However, we are mostly interested in
investigating the role of the $\Delta$ in all partial waves and
therefore keep the cut--off $\Lambda$ fixed at 875~MeV but refit
the LECs $C_i$ (see table~\ref{tab:LECs} for their values). More
precisely, a best global fit leads to a very similar cut--off value as
in NNLO (this is expected due to the cut--off sensitivity of
  the D--waves) but for better comparison we discuss here the results
obtained with exactly the same value for $\Lambda$. Note also that the
precision of the fits is better than NLO in the theory without
$\Delta$ but somewhat worse than the corresponding NNLO fits. This is
due to the absence of contributions from higher mass states encoded
in the LECs $c_{1,3,4}$ not present in the NNLO--$\Delta$ approach 
discussed here. It goes beyond
the scope of this paper to systematically include the effects of this
important resonance in the framework of the EFT expansion
as detailed in ref.\cite{HHK}. In fact, a study of pion--nucleon
scattering in that framework is not yet available and thus the
corresponding LECs are not determined. Formally, we follow the Munich
group~\cite{norb2} (for details, see app.\ref{sec:delta}). Again, it
is important to stress that we iterate our potential to all orders.

\medskip

\noindent
Let us now discuss the results of the NNLO--$\Delta$ approach. We
refrain from showing all partial waves but rather discuss some
particular examples, collected in fig.\ref{fig:delta}. The two
S--waves shown in that figure are not very different from the NNLO
result, although the description of $^3S_1$ is slighly worse at higher
energies. All P--waves are very similar in NNLO and NNLO--$\Delta$,
the most visible difference appears in $\epsilon_1$, as can be seen
in the figure. The most dramatic effects appear in the D--waves. This
is expected since these are parameter--free predictions and we had 
already pointed out the cut--off sensitivity in sec.~\ref{sec:DF}.
Interestingly, the description of $^1D_2$ is almost identical in the
two approaches, consequently any important isobar effect in this
partial wave can be well represented by contact interactions with
their strength given by the coupling of the $\Delta$ to the $\pi N$
system. In $^3D_3$ (also shown in fig.\ref{fig:delta}) the absence of
the scalar--isoscalar two--pion correlations is clearly visible. Our
result thus confirms a finding made already in the Bonn potential,
namely that this particular partial wave is essential dominated by
correlated TPE. We also note that the description of $^3D_1$ and
$^3D_2$ is worse in NNLO--$\Delta$ than in NNLO.
For the F--waves, the differences between the two approaches are very
small, with the exception of $^3F_4$, which is improved in the presence of
the isobar. The most significant effect in the higher partial waves
shows up in $^3G_5$ as shown in fig.\ref{fig:delta}. Clearly,
two--pion correlations not related to the $\Delta$ play an important
role to bring the prediction for this partial wave in agreement with
the data. The deuteron properties for an exponential regulator with
$\Lambda =1.05\,$GeV are listed in table~\ref{tab:D}. Most observables
come out in between the NLO and NNLO results for the deltaless
approach. The sole  exception is the quadrupole moment, which is improved.
We remark again that only after a fully systematic inclusion of the $\Delta$, one
should draw quantitative conclusions. It appears, however, that simply
using resonance saturation to encode the phyiscs of the isobar in the
dimension two pion--nulceon LECs is a sufficient procedure in the
two--nucleon sector. A truely quantitative inclusion of the delta
should be done in the framework of the ``small scale
expansion''~\cite{HHK}. In that case, the leading delta effects would
already appear at NLO since the nucleon--delta mass splitting is
counted as a small parameter. Such an investigation is, however,
beyond the scope of this article.
We also note that explicit isobar degrees of freedom were
considered in ref.\cite{ubi}. A direct comparison with their work is,
however, not possible since no separate result for fits with and
without delta are given.

\section{Summary}
\def\theequation{\arabic{section}.\arabic{equation}}
\setcounter{equation}{0}
\label{sec:summ}

In this paper, we have calculated properties of the two--nucleon
system based on a chiral effective field theory. The underlying
formalism was already presented in~I. The results of this
investigation can be summarized as follows:

\begin{enumerate}
\item[1)] Based on a modified Weinberg power counting (as explained
  in~I), we have constructed a chiral two--nucleon potential at 
  NNLO. It consists of one-- and two--pion exchange diagrams,
  including dimension two insertions from the effective pion--nucleon
  Lagrangian. The corresponding LECs have been taken from an
  investigation of $\pi N$ scattering~\cite{paul}. In addition, there
  are two and seven four--nucleon contact interactions at LO and NLO,
  respectively. The coupling constants of these terms must be fixed by
  a fit to data. 
\item[2)] For large momenta, the potential becomes unphysical and has
   to be regularized. We perform this regularization on the level of the
   Lippmann--Schwinger equation, as explained in sect.\ref{sec:pot}
   using either a sharp or an exponential regulator function. At NLO,
   physics does not depend on the cut--off in the range between 400
   and 650~MeV. At NNLO, this range is larger and extends from 650 to
   1000~MeV. This can be understood from the chiral TPEP, which at
   NNLO includes $\pi\pi$ correlations. These introduce a new mass
   scale well above twice the pion mass.
\item[3)] We have shown that the contact interactions can be combined in such
  a way that each combination feeds into one partial wave. More
  precisely, the nine four--nucleon couplings can be determined
  uniquely by a fit to the two S--waves, four P--waves and the mixing parameter
  $\epsilon_1$ for nucleon laboratory energies below 100~MeV. As
  expected from the power counting underlying the EFT, the fits
  improve when going from LO to NLO to NNLO, compare fig.\ref{figfit}.  
\item[4)] At NNLO, the resulting S--waves are of very high precision
  (for nucleon laboratory energies below 300~MeV),
  see e.g. tables~\ref{tab:3S1},\ref{tab:1S0} and fig.\ref{SW}. 
  The so--called range parameters collected in table~\ref{tab:range}
  agree with what is found in the phase shift analysis. The P--waves
  are mostly well described, in particular the mixing parameter
  $\epsilon_1$ is in good agreement with the phase shift analysis. 
  We also note that above nucleon cms momenta of about 150~MeV, our
  NLO and NNLO results are far better than the one obtained in the
  KSW scheme at NLO and NNLO.
\item[5)] All other partial waves are free of parameters. The
  D--waves, in particular $^3D_1$ and $^3D_3$ are very well described.
  We have also discussed the cut--off sensitivity of these results.
  The NNLO TPEP is too strong in the triplet F--waves. This is
  expected to be
  cured at N$^3$LO due to the appearance of dimension four contact
  terms. For the peripheral waves, we recover the results of the
  Munich group~\cite{norb}, namely that in most cases OPE works well
  but chiral NNLO TPEP clearly improves the description of some
  partial waves like e.g. $^3G_5$, $^3H_5$ or $^3I_7$.
\item[6)] The deuteron properties are mostly well described, at NLO
  and NNLO, compare table~\ref{tab:D}. At NNLO, the deuteron wave
  functions shows some interesting structure due to the appearance of
  two very deeply bound states. These are an artefact of the NNLO
  approximation. They have no influence on low energy properties and 
  can be completely projected out  from the theory. Our precise
  deuteron wavefunctions can be used for pion photoproduction,
  pion--deuteron scattering or Compton scattering off deuterium
  (still, the hybrid approach proposed by Weinberg~\cite{weind} 
  remains a useful tool).
\item[7)] We have also considered an approach with explicit $\Delta$
  degrees of freedom in the TPEP. This NNLO--$\Delta$ approach is
  very similar to the NNLO results in the theory without isobars, with
  the exception of the partial waves that are sensitive to pionic
  scalar--isoscalar correlations like e.g. $^3D_3$. We conclude that
  the inclusion of the $\Delta$ via resonance saturation of the
  dimension two $\pi N$ LECs captures the essential physics of the
  isobar in the two--nucleon system.  We note, however, that a more
  systematic study of pion--nucleon scattering in an EFT including the
  $\Delta$ is needed to further quantify these statements.
\end{enumerate}
Our findings do not only show that the scheme originally proposed by
Weinberg works quantitatively, it even works much better than it was
expected. It extends the succesfull applications of effective field
theory (chiral perturbation
theory) in the pion and pion--nucleon sectors to systems with more than
one nucleon. Clearly, one should now reconsider processes, which have
been evaluated using Weinberg's hybrid approach~\cite{weind} ($\pi-d$ 
scattering~\cite{weind,bblm}, $\gamma d \to \pi^0 d$~\cite{bblmv}, $\gamma d
\to \gamma d$~\cite{bpmv}) and extend these considerations to systems
with more than two nucleons. First steps in that direction within the 
potential approach can be found in recent works~\cite{H1,H2}.
In addition, a fresh look at charge symmetry
and charge independence breaking is called for (for earlier studies, see
e.g. refs.\cite{birai,birar,em}). Work along these lines is underway.
Furthermore, the precise relation to the KSW scheme, which has been shown
to be successfull at low energies, has to be worked out.

\bigskip

\section*{Acknowledgements}

We are grateful to D.~Bugg, J.~Haidenbauer, N.~Kaiser, H.~Kamada, R.~Machleidt,
M.~Rentmeester, I.~Stewart, and V.~Stoks for useful comments and communications.

\bigskip

\appendix
\def\theequation{\Alph{section}.\arabic{equation}}
\setcounter{equation}{0}
\section{The renormalization procedure}
\label{app:ren}

In this appendix we spell out some details about the renormalization
procedure mentioned briefly in sect.\ref{sec:pot}. We will use the
divergent loop integrals $J_{0,2,4}$ defined in eq.(\ref{lfcts}) to
express all divergences and remove these by proper subtractions and
redefinitions of physical quantities.

\medskip
\noindent We consider first
the one--loop corrections to the OPEP. These were given in eq.(4.33)
of~I. We rewrite this expression here using different conventions for
the isospin operators and the pion decay constant
\beqa
\label{OPEPloop}
V^{(2)}_{1 \pi,\, {\rm 1-loop} } &=&  
\frac{g_A^4}{(2 f_\pi)^4} \,( \vec \tau_1 \cdot \vec \tau_2) \, 
\frac{1}{\omega_q^2} \,  \int \,\frac{d^3 l}{(2 \pi )^3} \,
\frac{1}{\omega_l^3} \, \Bigg\{ ( \vec l \cdot \vec q \, )  
\Big( ( \vec \sigma_1
\cdot \vec l \, ) \, ( \vec \sigma_2 \cdot \vec q \, ) 
+ ( \vec \sigma_1 \cdot \vec q \, ) 
\, ( \vec \sigma_2 \cdot \vec l \, ) 
\Big)  \nonumber \\
&& \qquad \qquad  \qquad \qquad  \qquad \qquad \qquad \qquad
 +   2 {\vec l \,}^2 \, ( \vec \sigma_1 \cdot \vec q \, ) \,
( \vec \sigma_2 \cdot \vec q \, ) \Bigg\} \,\,\, ,\no  \\
&=& \frac{g_A^4}{(2 f_\pi)^4} \, ( \vec \tau_1 \cdot \vec \tau_2) \,
 \frac{8}{3} \, ( \vec \sigma_1 \cdot \vec q \, ) \,
 ( \vec \sigma_2 \cdot \vec q \, ) \, \frac{1}{\omega_q^2}\,
\int \,\frac{d^3 l}{(2 \pi )^3}
 \, \frac{l^2}{\omega_l^3} \\
&=& \frac{g_A^4}{48\pi^2 f_\pi^4} \, ( \vec \tau_1 \cdot \vec \tau_2) \, 
( \vec \sigma_1 \cdot \vec q \, ) \, ( \vec \sigma_2 \cdot \vec q \, )
\, \frac{1}{\omega_q^2}\, \biggl\{  5\Mpz + 3\Mpz
 \ln \frac{\Mpz}{4\epsilon^2}  + 4J_2 - 6\Mpz J_0\biggr\}~, \no
\eeqa
in terms of the two divergent loop functions $J_{0,2}$ defined in eq.(\ref{lfcts}).
Therefore, this contribution has exactly the same form as the OPEP
(renormalized OPE),
\beq
V_{1\pi}^{(0)} + V_{1\pi}^{(2)} =  V_{1\pi}^{\rm r}~,
\eeq
provided we redefine the coupling constant $g_A^0$ (the superscript
``0'' denotes the leading term in the chiral expansion) in the following way:
\beq\label{gAr}
(g_A^r)^2 = (g_A^0)^2 - \frac{(g_A^0)^4}{12\pi^2 f_\pi^2} \biggl\{  5\Mpz + 3\Mpz
 \ln \frac{\Mpz}{4\epsilon^2}  + 4J_2 - 6\Mpz J_0\biggr\}~.
\eeq
Clearly, $g_A^r$ and $g_A^0$ differ by terms of second order in the
chiral dimension. Consequently, {\it all} NLO one--loop corrections
to OPEP can be taken care off by renormalization of $g_A^0$.

\medskip
\noindent In addition, there are the one--loop corrections to the
lowest order four--fermion interactions, cf. eq.(4.34) of~I.
Performing spin averaging, the corresponding contribution can be 
expressed in terms of the divergent loop functions $J_{0,2}$ as
\beqa
V_{{\rm NN, 1-loop}}^{(2)} &=& -\frac{g_A^2}{6\pi^2f_\pi^2} \, C_T^0 \,
(3-\vec \tau_1 \cdot \vec \tau_2 ) \, (\vec \sigma_1 \cdot \vec \sigma_2
)\, \int \, d^3 l  \, l^4 \, \omega_l^{-3} \no\\
&=& -\frac{g_A^2}{24\pi^2f_\pi^2} \, C_T^0 \,
(3-\vec \tau_1 \cdot \vec \tau_2 ) \, (\vec \sigma_1 \cdot \vec \sigma_2
)\,  \biggl\{ 5\Mpz + 3\Mpz \ln \frac{\Mpz}{4\epsilon^2}  + 4J_2 - 6\Mpz
 J_0\biggr\}~.\no\\ &&
\eeqa
or in a more compact notation
\beq\label{VNN1}
V_{{\rm NN, 1-loop}}^{(2)} = -3S \,  (\vec \sigma_1 \cdot \vec
\sigma_2 ) + S\,  (\vec \sigma_1 \cdot \vec \sigma_2 ) 
(\vec \tau_1 \cdot \vec \tau_2 )~,
\eeq
with 
\beq
S = \frac{g_A^2}{24\pi^2 f_\pi^2} \, C_T^0 \, \biggl\{  5\Mpz + 3\Mpz
 \ln \frac{\Mpz}{4\epsilon^2}  + 4J_2 - 6\Mpz J_0\biggr\}~.
\eeq
Antisymmetrization allows to map the two spin--isospin operators
appearing in eq.(\ref{VNN1}) onto the two non--derivative operators
used in $V^{(0)}_{NN}$, cf. eq.(\ref{Vcon}).
Therefore, the effect of the one--loop corrections to the lowest order
four--nucleon contact interactions can be completely absorbed by
renormalizing the constants $C_S^0$ and $C_T^0$,
\beq\label{Cr1}
C_S^r = C_S^0 - 3S~, \quad C_T^r = C_T^0 - 3S~.
\eeq
We note that further renormalization of these couplings is due to the
two--pion exchange, as dicussed below. 

\medskip \noindent
A similar procedure also works for the one--loop TPE graphs, cf. eq.(4.30) of~I.
In fact, the NLO TPEP renormalizes the various dimension zero and two
coupling constants $C_i$. So as not to repeat the argument, we simply
give the relevant momentum space integrals in terms of $J_{0,2}$ and $J_4$,
\beqa
\int \,\frac{d^3 l}{(2 \pi )^3} \, \frac{1}{\omega_+ \omega_-
  (\omega_+ + \omega_-)} &=& -\frac{1}{8\pi^2} \, \biggl\{ \frac{s}{q}
\ln \frac{s+q}{s-q} + \ln \frac{\Mpz}{\epsilon^2} - 2J_0 \biggr\}~,\\
\int \,\frac{d^3 l}{(2 \pi )^3} \, \frac{l^2}{\omega_+ \omega_-
  (\omega_+ + \omega_-)} &=& \frac{1}{24\pi^2} \, \biggl\{ \frac{2s^3}{q}
\ln \frac{s+q}{s-q} + (14\Mpz-q^2) + 2(9\Mpz+q^2)\ln
\frac{\Mpz}{\epsilon^2} \no\\
&& \quad\qquad\qquad + 6J_2 - 4(9\Mpz+q^2) J_0 \biggr\}~,\\
\int \,\frac{d^3 l}{(2 \pi )^3} \, \frac{l^4 + q^4}{\omega_+ \omega_-
  (\omega_+ + \omega_-)} &=& \frac{1}{24\pi^2} \, \biggl\{
\frac{1}{5qs}(512M_\pi^6 + 384M_\pi^4 q^2 + 156\Mpz q^4 +23q^6) \ln
\frac{s+q}{s-q} \no\\
&+&\frac{1}{10} (-898M_\pi^4+192\Mpz q^2 + q^4) 
-\frac{1}{5} (450M_\pi^4+50\Mpz q^2 + 23q^4)\ln\frac{\Mpz}{\epsilon^2} \no\\
&+& 6J_4 -4(9\Mpz+q^2) J_2 + \frac{2}{5}(450M_\pi^4 + 50\Mpz q^2
  +23 q^4) J_0 \biggr\}~,\no \\ && \\  
\int \,\frac{d^3 l}{(2 \pi )^3} \, \frac{(q\cdot l)^2}{\omega_+ \omega_-
  (\omega_+ + \omega_-)} &=& \frac{1}{8\pi^2} \, \biggl\{ 
2\Mpz q^2  \ln \frac{\Mpz}{\epsilon^2} - \frac{q^4}{5} + \frac{10}{3}
  q^2 \Mpz + \frac{2}{3} q^2 J_2 - 4\Mpz q^2 J_0 \biggr\}~,\no\\ && \\
\int \,\frac{d^3 l}{(2 \pi )^3} \, \frac{(\omega_+ - \omega_-)^2}{\omega_+ \omega_-
  (\omega_+ + \omega_-)} &=& \frac{1}{6\pi^2} \, \biggl\{ -\frac{s^3}{q}
\ln \frac{s+q}{s-q} -q^2 \ln \frac{\Mpz}{\epsilon^2} +8\Mpz + 2q^2J_0
  \biggr\}~,
\eeqa
with $s= \sqrt{4\Mpz +q^2}$ and $\omega_\pm$ given in eq.(4.24) of~I. All other
integrals appearing in the 1--loop NLO TPEP can be deduced from these
  expressions by taking proper linear combinations or differentiation
with respect to $\Mpz$. Putting pieces together, the expression
for $V^{(2)}_{2\pi, {\rm 1-loop}}$ takes the form
\beq\label{V2p}
V^{(2)}_{2\pi, {\rm 1-loop}} = V^{\rm TPEP}_{\rm NLO} + (S_1 + S_2 \,q^2) \, (\vec
  \tau_1 \cdot  \vec \tau_2 ) + S_3 \, \biggl[ ( \vec \sigma_1 \cdot \vec q \, ) \,
 ( \vec \sigma_2 \cdot \vec q \, ) - ( \vec \sigma_1 \cdot\vec
 \sigma_2 ) \, q^2 \biggr]~,
\eeq 
with
\beqa 
S_1 &=& \frac{1}{384 \pi^2 f_\pi^4} \biggl\{ -18\Mpz (5g_A^4 -2g_A^2)
\ln \frac{\Mp}{\varepsilon} - \Mpz (61g_A^4-14g_A^2+4) \no\\
&& \qquad\qquad\qquad \qquad\qquad\qquad + 18\Mpz (5g_A^4 -2g_A^2) J_0 - 3(3g_A^4
-2g_A^2) J_2 \biggr\}~,\\
S_2 &=& \frac{1}{384 \pi^2 f_\pi^4} \biggl\{ (-23g_A^4 +10g_A^2 +1)
\ln \frac{\Mp}{\varepsilon} -  (13g_A^4 + 2g_A^2) 
+ (23g_A^4 -10g_A^2-1) J_0 \biggr\},\\
S_3 &=& -\frac{3g_A^4}{64 \pi^2 f_\pi^4} \biggl\{ \ln
\frac{\Mp}{\varepsilon} + \frac{1}{3} + J_0 \biggr\}~,
\eeqa
and the non--polynomial part $V^{\rm TPEP}_{\rm NLO}$ is given in
eq.(\ref{TPEP}). The polynomial terms clearly renormalize the
coupling constants of the dimension zero and two four--nucleon contact
terms. Again, we perform antisymmetrization to map the terms appearing
in eq.(\ref{V2p}) onto the basis used in $V_{NN}^{(0,2)}$,
cf. eq.(\ref{Vcon}). Including the contribution from eq.(\ref{Cr1}),
the complete renormalization of the four--nucleon couplings takes the
form
\beqa
C_T^r &=& C_T^0 - 3S - S_1~, \quad C_S^r = C_S^0 - 3S - 2S_1~, \no\\
C_1^r &=& C_1^0 - S_2~, \quad  C_2^r = C_2^0 -4 S_2~, 
\quad C_3^r = C_3^0 - S_3~,\no\\
C_4^r &=& C_4^0 - 4S_2~, \quad  C_5^r = C_5^0~, \quad C_6^r = C_6^0 +
S_3~,\quad C_7^r = C_7^0~.
\eeqa
Note that $C_5$ and $C_7$ do not get renormalized to this order.
All coupling constants appearing in the main text are understood as
the renormalized quantities discussed here.

\def\theequation{\Alph{section}.\arabic{equation}}
\setcounter{equation}{0}
\section{Inclusion of the $\Delta (1232)$}
\label{sec:delta}
The inclusion of the $\Delta (1232)$ in the TPEP has been worked out
in ref.\cite{norb2}. For completeness, we collect here the pertinent
formulae. There are three distint contributions.

\medskip

\noindent $\Delta$--excitation in the triangle graphs:
\beq
V^{\rm TPEP}_{\Delta, \, {\rm triangle}} = -\frac{g_A^2}{192 \pi^2 f_\pi^4}\,
(\vec \tau_1 \cdot \vec \tau_2) \, \biggl\{ (6E-s^2) L(q) +
12 \Delta^2 E D(q) \biggr\}~,
\eeq
with
\beqa
L(q) &=& \frac{s}{q} \ln \frac{s+q}{2M_\pi}~, \\
D(q) &=& \frac{1}{\Delta} \, \int_{2M_\pi}^\infty
\frac{d\mu}{\mu^2+q^2} \, \arctan \frac{\sqrt{\mu^2-4M_\pi^2}}{2\Delta}~,\\
\Delta &=& m_\Delta - m = 293~{\rm MeV}~,\\
E &=& 2M_\pi^2 + q^2 -2\Delta^2~.
\eeqa 

\medskip

\noindent Single $\Delta$--excitation in the box graphs:
\beqa
V^{\rm TPEP}_{\Delta, \, {\rm box}-1} &=& -\frac{3g_A^4}{32 \pi
  f_\pi^4 \Delta}\, (2M_\pi^2 +q^2)^2 \,  A(q)\no\\
&-& \frac{g_A^4}{192 \pi^2 f_\pi^4}\, 
(\vec \tau_1 \cdot \vec \tau_2) \,
\biggl\{(12\Delta^2-20 M_\pi^2-11q^2) L(q) + 6E^2D(q) \biggr\}
\no\\
&-& \frac{3g_A^4}{128 \pi^2 f_\pi^4}\,  
\biggl( (\vec \sigma_1 \cdot \vec q\,)(\vec \sigma_2 \cdot \vec q\,) 
-q^2 (\vec \sigma_1 \cdot\vec \sigma_2 )\biggr) \,
\biggl\{ -2L(q) + (s^2-4\Delta^2) D(q) \biggr\} \no\\
&-& \frac{g_A^4}{128 \pi f_\pi^4 \Delta}\, (\vec \tau_1 \cdot \vec
\tau_2) \, \biggl( (\vec \sigma_1 \cdot \vec q\,)(\vec \sigma_2 \cdot \vec q\,) 
-q^2 (\vec \sigma_1 \cdot\vec \sigma_2 )\biggr) \, s^2 \, A(q)~,
\eeqa
with $A(q)$ given in eq.(\ref{Lq}).

\medskip

\noindent Double $\Delta$--excitation in the box graphs:
\beqa
V^{\rm TPEP}_{\Delta, \, {\rm box}-2} &=& -\frac{3g_A^4}{64 \pi^2
  f_\pi^4}\, \biggl\{-4\Delta^2 L(q) + E[ H(q) + (E+8\Delta^2)D(q) ] \biggr\}
\no\\
&-& \frac{g_A^4}{384 \pi^2 f_\pi^4}\, 
(\vec \tau_1 \cdot \vec \tau_2) \,
\biggl\{(12E- s^2) L(q) + 3E [H(q)+ (8\Delta^2 -E)D(q)] \biggr\}
\no\\
&-& \frac{3g_A^4}{512 \pi^2 f_\pi^4}\,  
\biggl( (\vec \sigma_1 \cdot \vec q\,)(\vec \sigma_2 \cdot \vec q\,) 
-q^2 (\vec \sigma_1 \cdot\vec \sigma_2 )\biggr) \,
\biggl\{ 6L(q) + (12\Delta^2- s^2) D(q) \biggr\} \\
&-& \frac{g_A^4}{1024 \pi^2 f_\pi^4 }\, (\vec \tau_1 \cdot \vec
\tau_2) \, \biggl( (\vec \sigma_1 \cdot \vec q\,)(\vec \sigma_2 \cdot \vec q\,) 
-q^2 (\vec \sigma_1 \cdot\vec \sigma_2 )\biggr) \, \biggl\{
2L(q) + (4\Delta^2 + s^2 ) D(q) \biggr\}~,\no
\eeqa
with
\beq
H(q) = \frac{2E}{s^2-4\Delta^2} \biggl[ L(q) - L (2\sqrt{\Delta^2 -
  M_\pi^2}) \biggr] \quad .
\eeq
Finally, let us show how to arrive at the  coupling constant relation
given in eq.(\ref{gpiND}). In spin--isopsin SU(4) (or more generally,
SU(6)), one obtains
\beq
g_{\Delta N\pi} = \frac{3}{\sqrt{2}} \, g_{\pi N}~.
\eeq
Using now  the Goldberger--Treiman relation
\beq
g_{\pi N}= g_A \, \frac{m}{F_\pi}~,
\eeq
one arrives at eq.(\ref{gpiND}).
The numerical value of $g_{\Delta N\pi}$ follows from using $g_A
=1.26$. Note that this value for the coupling constant leads
to a $\Delta$--width of about 117~MeV.

\def\theequation{\Alph{section}.\arabic{equation}}
\setcounter{equation}{0}
\section{Formulae for the deuteron properties}
\label{app:deut}
Here, we collect the formulae needed to calculate the deuteron
properties. Denote by $u(r)$ and $w(r)$ the S-- and D--wave coordinate
space wave functions. We denote the momentum space representations of
$u(r)/r$ and $w(r)/r$ by $\tilde{u}(p)$ and $\tilde{w}(p)$, respectively. We have
\beqa
{\rm Normalization}&:& \int_0^\infty dp \, p^2 \, [\tilde{u}(p)^2
+ \tilde{w}(p)^2] = \int_0^\infty dr  \, [{u}(r)^2+{w}(r)^2] =
1~,\\
{\rm D-state~probability}&:&  \int_0^\infty dp \, p^2 \,
\tilde{w}(p)^2 = \int_0^\infty dr \, w(r)^2~, 
\\ \label{Qd}
{\rm Quadrupole~moment}&:& Q_d = \frac{1}{20} \int_0^\infty dr \, r^2
\, w(r) \, [\sqrt{8} u(r) - w(r) ] \no\\
&& \quad\,\,\, = -\frac{1}{20}\int_0^\infty dp 
\biggl\{ \sqrt{8} \biggl[ p^2 \frac{d\tilde{u}(p)}{dp}
\frac{d\tilde{w}(p)}{dp} + 3p \tilde{w}(p) \frac{d\tilde{u}(p)}{dp}
\biggr] \no\\ && \qquad\qquad \qquad\qquad\qquad\qquad +
p^2\biggl(\frac{d\tilde{w}(p)}{dp}\biggr)^2 +6 \tilde{w}(p)^2
\biggr\}~,\\
{\rm Asymptotic~ S-state}&:& u(r) \to A_S \, {\rm e}^{-\gamma\, r} \quad
{\rm for}~r\to \infty~,\\
{\rm Asymptotic~D/S-ratio}~\eta &:& w(r) \to \eta \, A_S \, \biggl(
1+ \frac{3}{\gamma r} + \frac{3}{(\gamma r)^2} \biggr)\, {\rm
  e}^{-\gamma \, r}  \quad {\rm for}~r\to \infty~,\\
{\rm RMS~(matter)~radius}&:& r_d = \frac{1}{2}\,\biggl[ \,
\int_0^\infty dr \, r^2 \,  [{u}(r)^2+{w}(r)^2] \, \biggr]^{1/2}~,
\eeqa
with $\gamma = \sqrt{m \, |E_d|} =45.7\,$MeV (using $m = (m_p+m_n)/2)$.
Note that the momentum--space representation of $Q_d$ given in
eq.(\ref{Qd}) shows why one cannot use a sharp momentum--space
regulator to calculate this quantity. We also remark that the 
D--state probability is not an observable. Meson--exchange current
corrections to  $Q_d$ are not given.

\bigskip\bigskip\bigskip


\pagebreak


\section*{Figures}

\begin{figure}[H]
\psfrag{1S0}{$^1S_0$}
\psfrag{3S1}{$^3S_1$}
\psfrag{1P1}{$^1P_1$}
\psfrag{3P0}{$^3P_0$}
\psfrag{3P1}{$^3P_1$}
\psfrag{3P2}{$^3P_2$}%
\vspace{1.5cm}
\parbox{6.5cm}{\centerline{\psfig{file=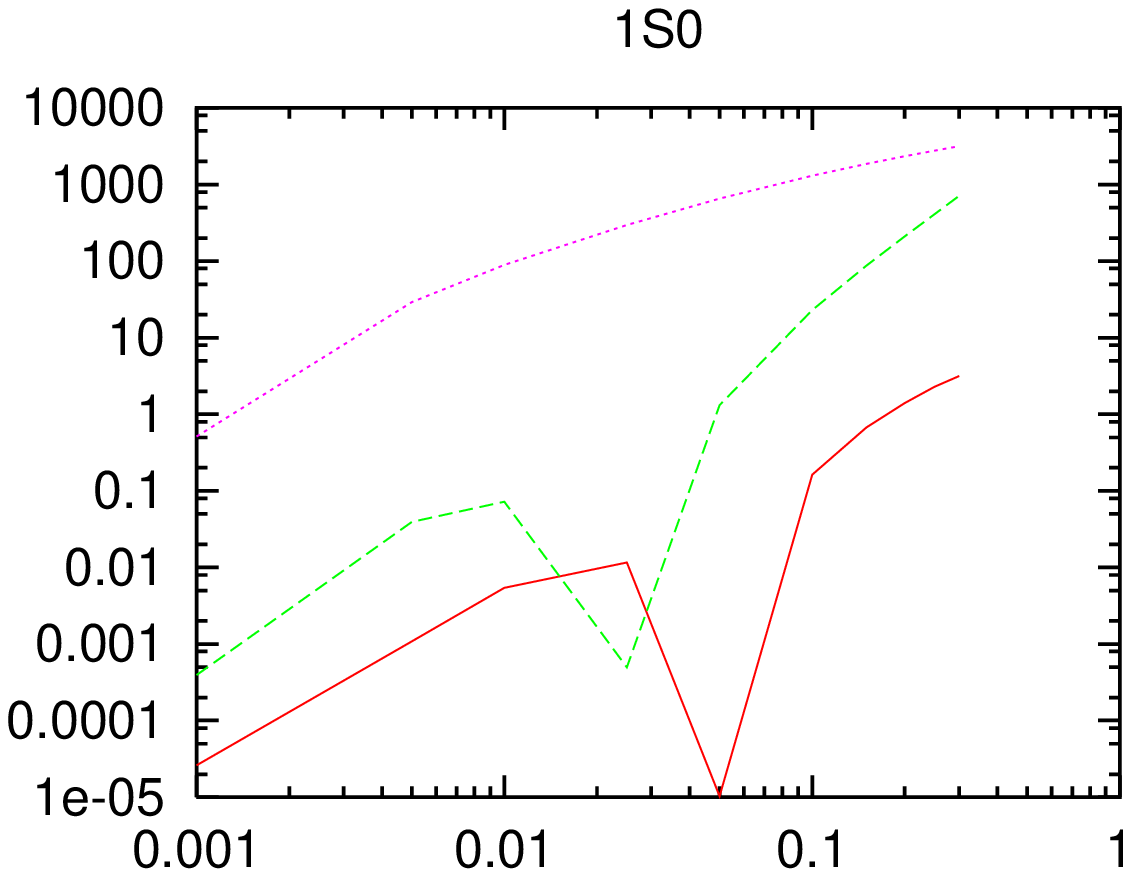,width=2.9in}}}
\hfill
\parbox{6.5cm}{
\centerline{\psfig{file=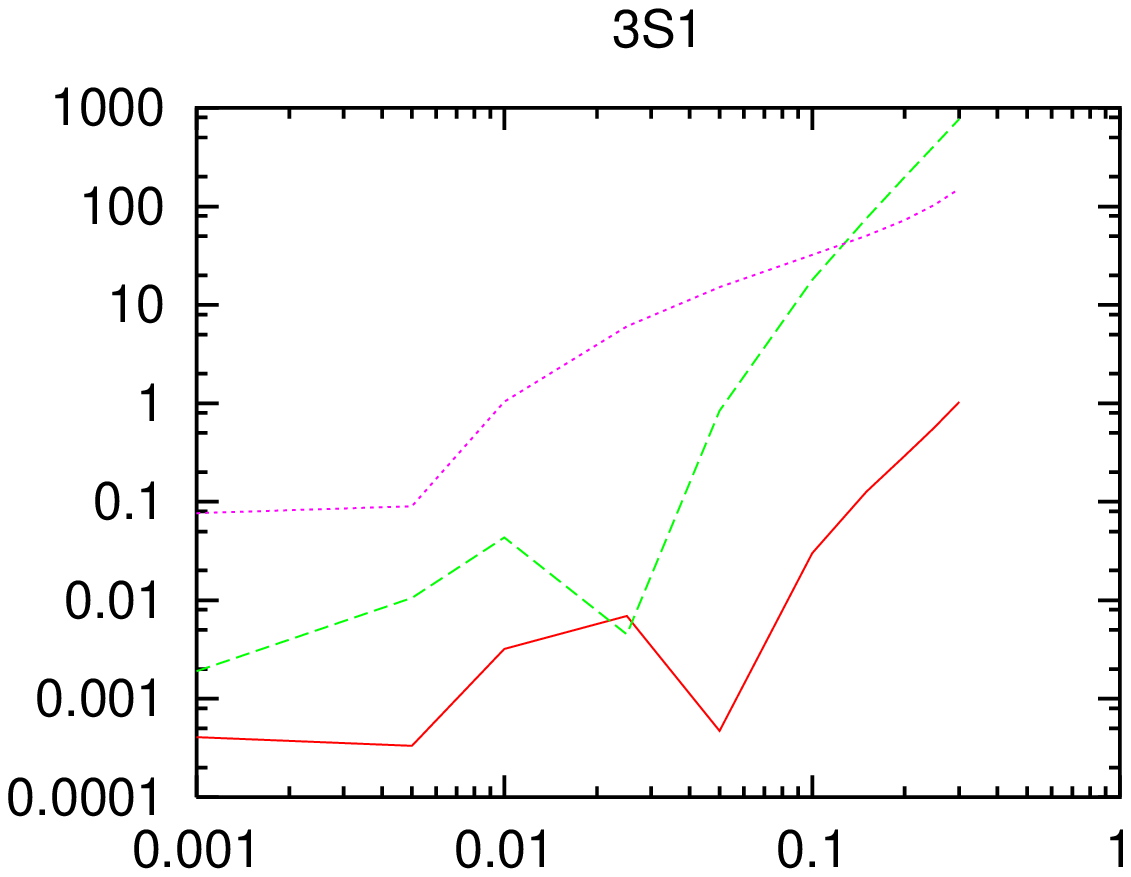,width=2.9in}}}
\parbox{6.5cm}{\centerline{\psfig{file=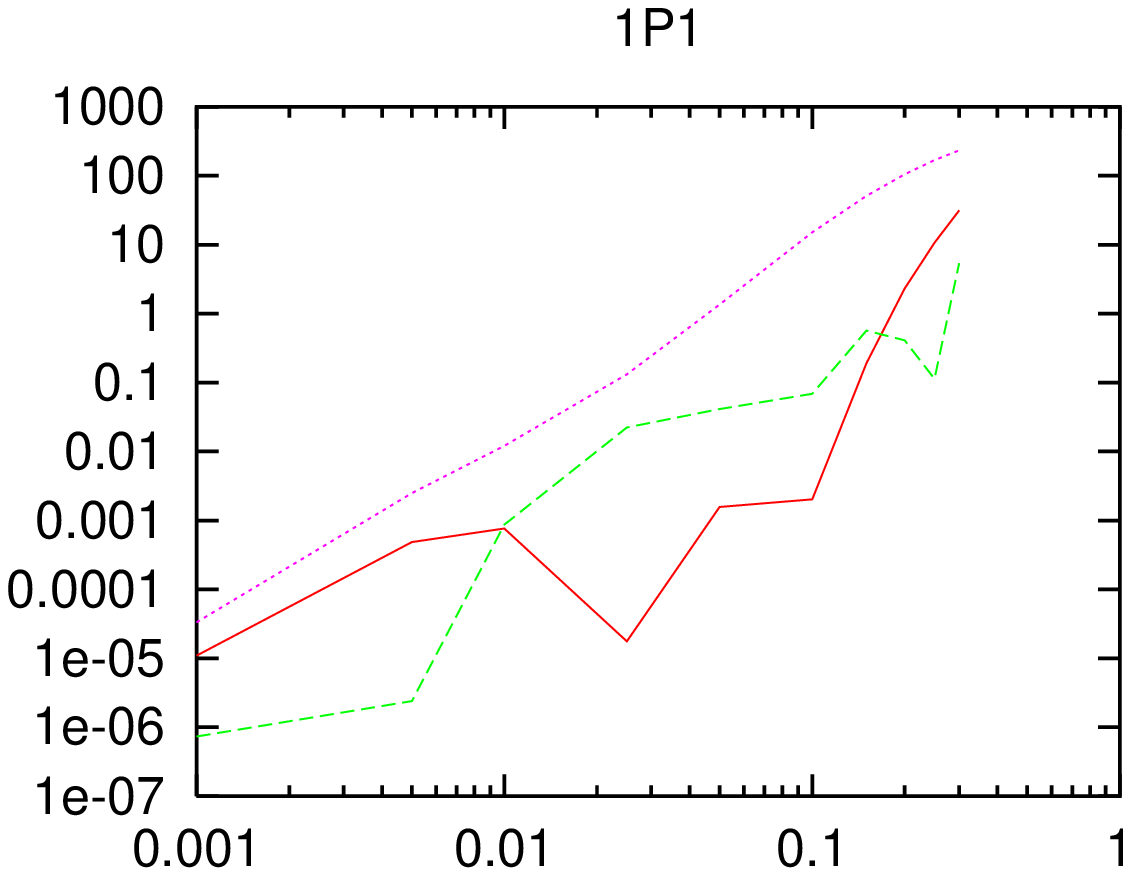,width=2.9in}}}
\hfill
\parbox{6.5cm}{
\centerline{\psfig{file=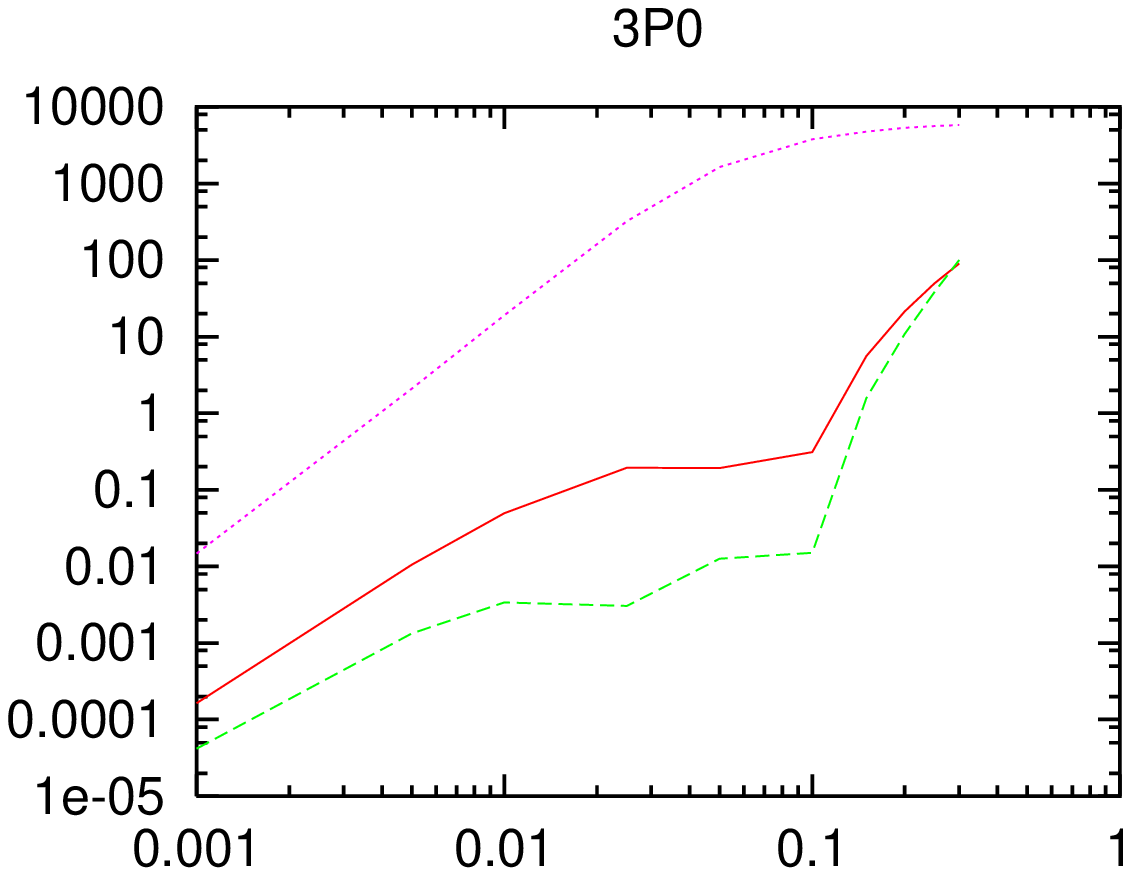,width=2.9in}}}

\vspace{0.2cm}
\parbox{6.5cm}{\centerline{\psfig{file=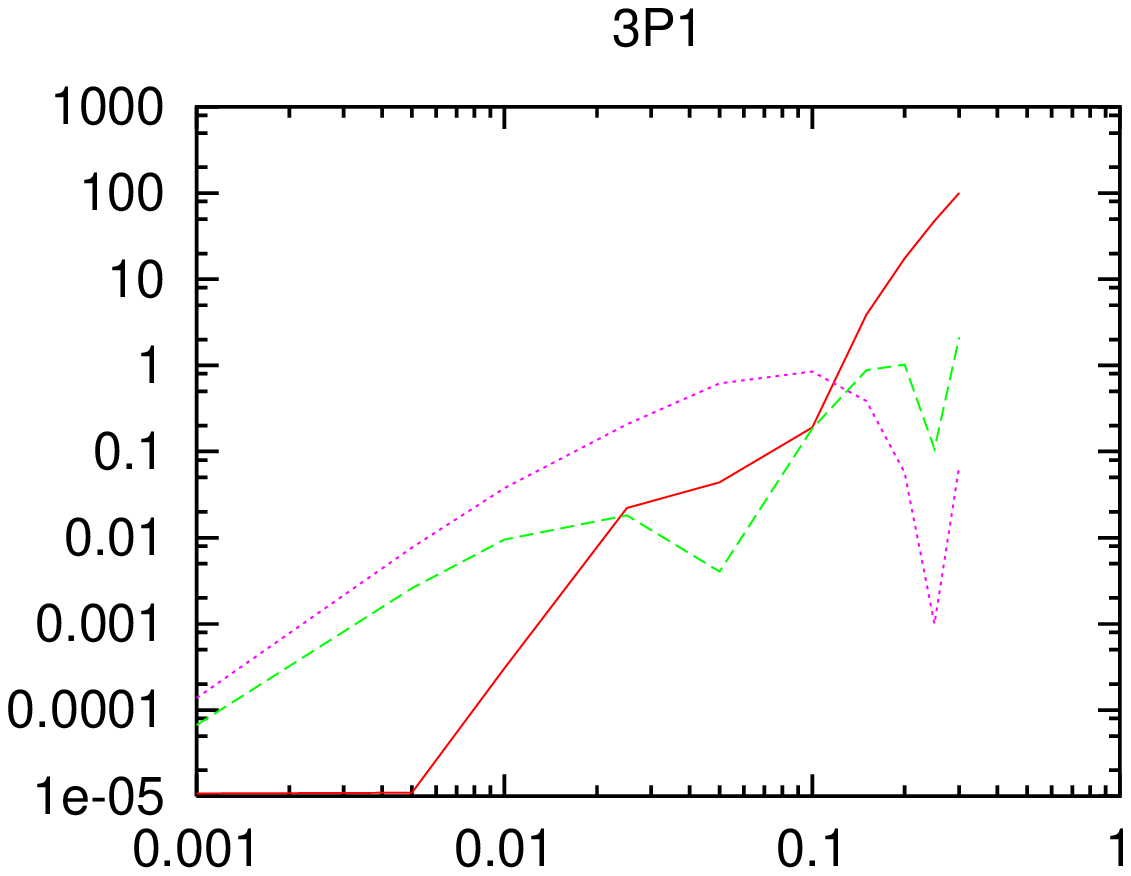,width=2.9in}}}
\hfill
\parbox{6.5cm}{
\centerline{\psfig{file=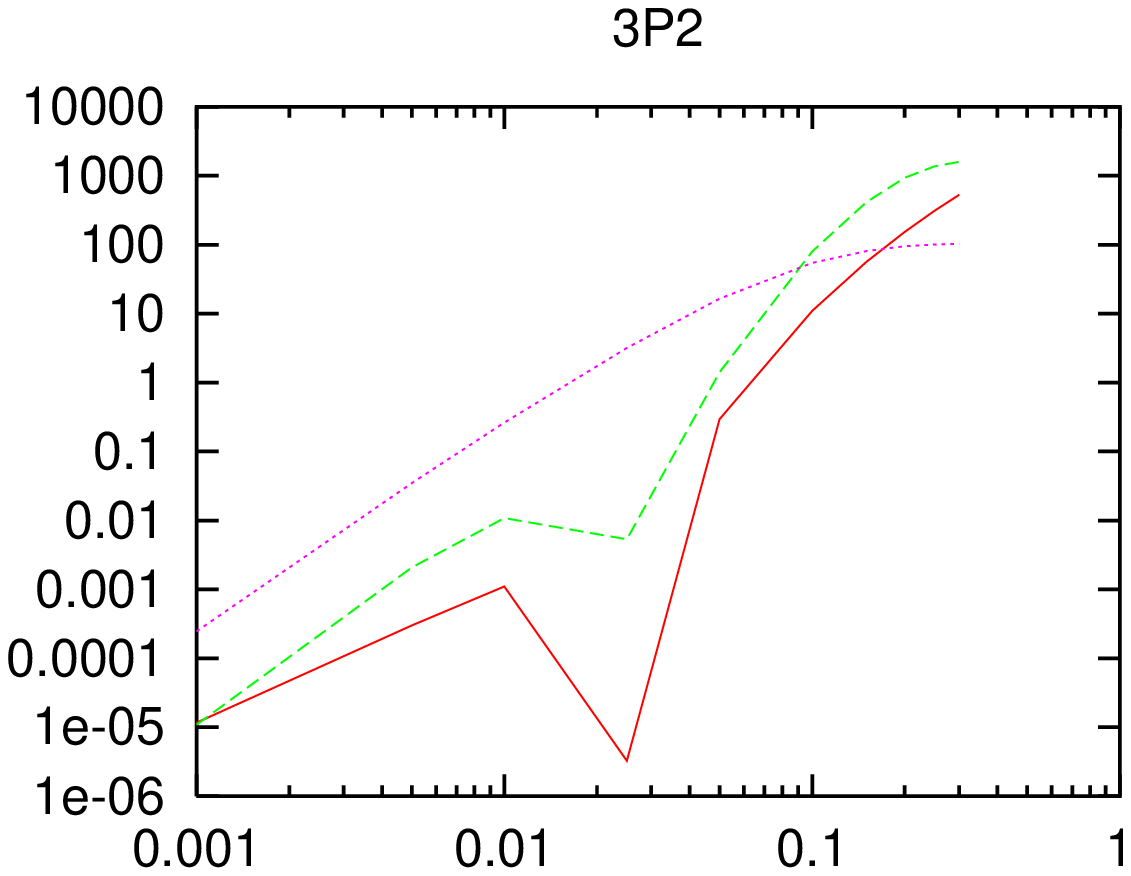,width=2.9in}}}
\vspace{1cm}
\caption{\label{figfit}
Quadratic deviations in the phase shifts, $(\delta^{\rm fit}
-\delta^{\rm PSA})^2$, versus $E_{\rm lab}$ in GeV. 
Data from the Nijmegen PSA have been fitted for $E_{\rm lab} \le 0.1\,$
GeV. The purple dotted, green dashed and red solid 
curves represent LO, NLO and NNLO results, respectively.}
\end{figure}
\pagebreak



\begin{figure}[htb]
   \vspace{2.9cm}
   \epsfysize=13cm
   \centerline{\epsffile{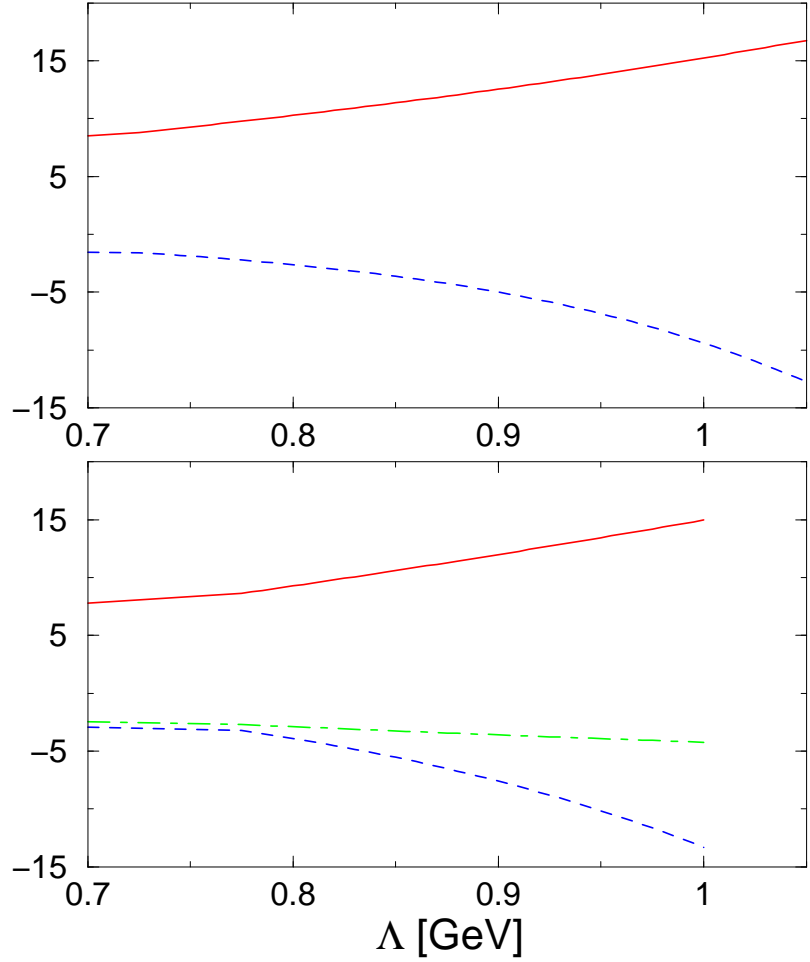}}
   \vspace{2cm}
   \centerline{\parbox{13cm}{\caption{\label{figCrun}
Running of the  four--nucleon coupling constants in the
$^1S_0$ (upper panel) and $^3S_1 - ^3D_1$ (lower panel) partial waves.
In the upper panel, the solid red (dashed blue) line refers to $C_{^1S_0}$
($\tilde{C}_{^1S_0}$). In the lower panel, the solid red (dashed blue) 
line refers to $C_{^3S_1}$ ($\tilde{C}_{^3S_1}$). The green
(dashed--dotted) line refers to the constant in the coupled $^3S_1 -^3D_1$
system. The units are the same as in the table \ref{tab:LECs}.  }}}
\end{figure}

\begin{figure}[htb]
   \vspace{2.9cm}
   \epsfysize=13cm
   \centerline{\epsffile{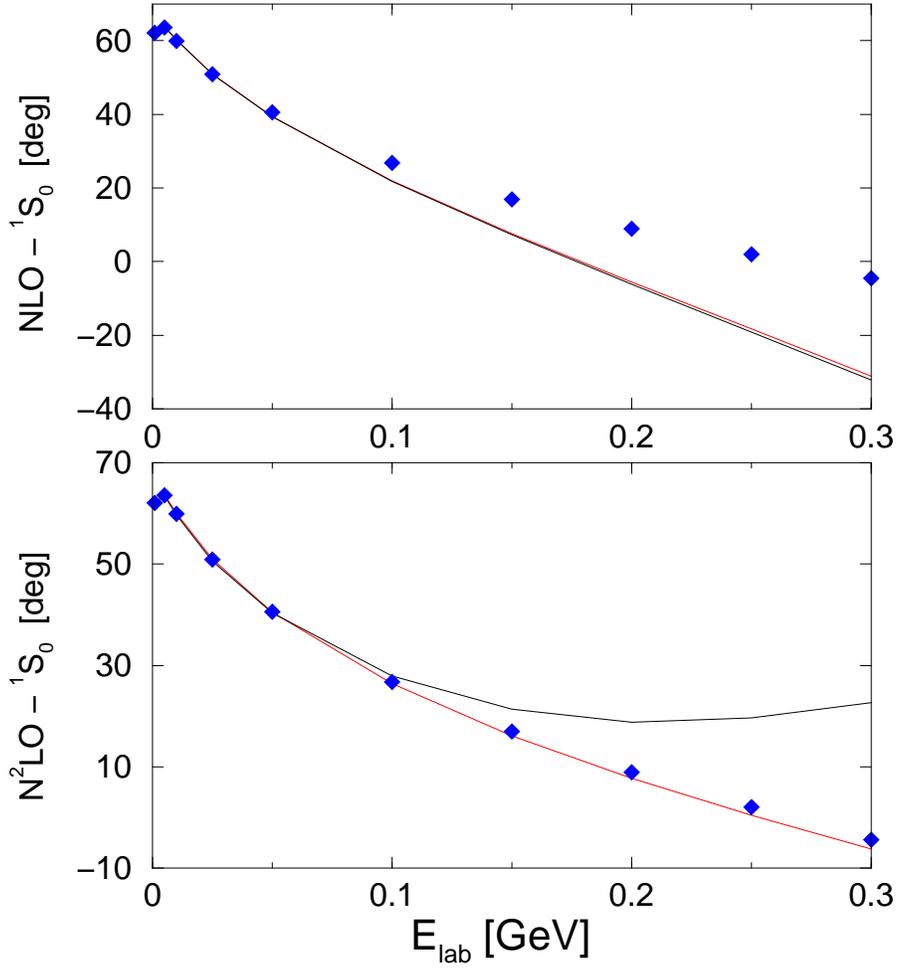}}
   \vspace{2cm}
   \centerline{\parbox{13cm}{\caption{\label{fig2sol}
Phase shifts for the two solutions in the $^1S_0$--wave as dicussed 
in the text. At NLO (upper panel), these are indistinguishable.
At NNLO (lower panel), one of the solutions (black line) shows an
unacceptable behaviour at higher momenta and is discarded.
  }}}
\end{figure}

\begin{figure}[bht]
\psfrag{1S0}{$^1S_0$ [deg]}
\psfrag{3S1}{$^3S_1$ [deg]}   
   \vspace{0.4cm}
   \centerline{\psfig{file=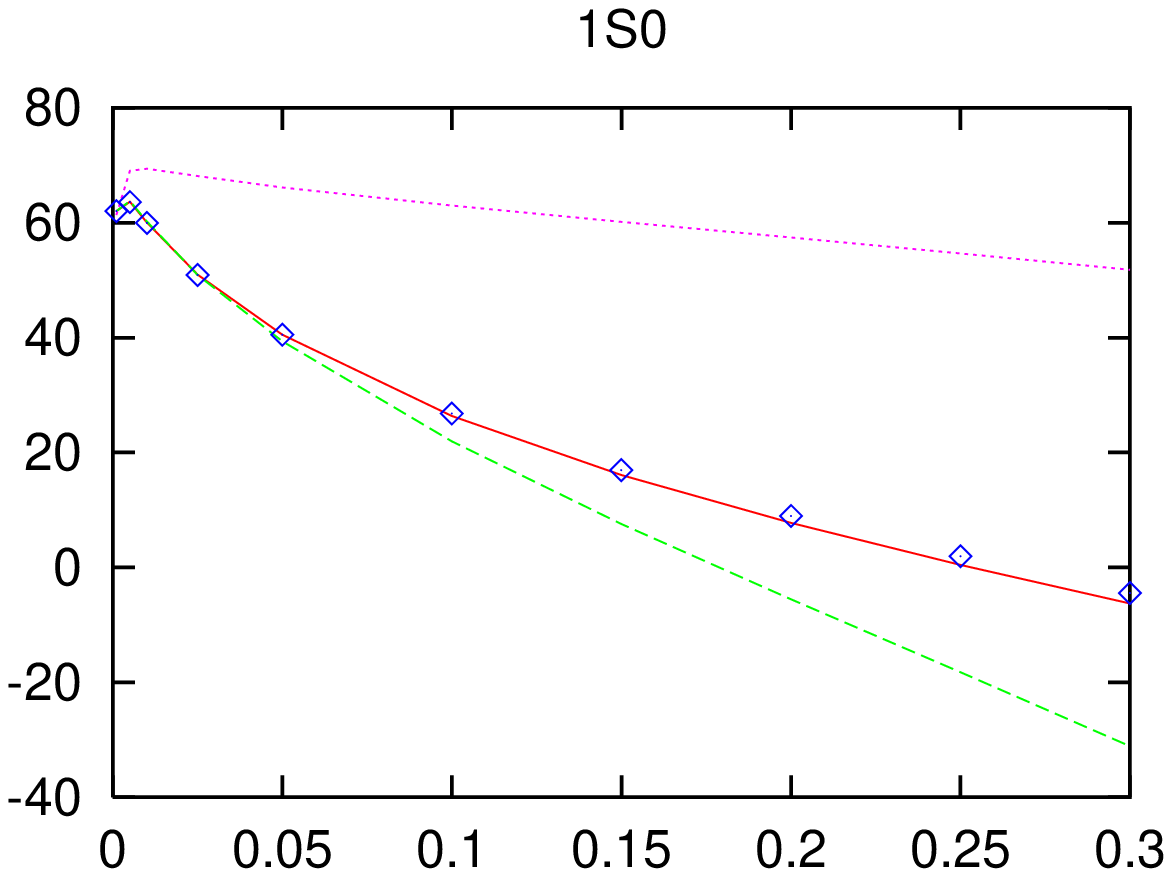,width=10cm}}
   \vspace{0.8cm}
   \centerline{\psfig{file=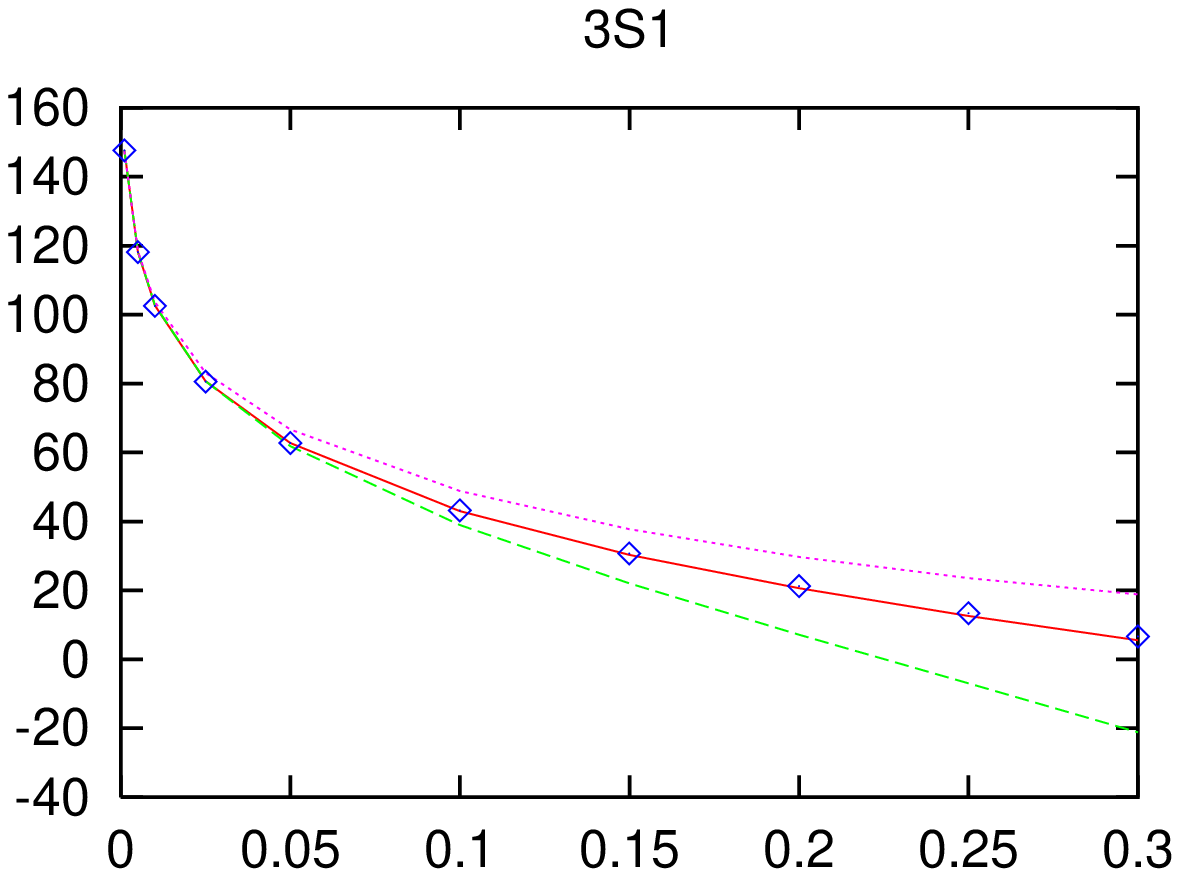,width=10cm}}
   \vspace{0.8cm}
   \centerline{\parbox{13cm}{\caption{\label{SW}
Predictions for the S--waves (in degrees) for 
nucleon laboratory energies $E_{\rm lab}$ below 300~MeV (0.3~GeV).
The purple dotted, green dashed and red solid 
curves represent LO, NLO and NNLO results, in order. The blue
   squares depict the Nijmegen PSA results.
In the upper and the lower panel, $^1S_0$ and $^3S_1$, respectively, are shown. 
  }}}
\end{figure}

\begin{figure}[htb]
\psfrag{1P1}{$^1P_1$ [deg]}
\psfrag{3P0}{$^3P_0$ [deg]}
\psfrag{3P1}{$^3P_1$ [deg]}
\psfrag{3P2}{$^3P_2$ [deg]}
\psfrag{E1}{$\epsilon_1$ [deg]}
\parbox{6.7cm}{\centerline{\psfig{file=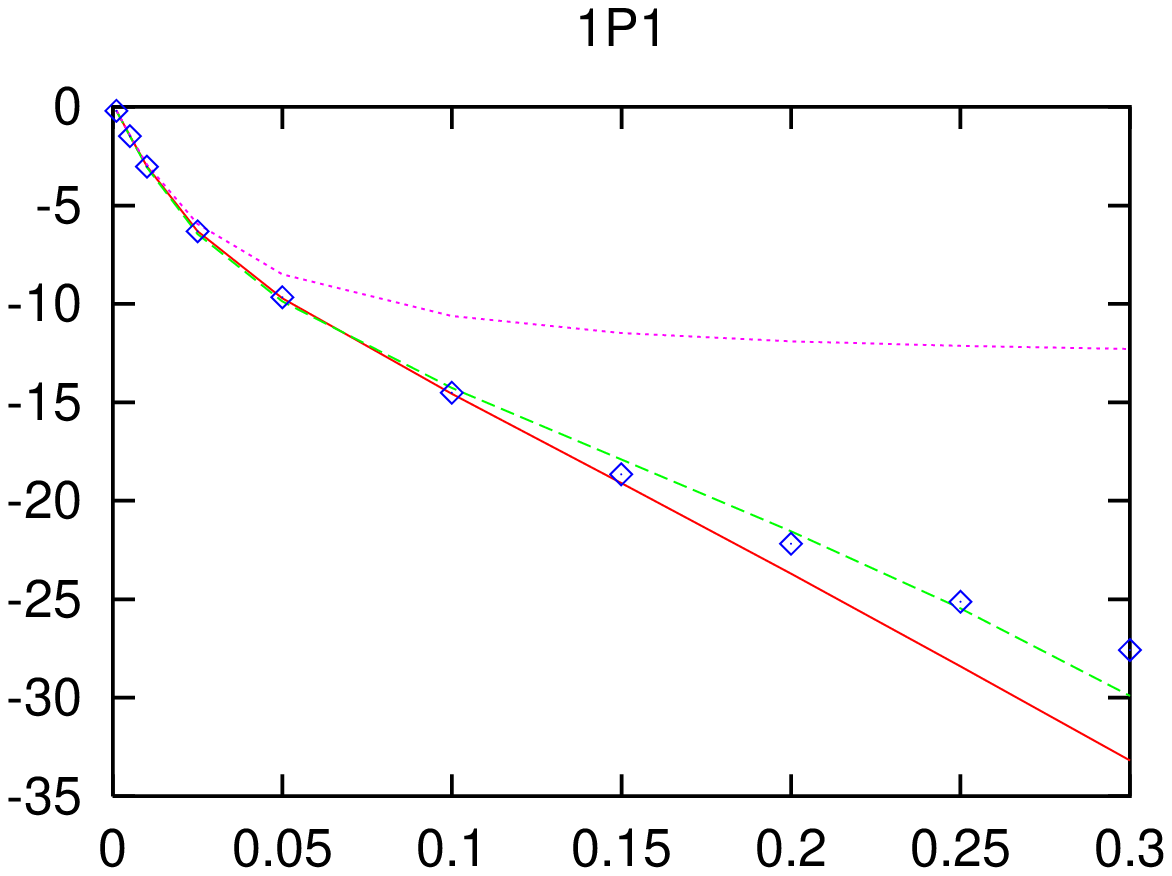,width=3.1in}}}
\hfill
\parbox{6.7cm}{
\centerline{\psfig{file=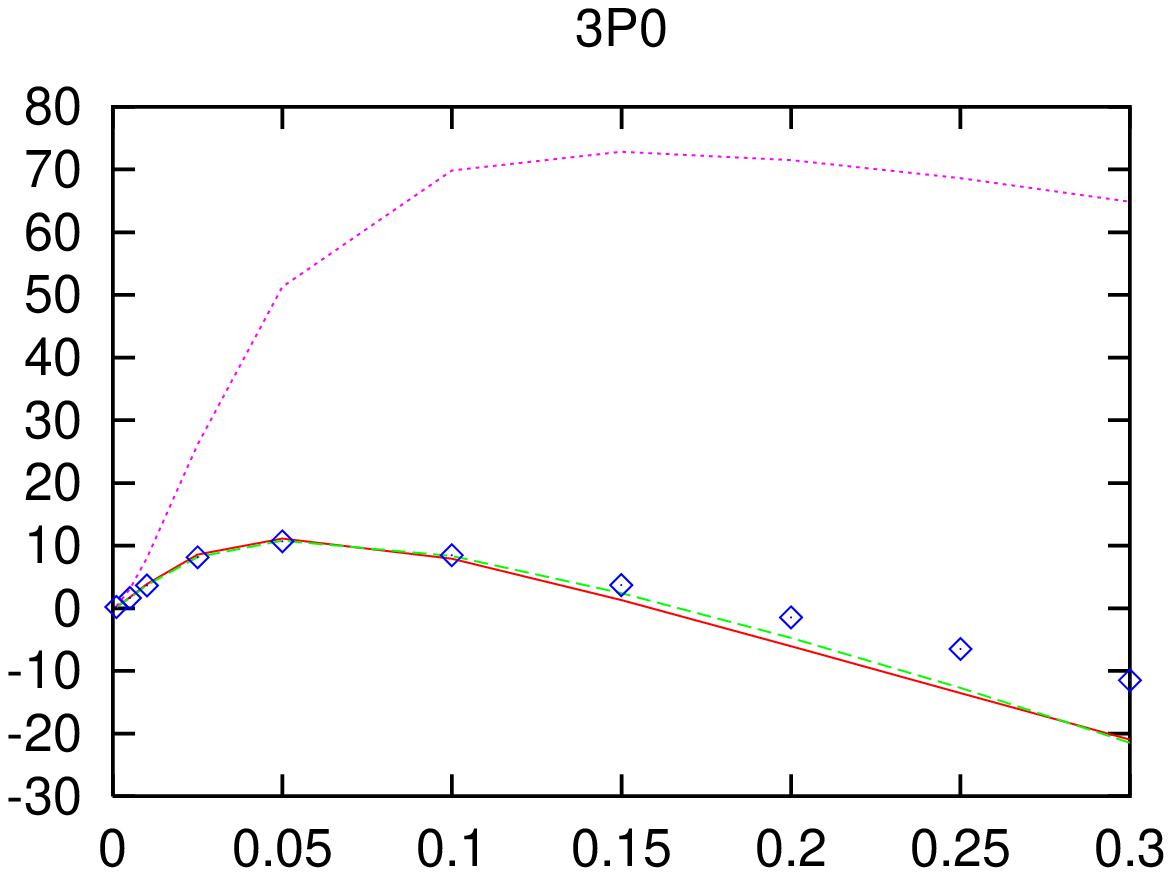,width=3.1in}}}

\vspace{0.2cm}
\parbox{6.7cm}{\centerline{\psfig{file=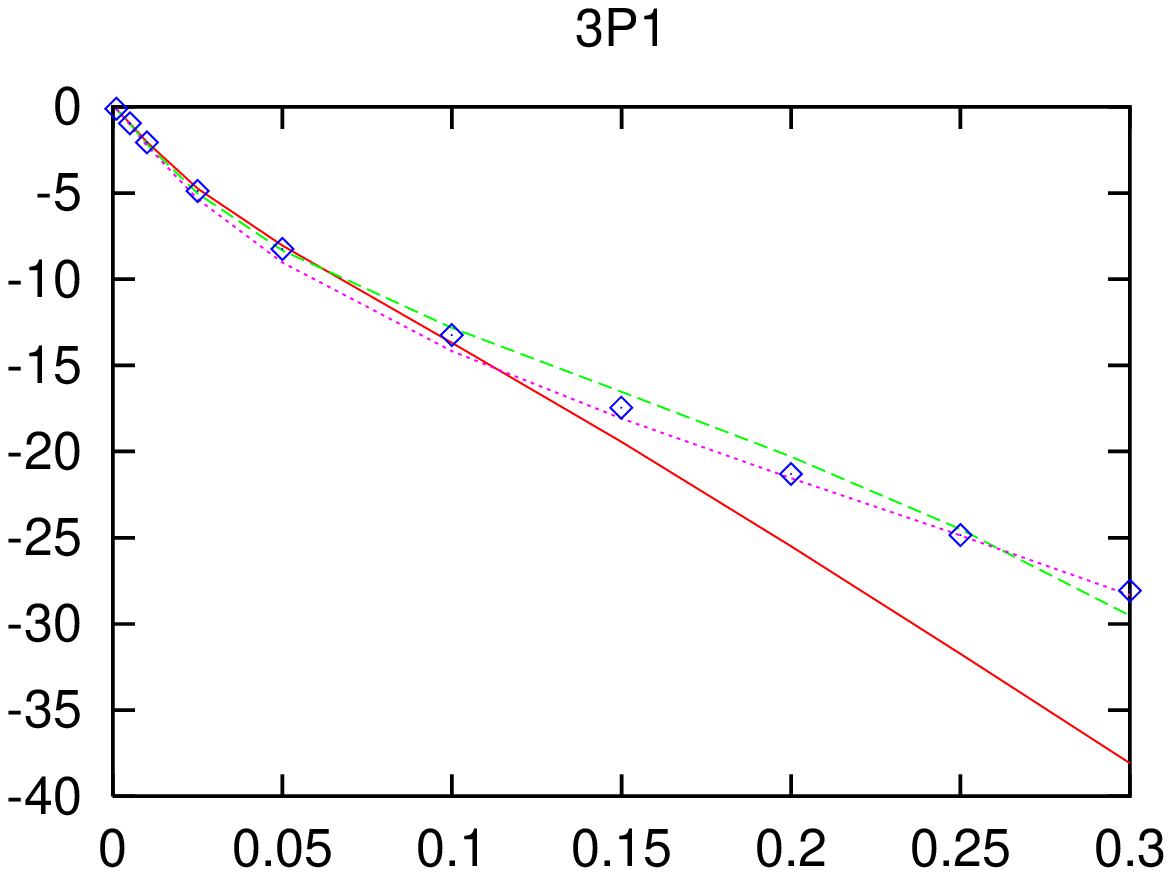,width=3.1in}}}
\hfill
\parbox{6.7cm}{
\centerline{\psfig{file=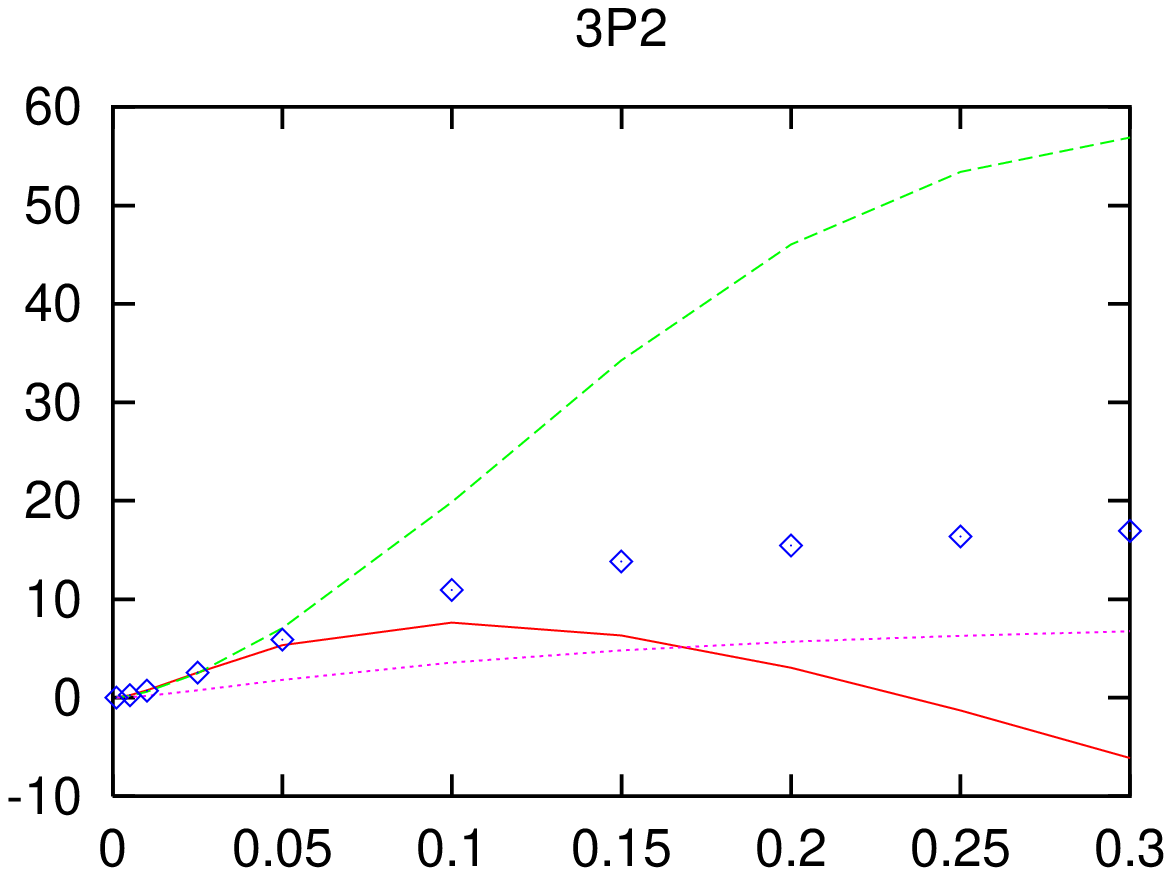,width=3.1in}}}

\vspace{0.2cm}
\centerline{\psfig{file=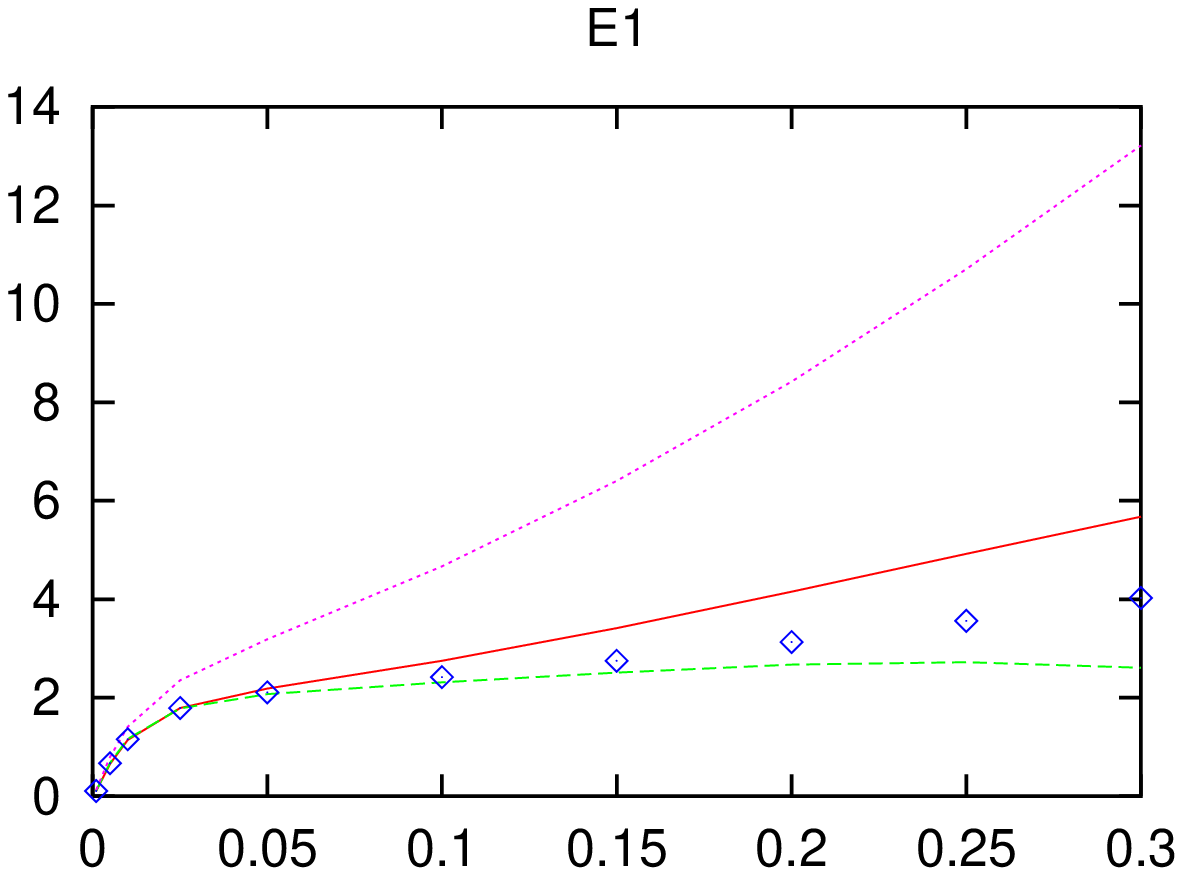,width=3.1in}}
\vspace{1cm}
\caption{\label{PW} 
Predictions for the P--waves  and the
mixing parameter $\epsilon_1$ (in degrees) for 
nucleon laboratory energies $E_{\rm lab}$ below 300~MeV.
For notations, see fig.\ref{SW}.
}
\end{figure}
\pagebreak

\begin{figure}[htb]
   \vspace{0.4cm}
   \epsfysize=12cm
   \centerline{\epsffile{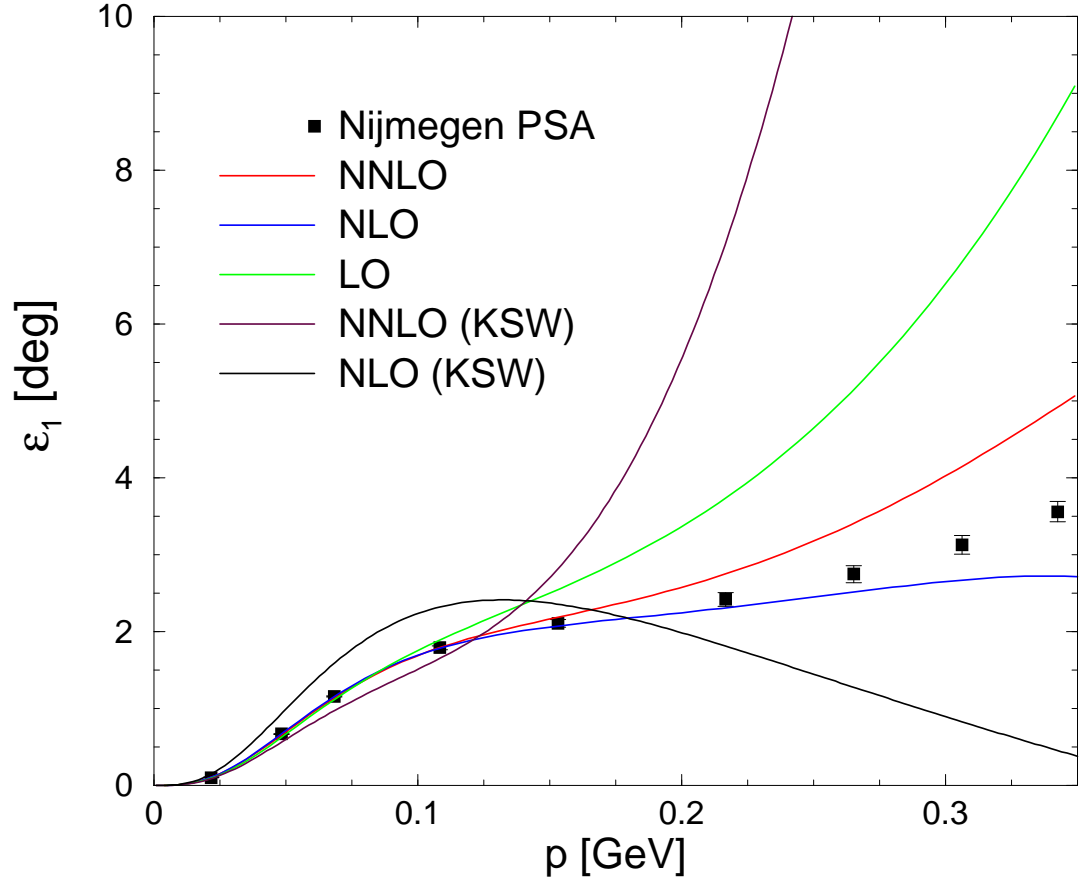}}
   \vspace{1.8cm}
   \centerline{\parbox{15cm}{\caption{\label{eps1}
Predictions for the mixing parameter $\epsilon_1$ for 
nucleon cms momenta $p$ below 350~MeV. The green, blue and red solid
curves represent our LO, NLO and NNLO results, in order. For
comparison, the NLO~\protect{\cite{KSW}} and NNLO~\protect{\cite{calt}}
results in the KSW scheme are also shown.
The black squares depict the Nijmegen PSA results.
  }}}
\end{figure}

\begin{figure}[htb]
\psfrag{1D2}{$^1D_2$ [deg]}
\psfrag{3D1}{$^3D_1$ [deg]}
\psfrag{3D2}{$^3D_2$ [deg]}
\psfrag{3D3}{$^3D_3$ [deg]}
\psfrag{E2}{$\epsilon_2$ [deg]}
\parbox{6.7cm}{\centerline{\psfig{file=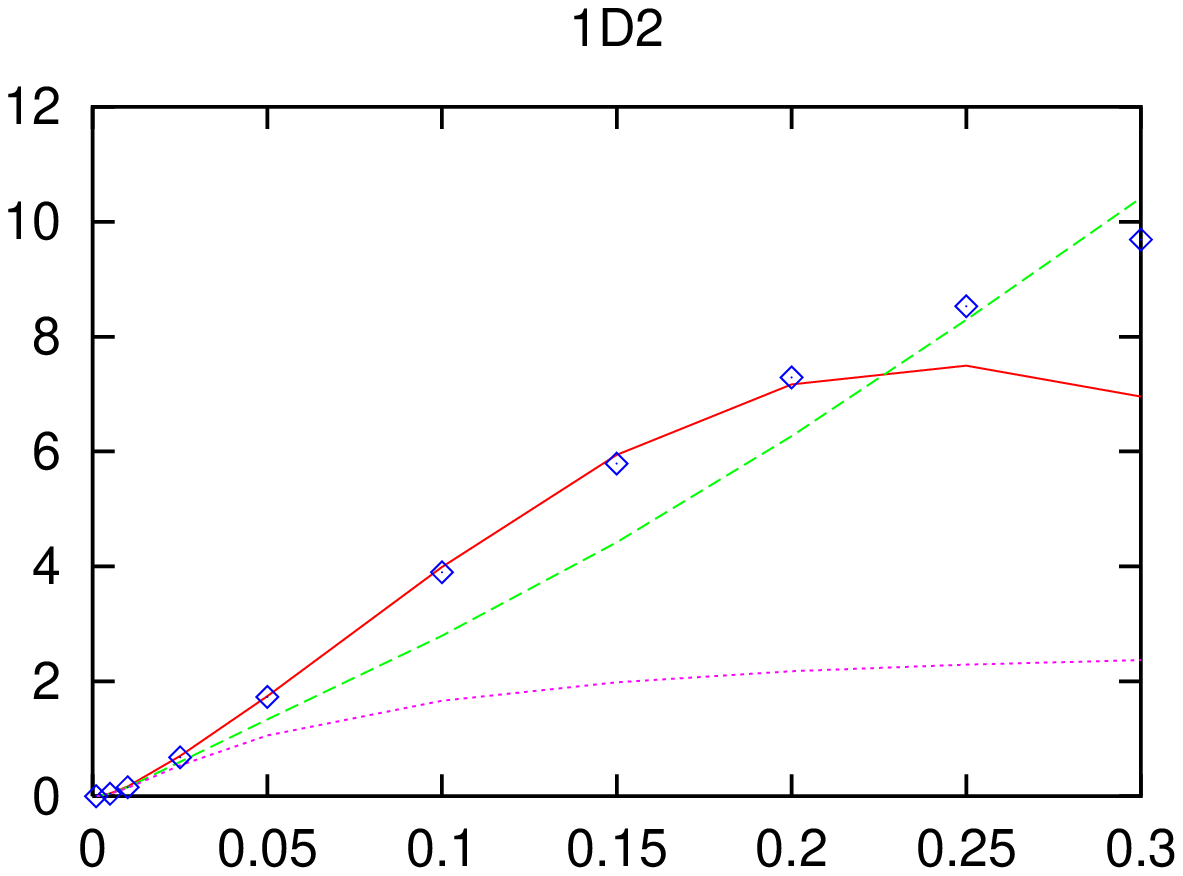,width=3.1in}}}
\hfill
\parbox{6.7cm}{
\centerline{\psfig{file=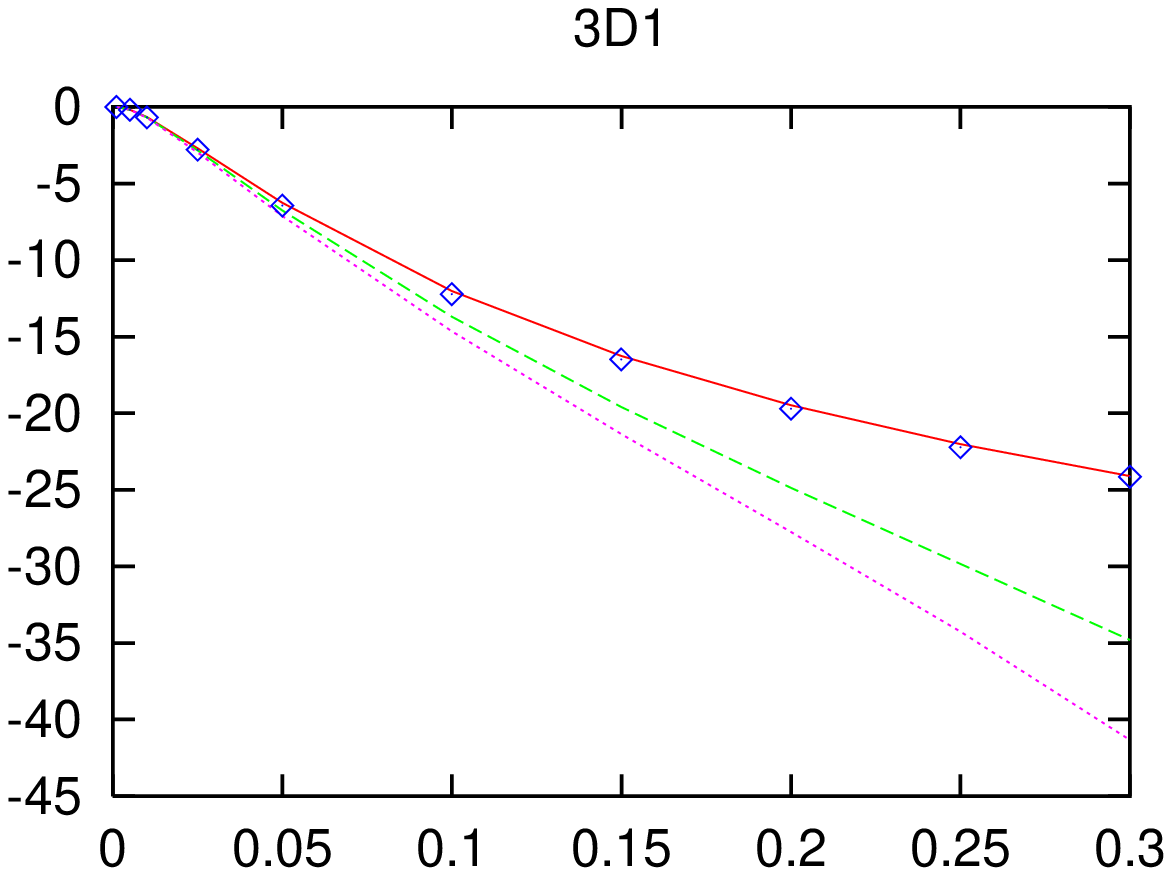,width=3.1in}}}

\vspace{0.2cm}
\parbox{6.7cm}{\centerline{\psfig{file=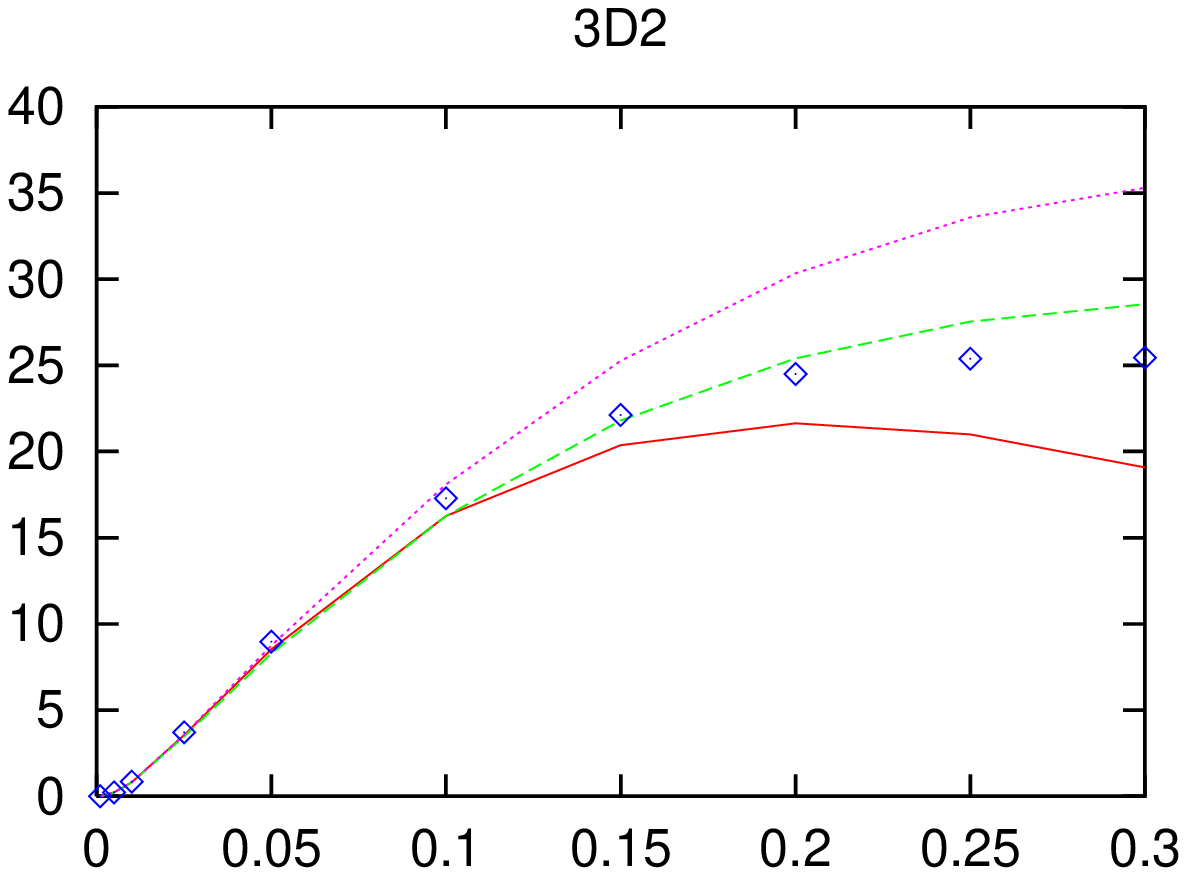,width=3.1in}}}
\hfill
\parbox{6.7cm}{
\centerline{\psfig{file=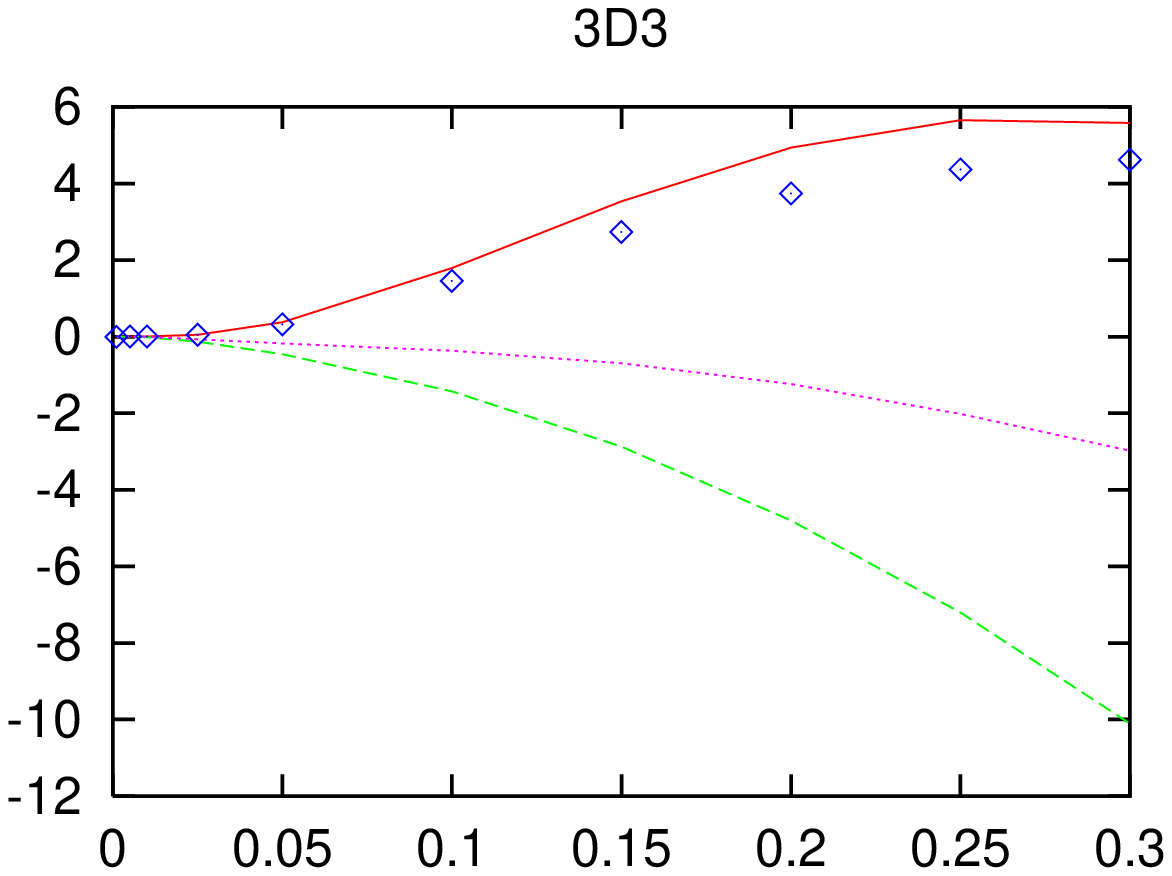,width=3.1in}}}

\vspace{0.2cm}
\centerline{\psfig{file=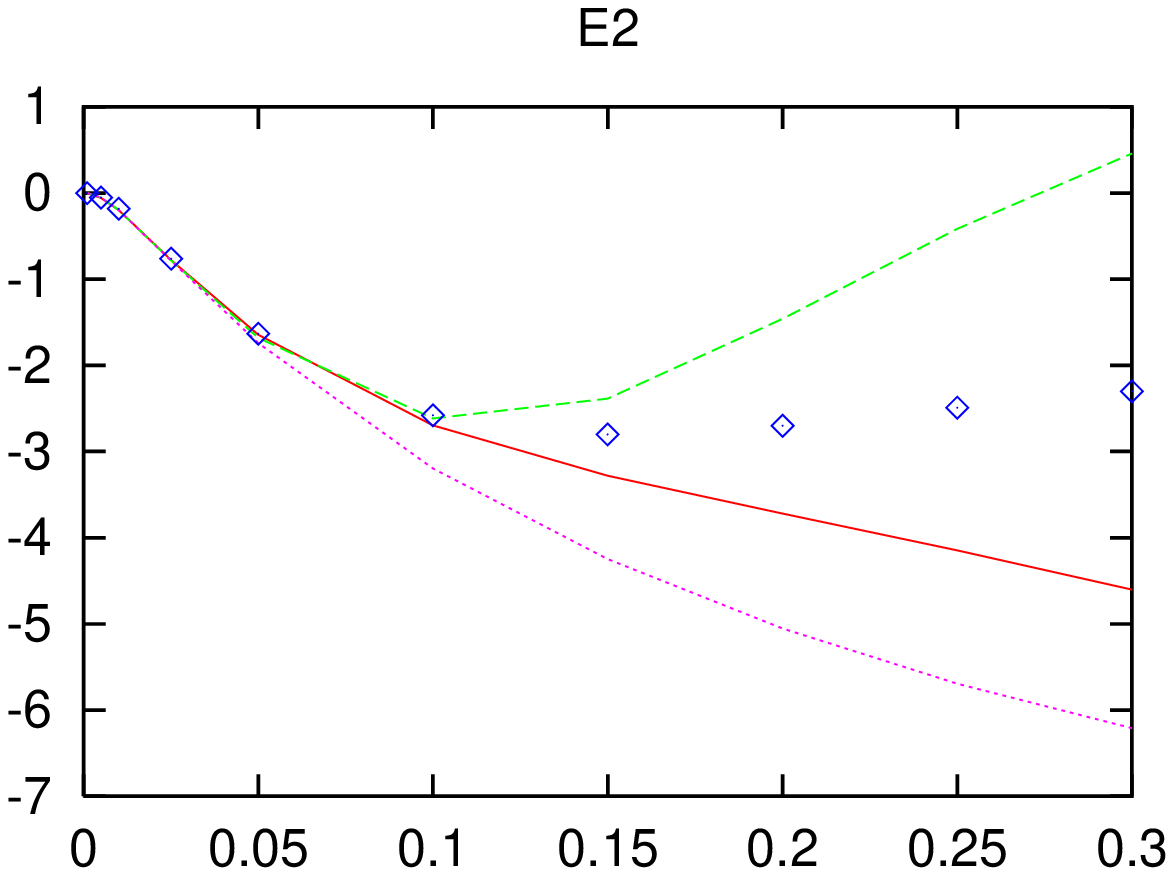,width=3.1in}}
\vspace{1cm}
\caption{\label{DW} 
Predictions for the D--waves  and the
mixing parameter $\epsilon_2$ (in degrees) for 
nucleon laboratory energies $E_{\rm lab}$ below 300~MeV.
For notations, see fig.\ref{SW}.
}
\end{figure}
\pagebreak

\begin{figure}[htb]
   \vspace{2.9cm}
   \epsfysize=14cm
   \centerline{\epsffile{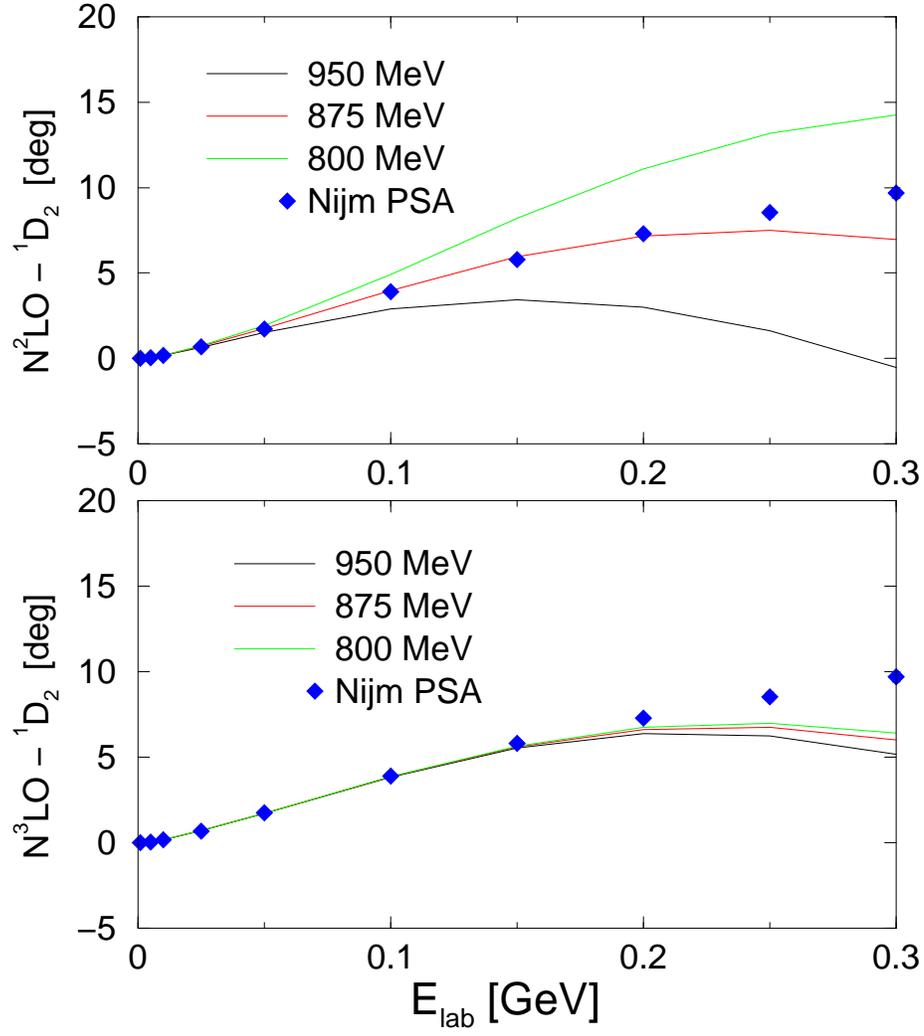}}
   \vspace{2cm}
   \centerline{\parbox{13cm}{\caption{\label{fig:DL}
The $^1D_2$ partial wave at NNLO for three different values of the
cut--off $\Lambda$ (upper panel) and at N$^3$LO (lower panel). The
   blue diamonds are the result of the Nijmegen partial wave analysis.
  }}}
\end{figure}

\begin{figure}[htb]
\psfrag{1F3}{$^1F_3$ [deg]}
\psfrag{3F2}{$^3F_2$ [deg]}
\psfrag{3F3}{$^3F_3$ [deg]}
\psfrag{3F4}{$^3F_4$ [deg]}
\psfrag{E3}{$\epsilon_3$ [deg]}
\parbox{6.7cm}{\centerline{\psfig{file=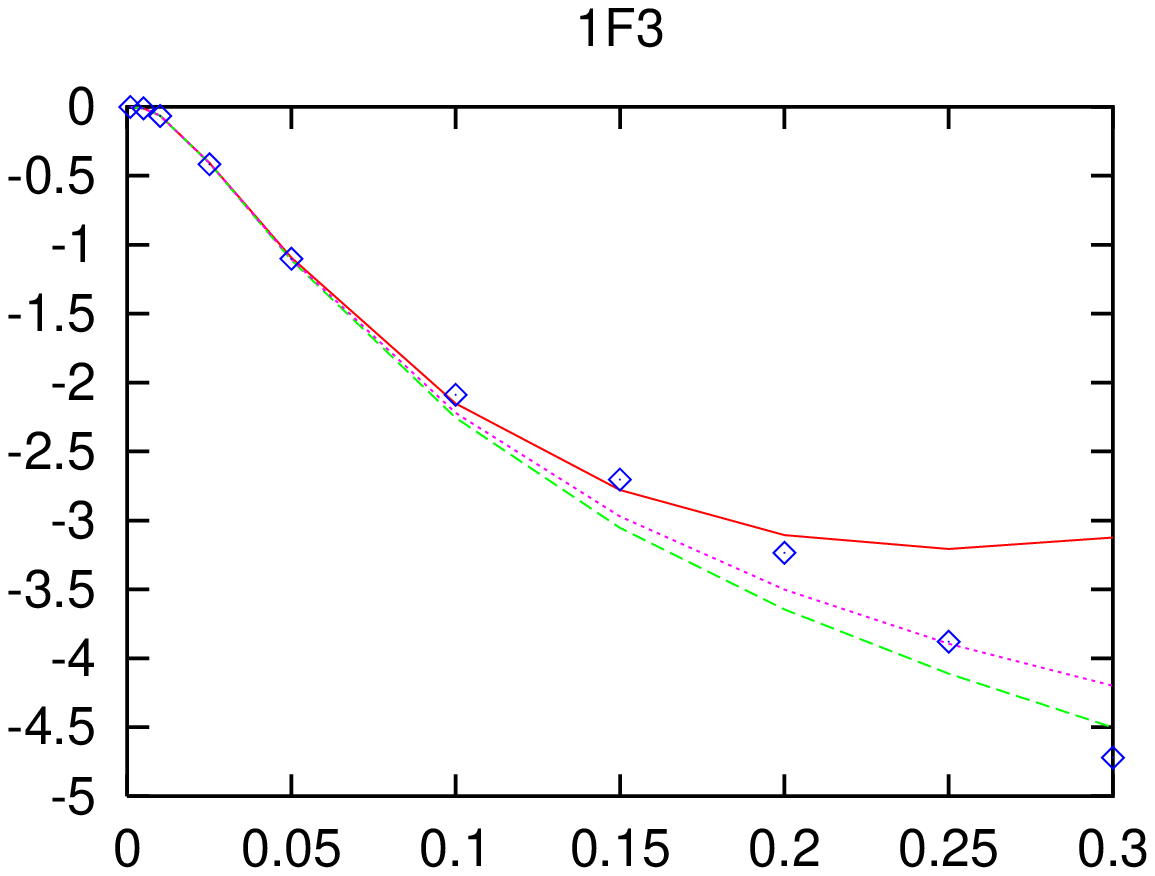,width=3.1in}}}
\hfill
\parbox{6.7cm}{
\centerline{\psfig{file=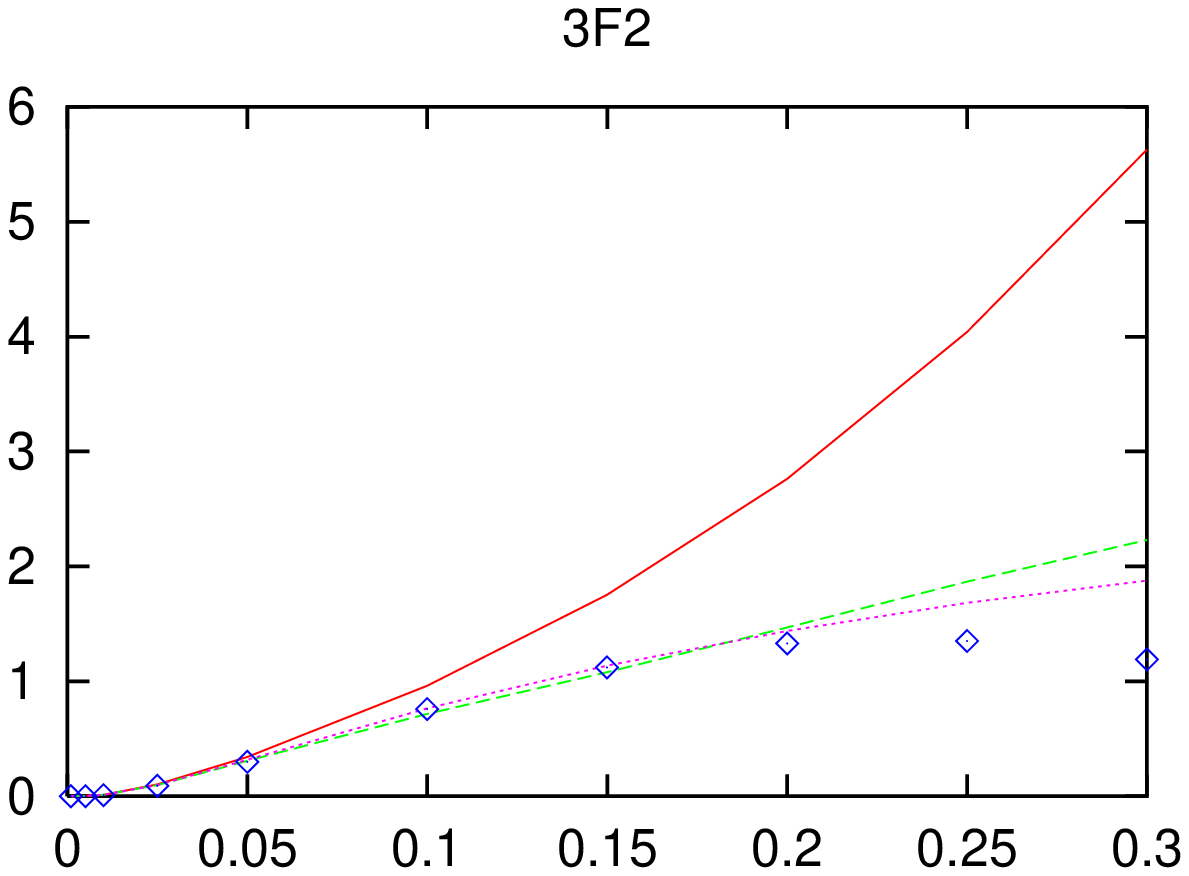,width=3.1in}}}

\vspace{0.2cm}
\parbox{6.7cm}{\centerline{\psfig{file=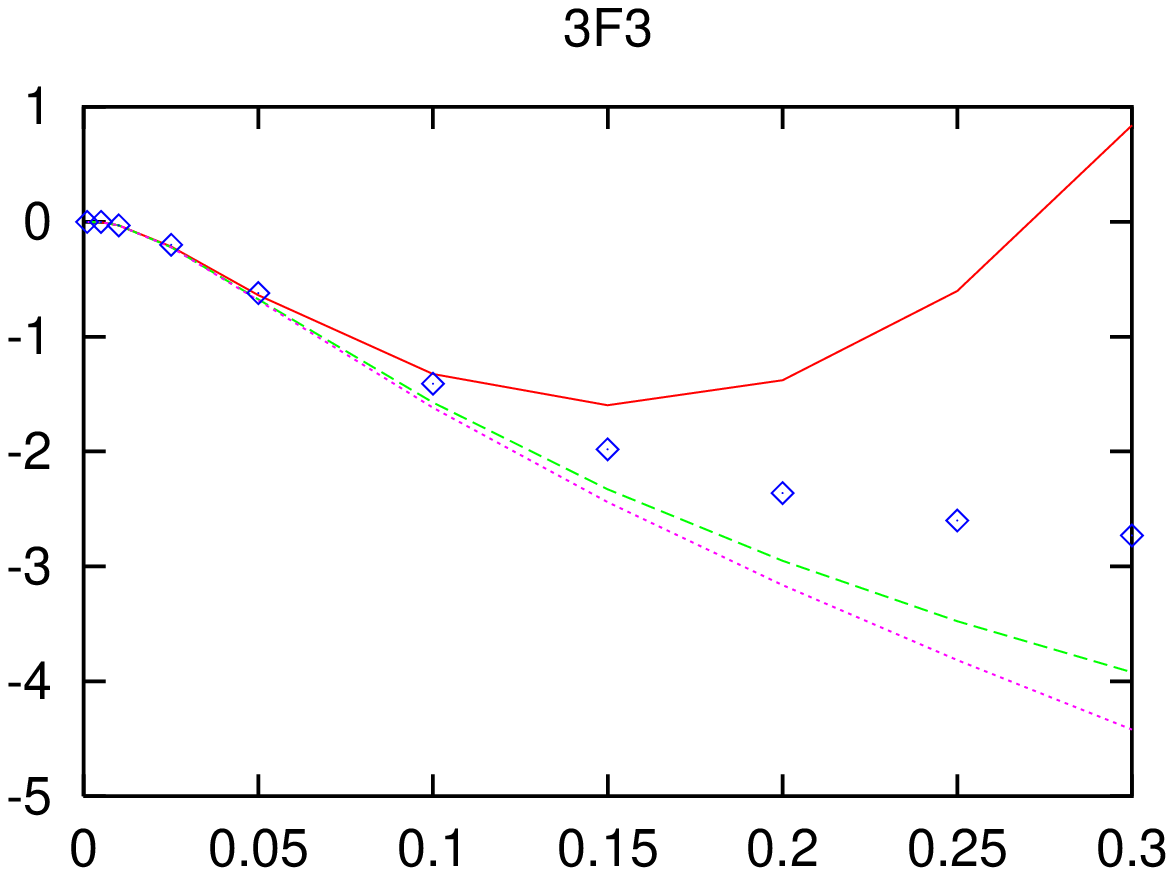,width=3.1in}}}
\hfill
\parbox{6.7cm}{
\centerline{\psfig{file=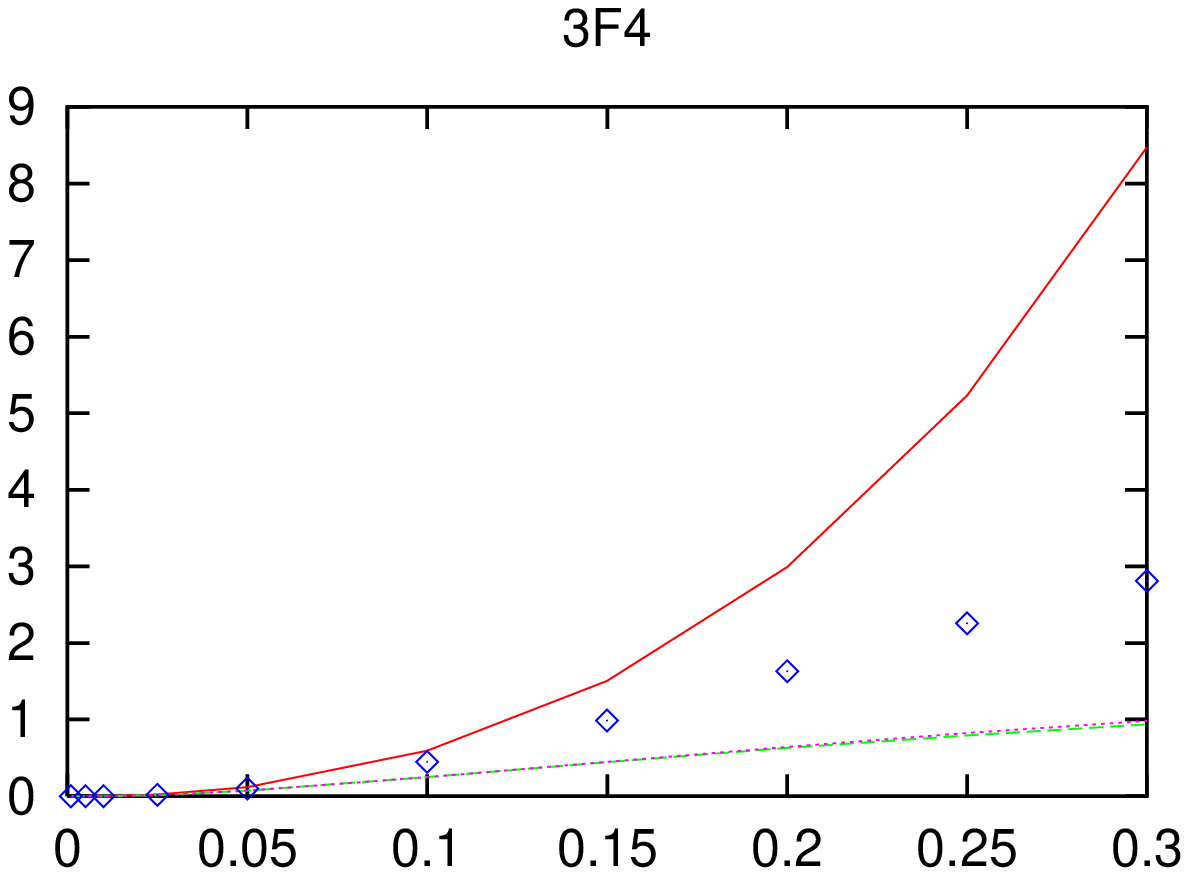,width=3.1in}}}

\vspace{0.2cm}
\centerline{\psfig{file=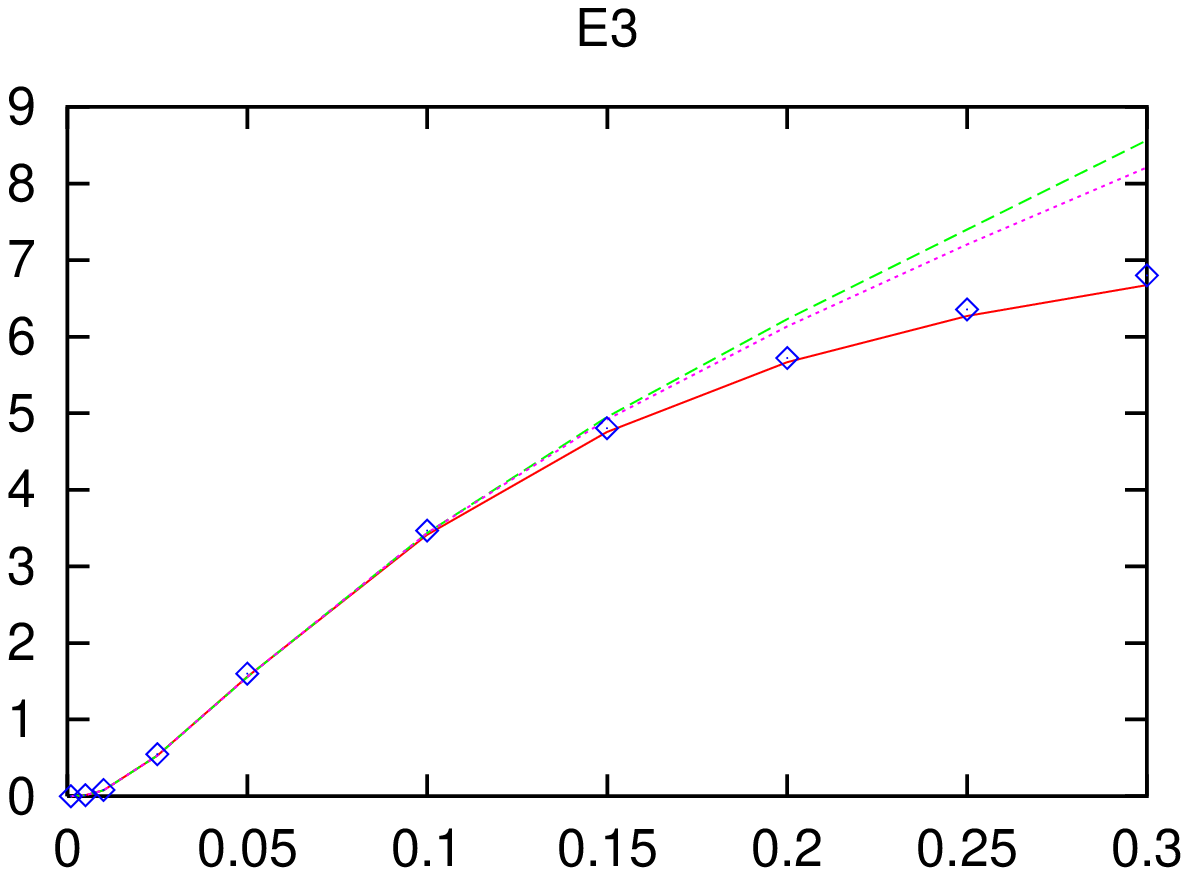,width=3.1in}}
\vspace{1cm}
\caption{\label{FW} 
Predictions for the F--waves  and the
mixing parameter $\epsilon_3$ (in degrees) for 
nucleon laboratory energies $E_{\rm lab}$ below 300~MeV.
For notations, see fig.\ref{SW}.
}
\end{figure}
\pagebreak

\begin{figure}[htb]
\psfrag{1G4}{$^1G_4$ [deg]}
\psfrag{3G3}{$^3G_3$ [deg]}
\psfrag{3G4}{$^3G_4$ [deg]}
\psfrag{3G5}{$^3G_5$ [deg]}
\psfrag{E4}{$\epsilon_4$ [deg]}
\parbox{6.7cm}{\centerline{\psfig{file=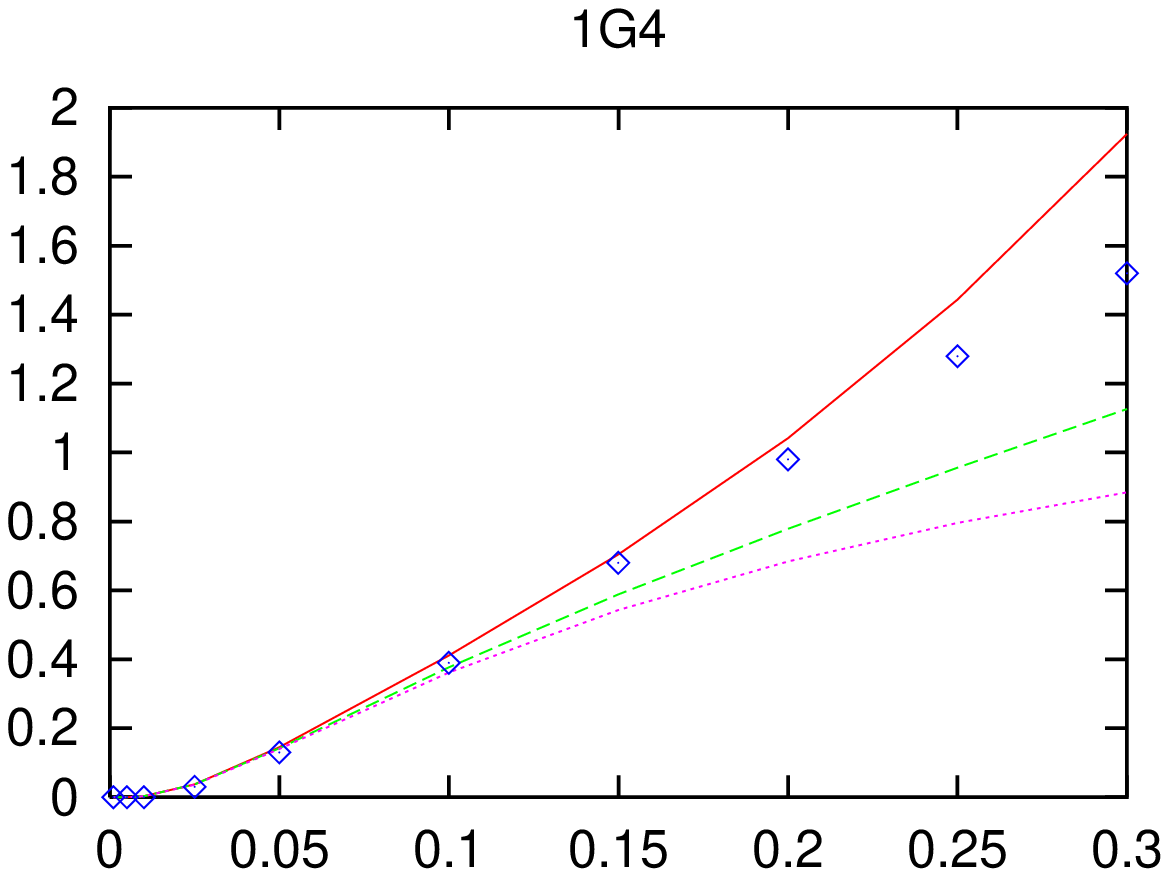,width=3.1in}}}
\hfill
\parbox{6.7cm}{
\centerline{\psfig{file=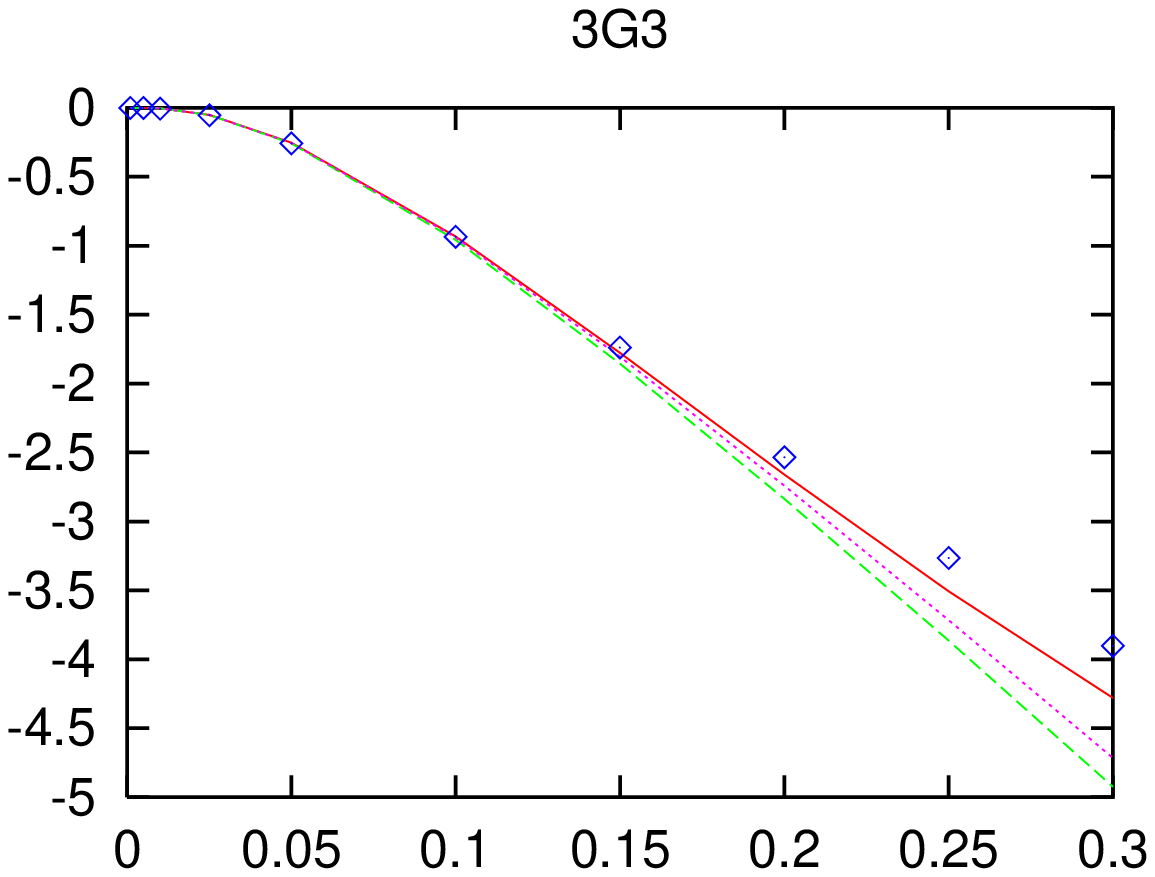,width=3.1in}}}

\vspace{0.2cm}
\parbox{6.7cm}{\centerline{\psfig{file=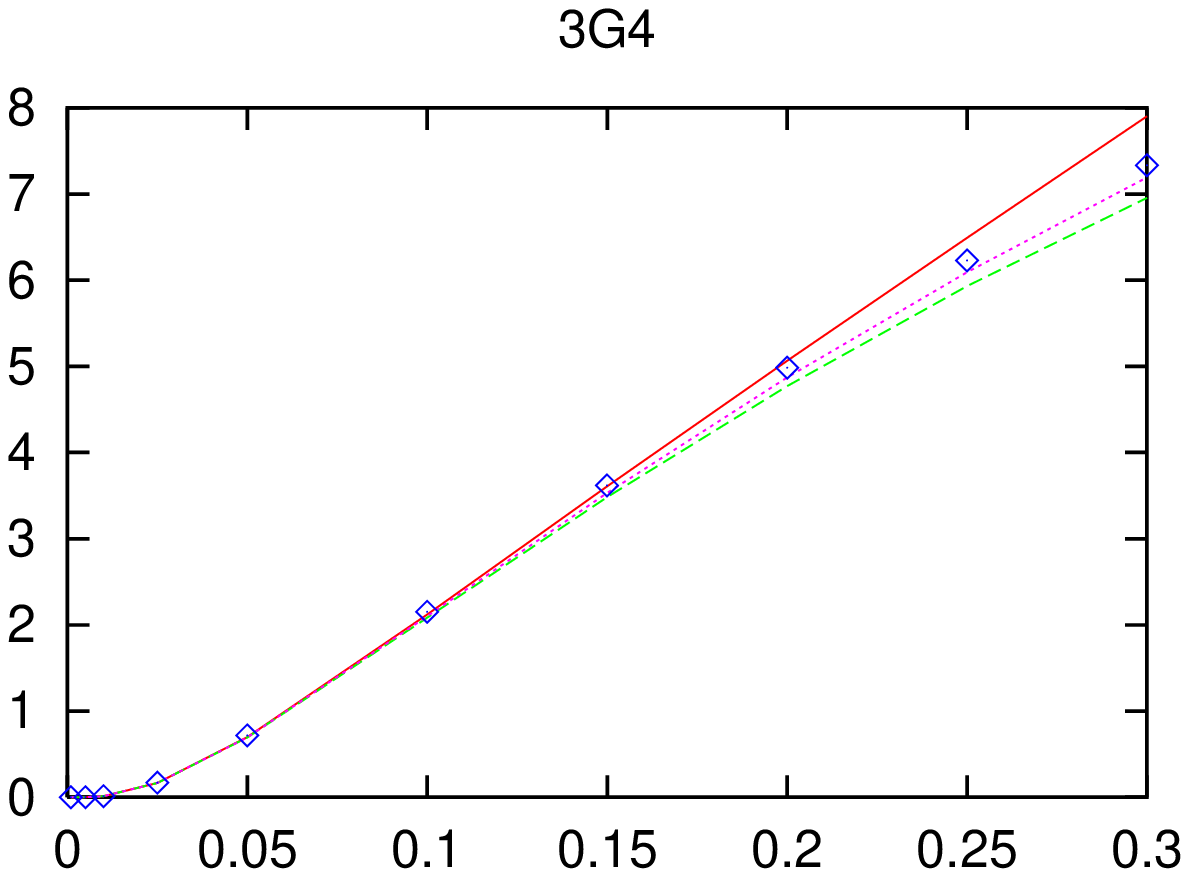,width=3.1in}}}
\hfill
\parbox{6.7cm}{
\centerline{\psfig{file=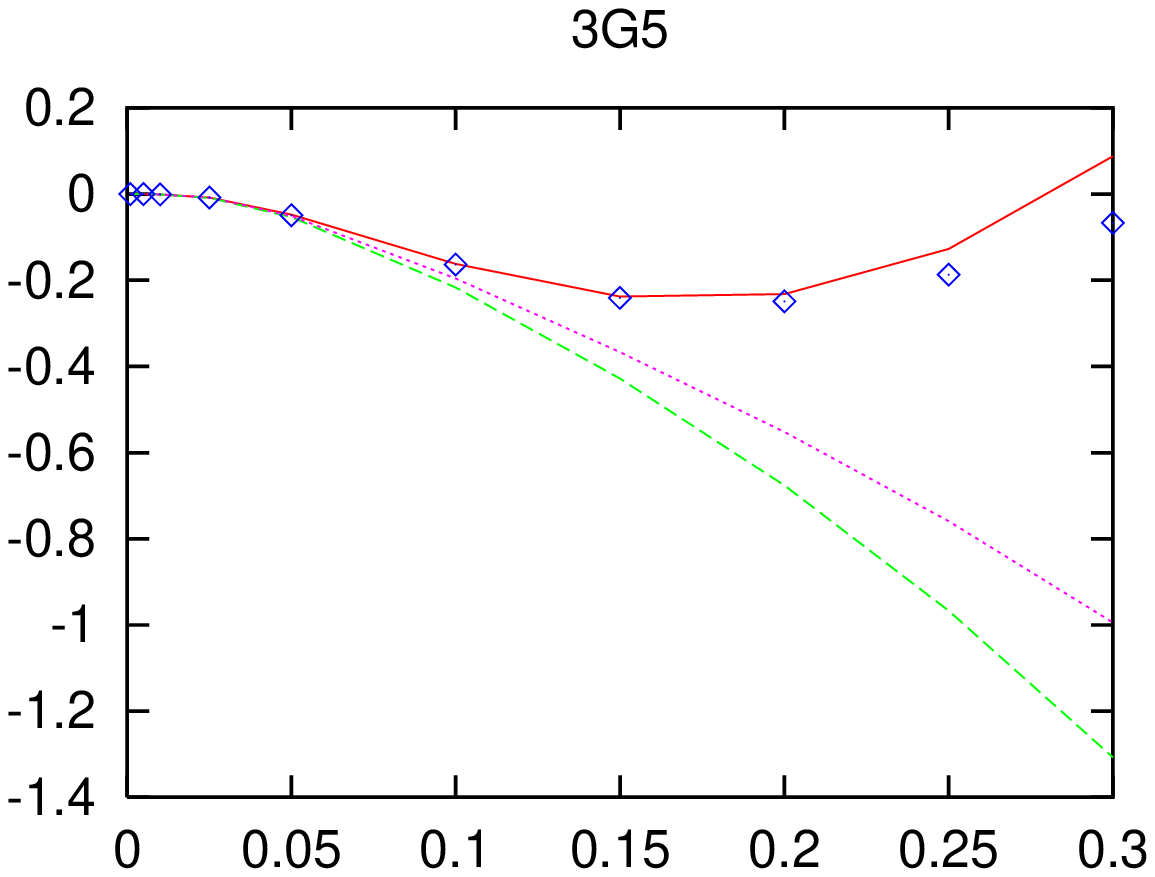,width=3.1in}}}

\vspace{0.2cm}
\centerline{\psfig{file=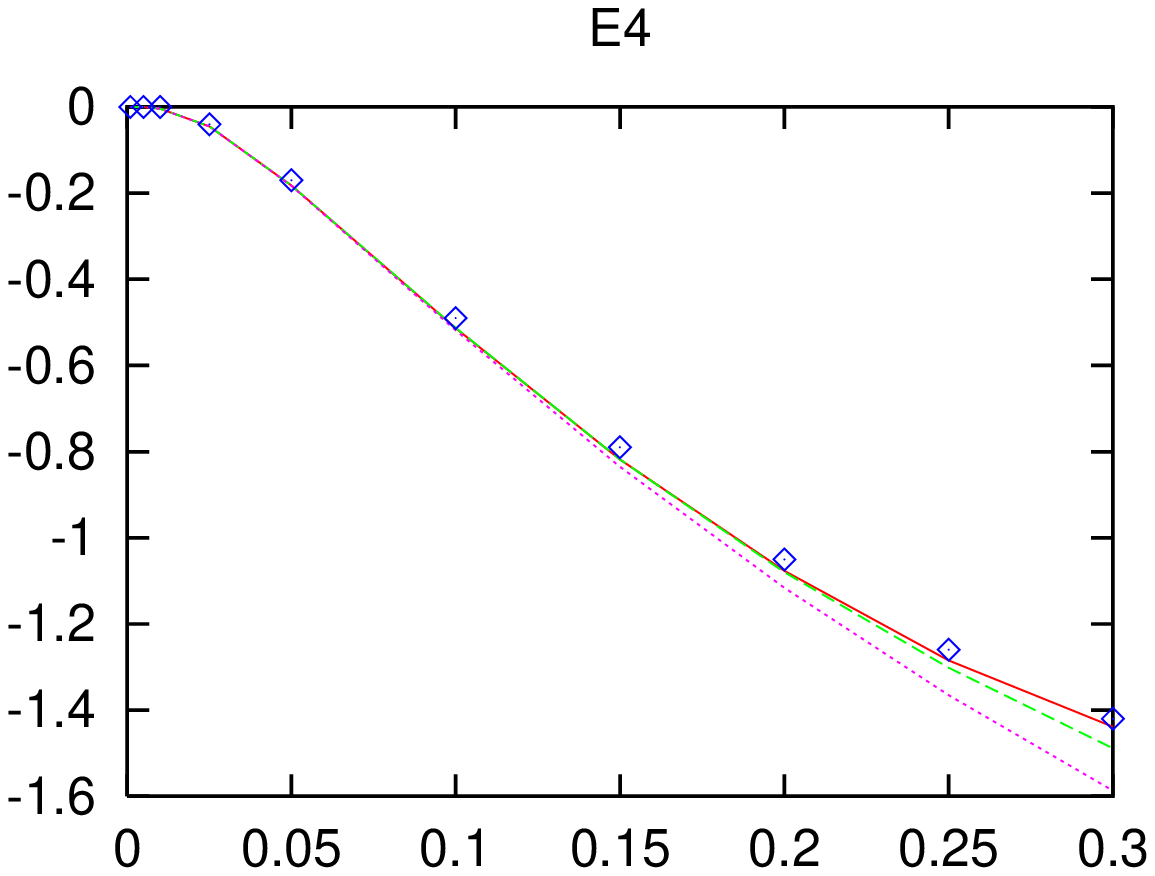,width=3.1in}}
\vspace{1cm}
\caption{\label{GW} 
Predictions for the G--waves  and the
mixing parameter $\epsilon_4$ (in degrees) for 
nucleon laboratory energies $E_{\rm lab}$ below 300~MeV.
For notations, see fig.\ref{SW}.
}
\end{figure}
\pagebreak

\begin{figure}[htb]
\psfrag{1H5}{$^1H_5$ [deg]}
\psfrag{3H4}{$^3H_4$ [deg]}
\psfrag{3H5}{$^3H_5$ [deg]}
\psfrag{3H6}{$^3H_6$ [deg]}
\psfrag{E5}{$\epsilon_5$ [deg]}
\parbox{6.7cm}{\centerline{\psfig{file=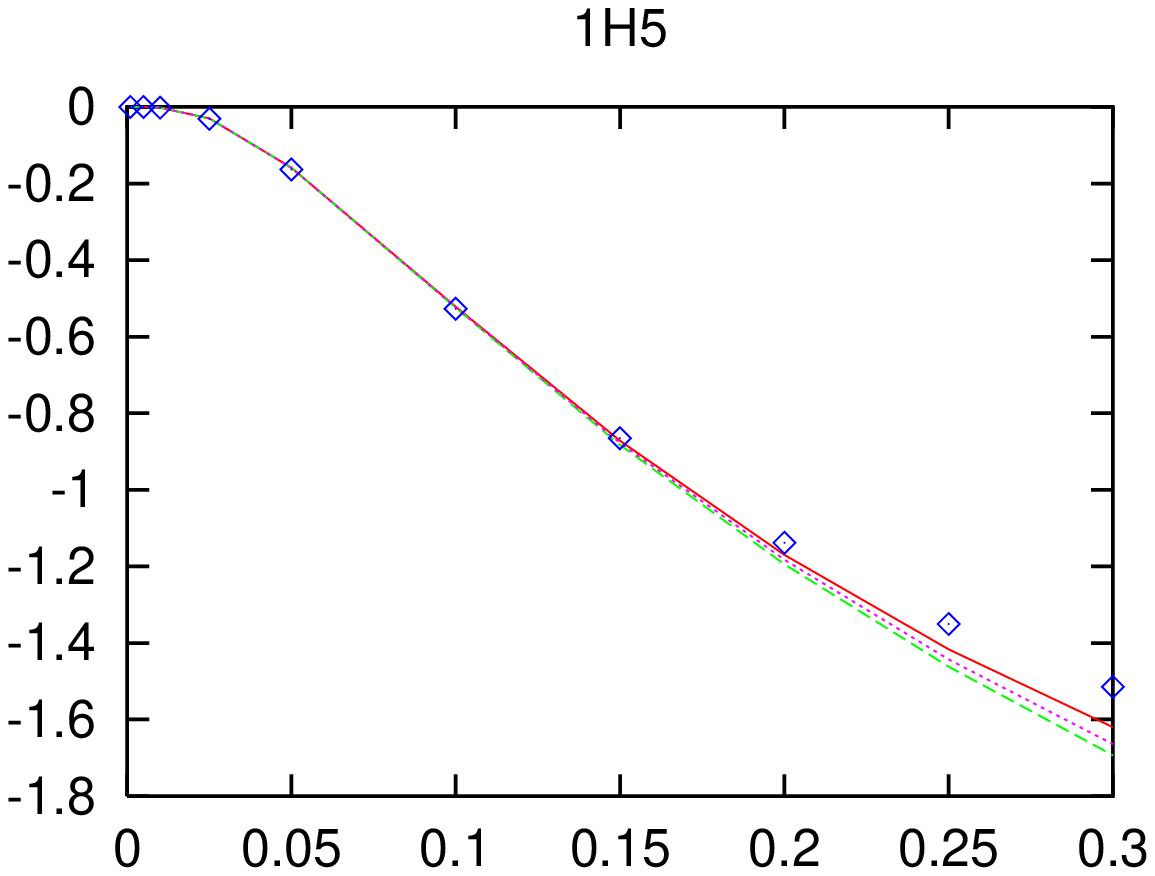,width=3.1in}}}
\hfill
\parbox{6.7cm}{
\centerline{\psfig{file=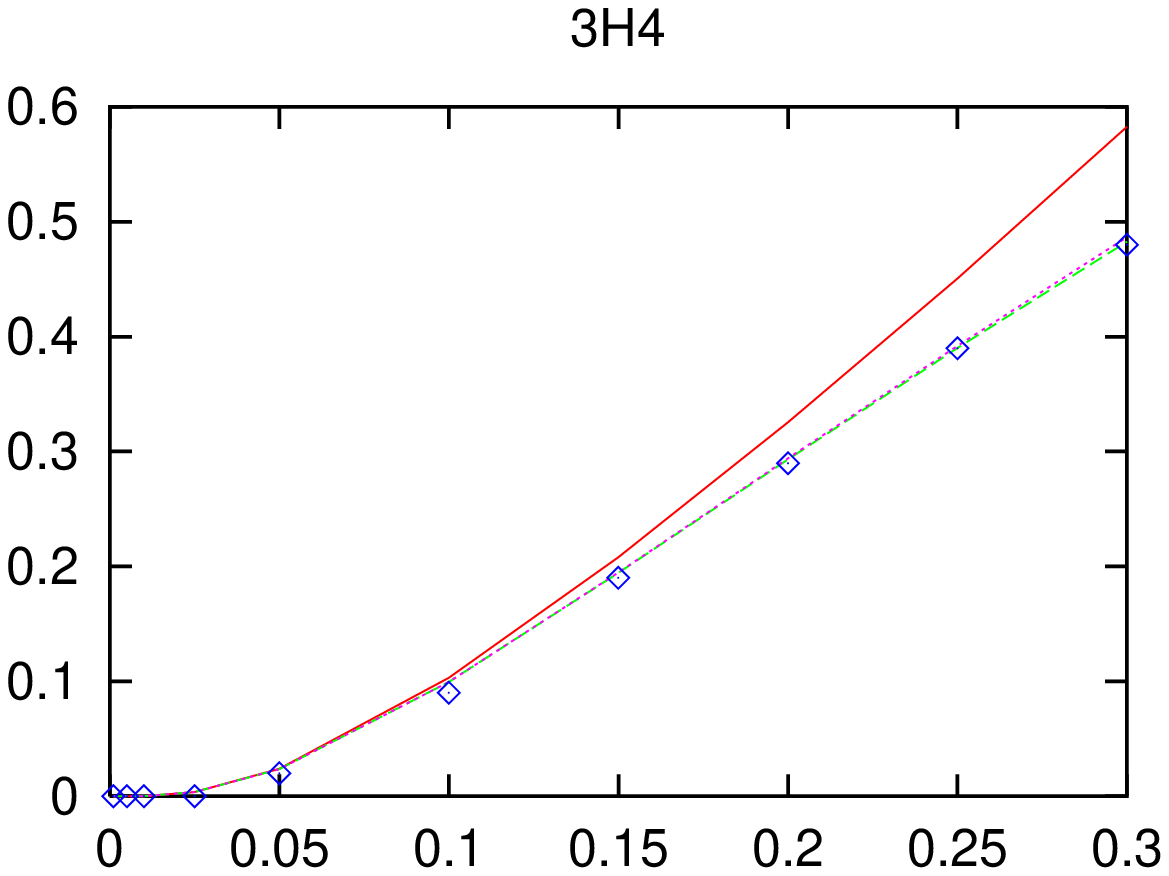,width=3.1in}}}

\vspace{0.2cm}
\parbox{6.7cm}{\centerline{\psfig{file=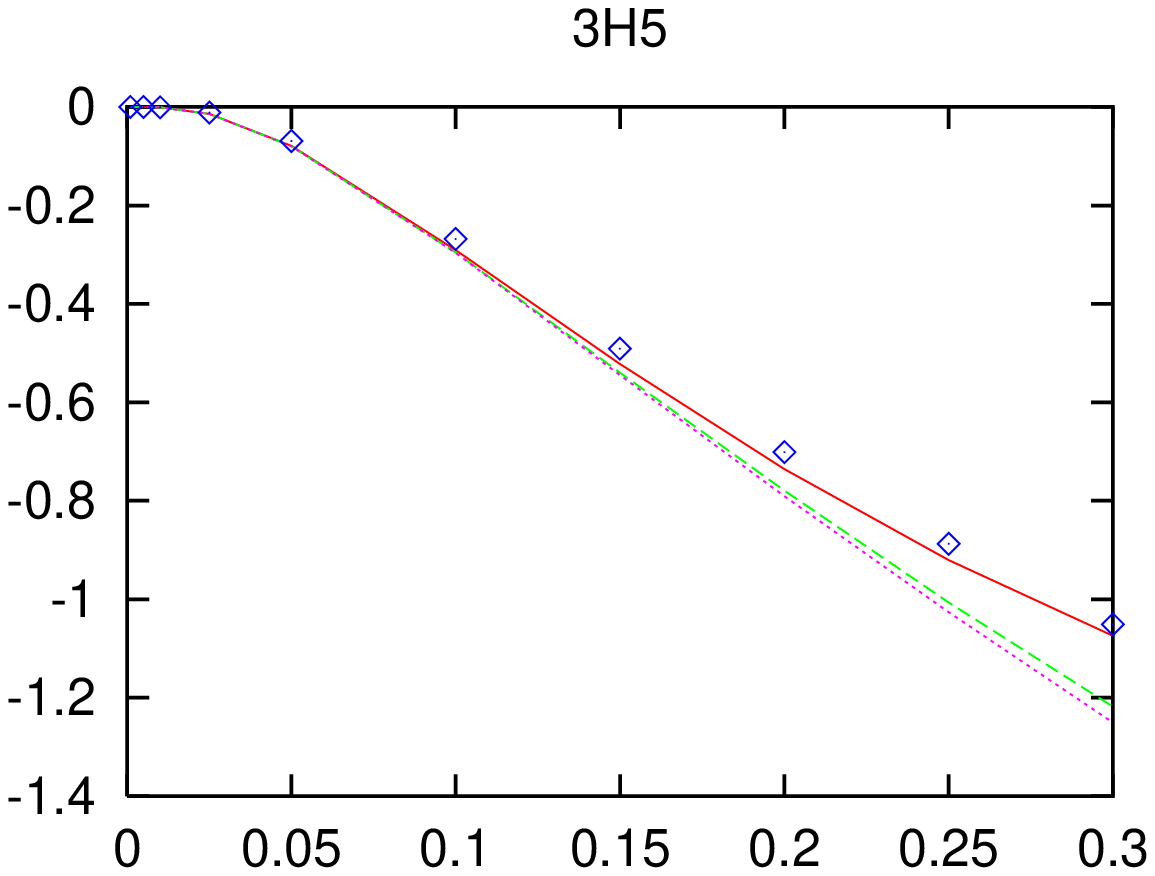,width=3.1in}}}
\hfill
\parbox{6.7cm}{
\centerline{\psfig{file=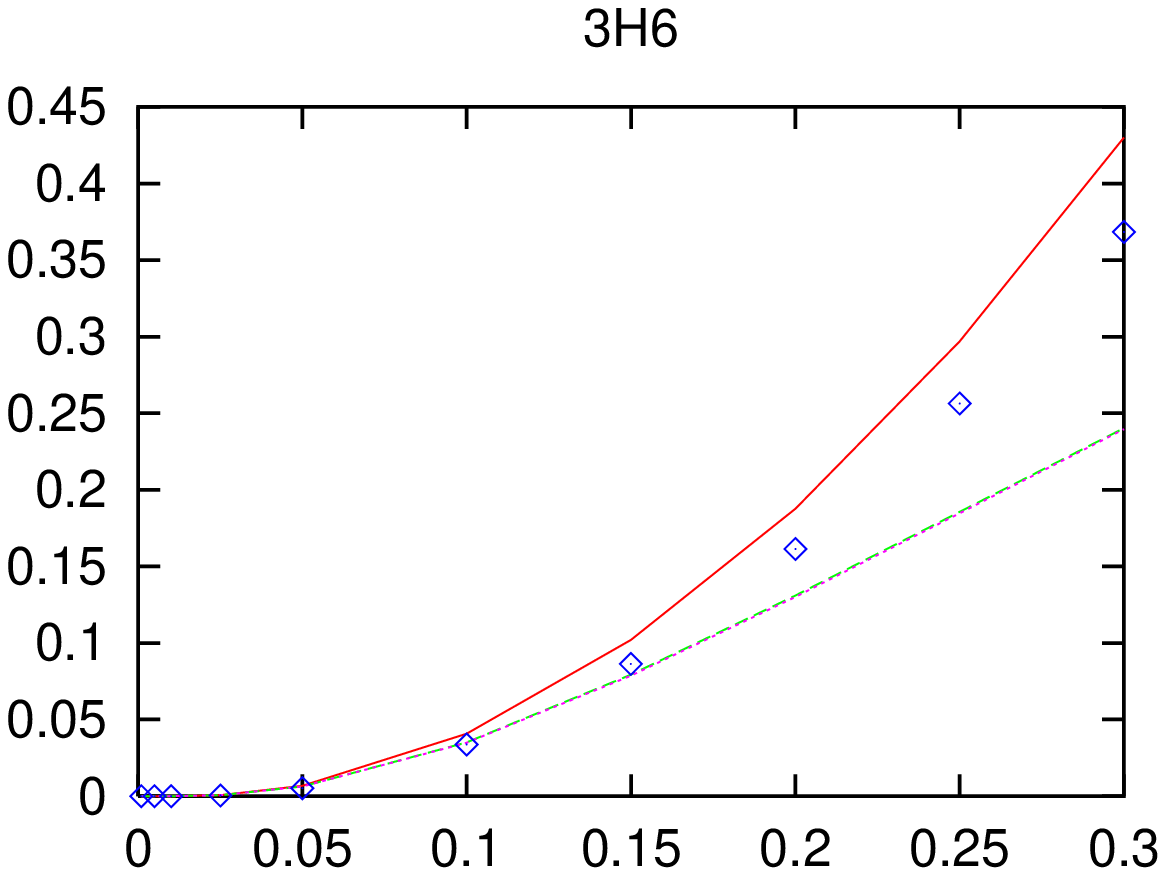,width=3.1in}}}

\vspace{0.2cm}
\centerline{\psfig{file=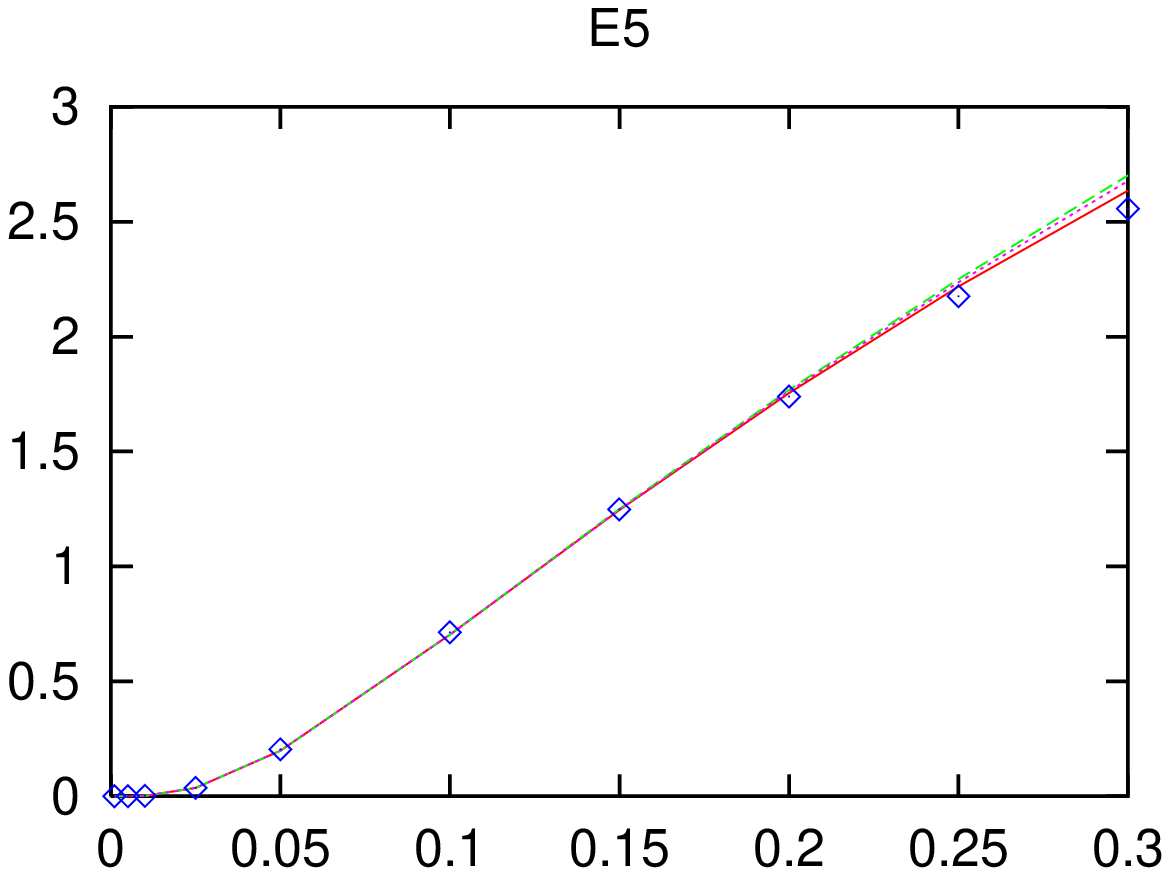,width=3.1in}}
\vspace{1cm}
\caption{\label{HW} 
Predictions for the H--waves  and the
mixing parameter $\epsilon_5$ (in degrees) for 
nucleon laboratory energies $E_{\rm lab}$ below 300~MeV.
For notations, see fig.\ref{SW}.
}
\end{figure}
\pagebreak

\begin{figure}[htb]
\psfrag{1I6}{$^1I_6$ [deg]}
\psfrag{3I5}{$^3I_5$ [deg]}
\psfrag{3I6}{$^3I_6$ [deg]}
\psfrag{3I7}{$^3I_7$ [deg]}
\psfrag{E6}{$\epsilon_6$ [deg]}
\parbox{6.7cm}{\centerline{\psfig{file=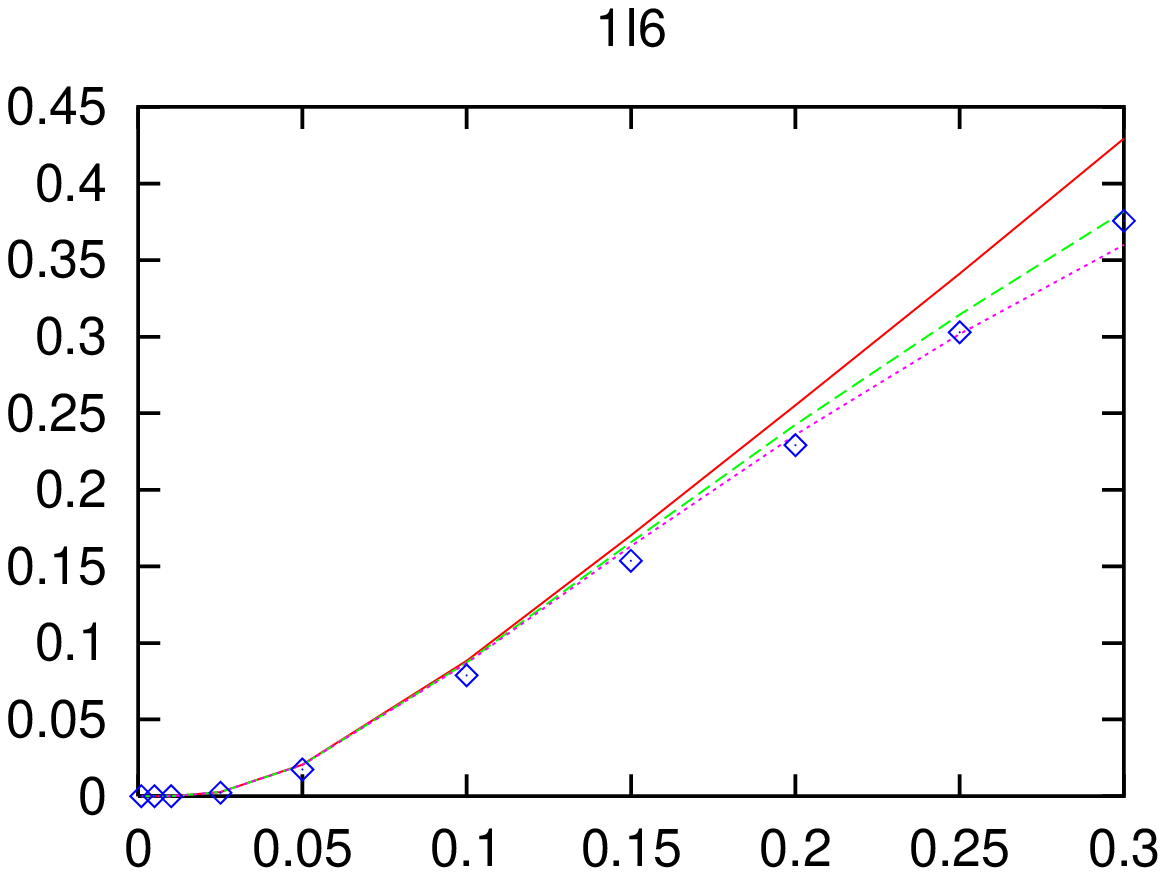,width=3.1in}}}
\hfill
\parbox{6.7cm}{
\centerline{\psfig{file=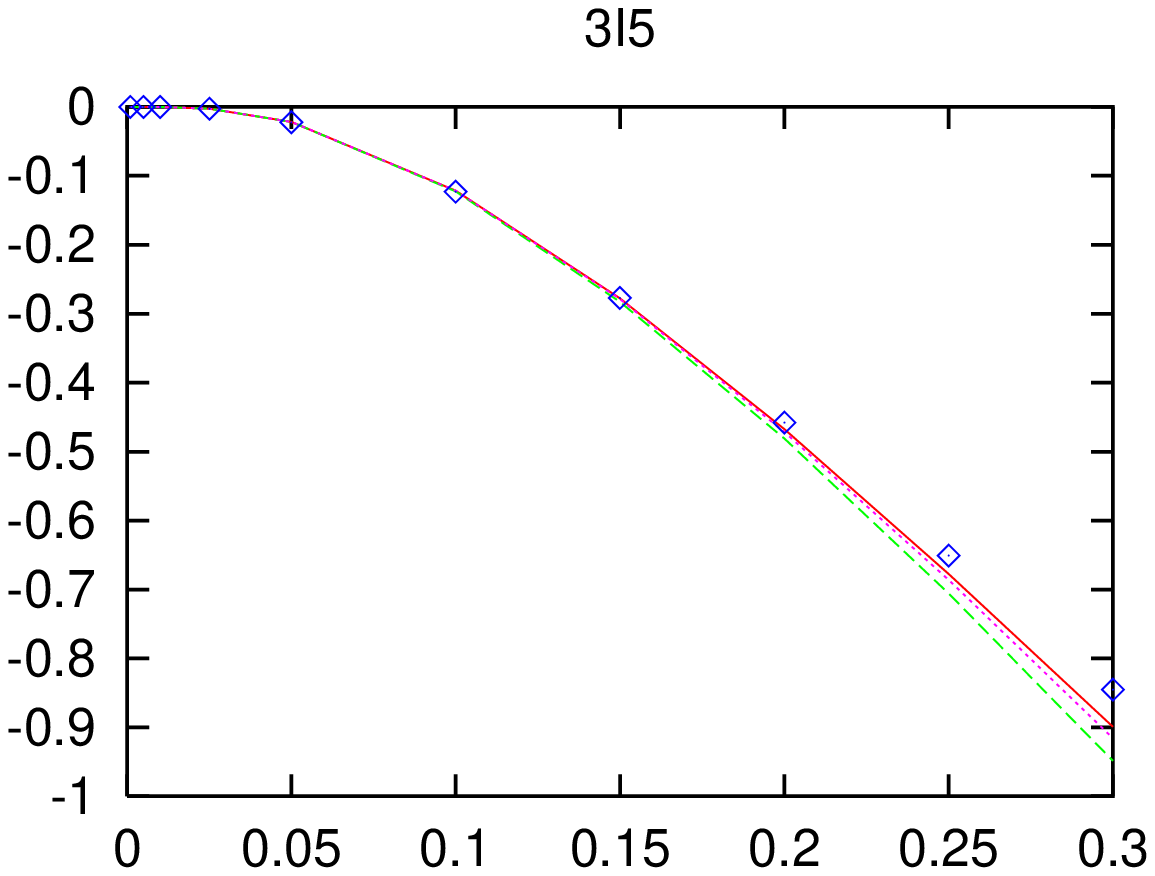,width=3.1in}}}

\vspace{0.2cm}
\parbox{6.7cm}{\centerline{\psfig{file=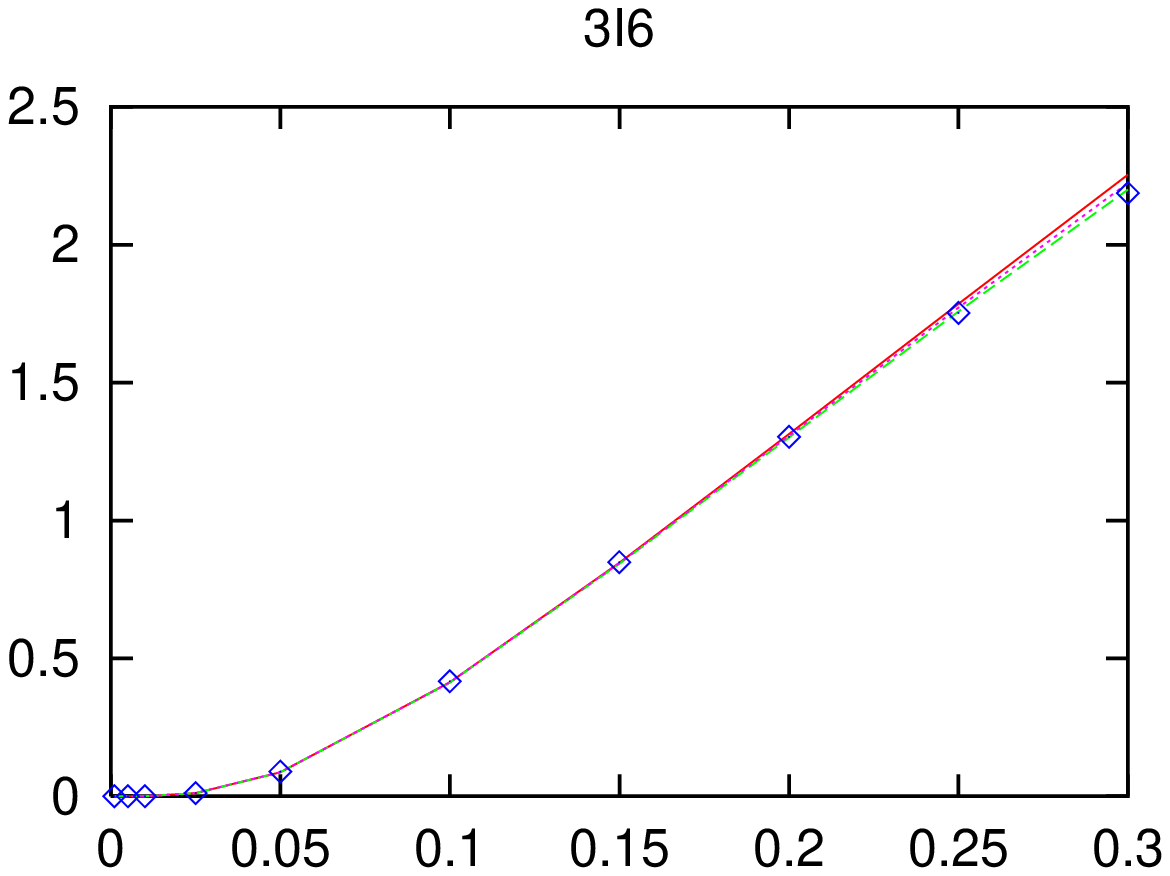,width=3.1in}}}
\hfill
\parbox{6.7cm}{
\centerline{\psfig{file=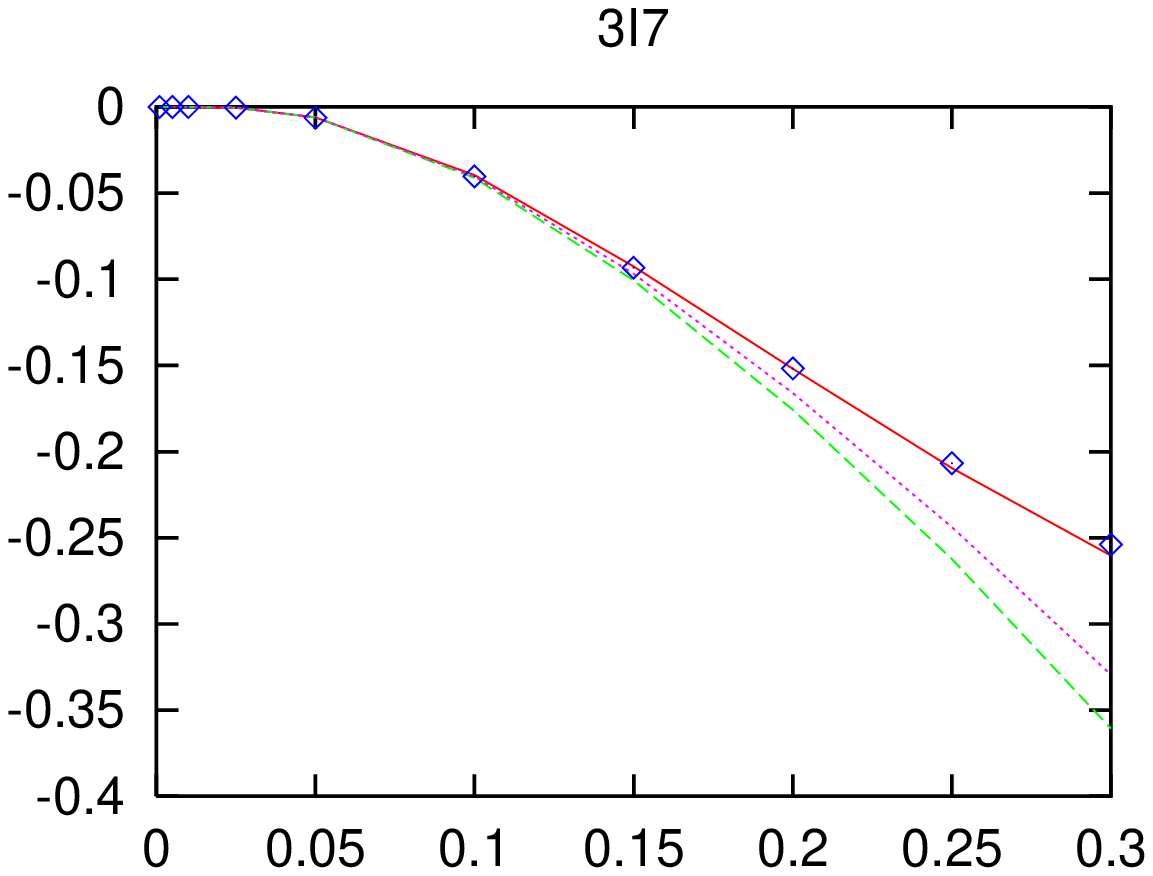,width=3.1in}}}

\vspace{0.2cm}
\centerline{\psfig{file=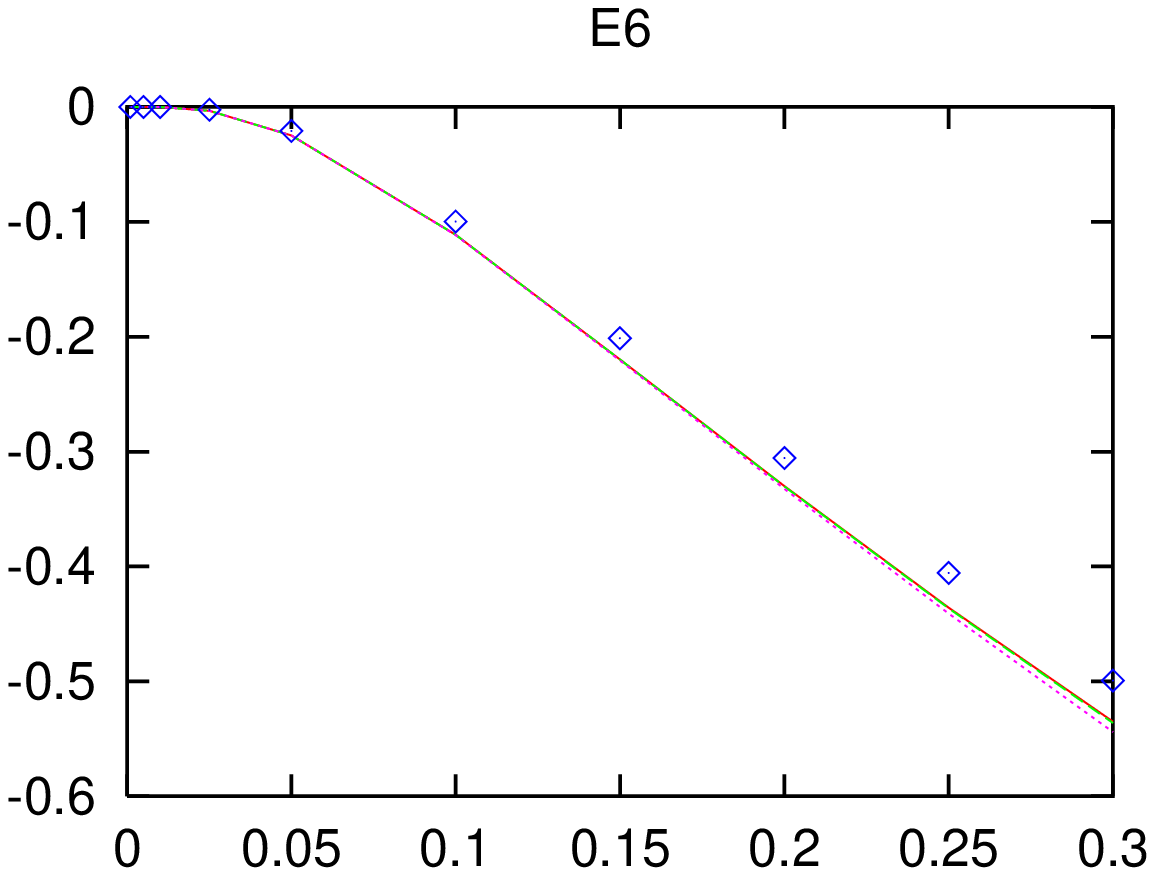,width=3.1in}}
\vspace{1cm}
\caption{\label{IW} 
Predictions for the I--waves  and the
mixing parameter $\epsilon_6$ (in degrees) for 
nucleon laboratory energies $E_{\rm lab}$ below 300~MeV.
For notations, see fig.\ref{SW}.
}
\end{figure}
\pagebreak

\begin{figure}[htb]
   \vspace{2.9cm}
   \epsfysize=14cm
   \centerline{\epsffile{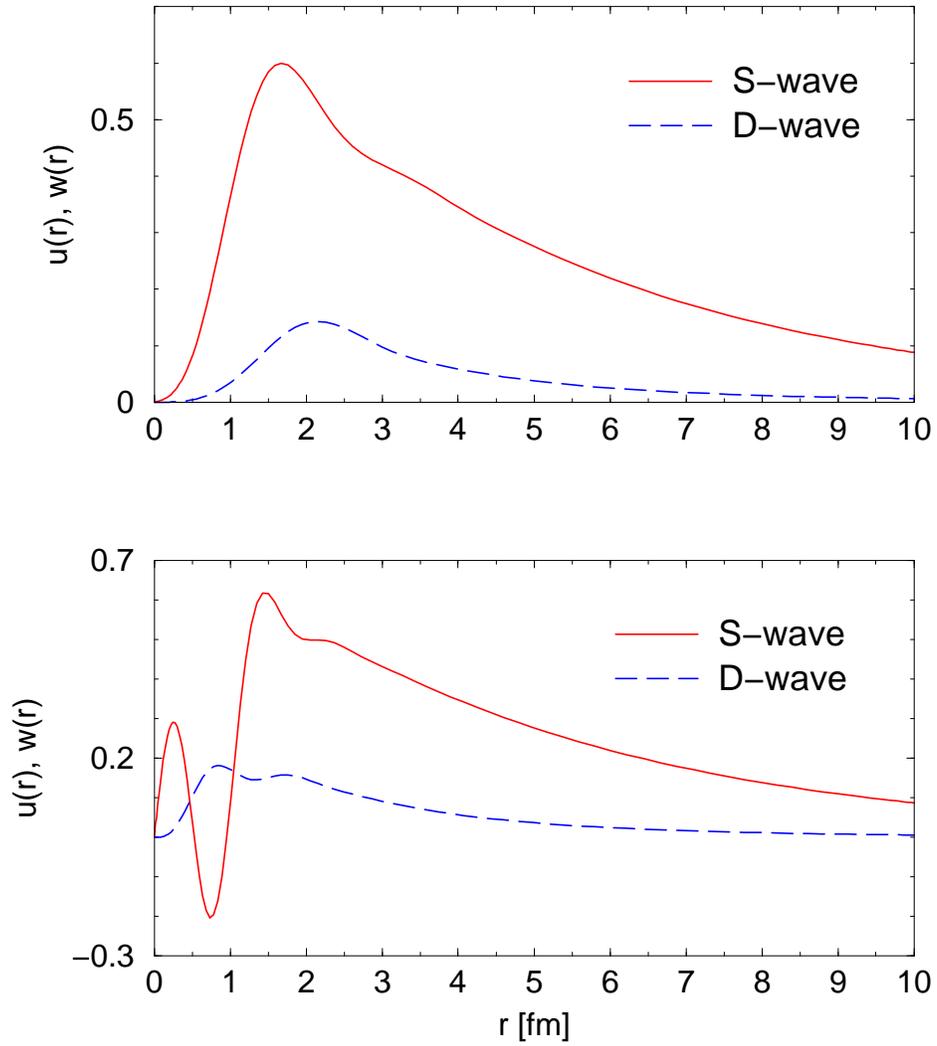}}
   \vspace{2cm}
   \centerline{\parbox{13cm}{\caption{\label{figdwf}
Coordinate space representations of the S-- (red solid line)
and D--wave (blue dashed line) deuteron
wave functions at NLO (upper panel) and NNLO (lower panel).
  }}}
\end{figure}
\pagebreak

\begin{figure}[htb]
   \vspace{2.9cm}
   \epsfysize=14cm
   \centerline{\epsffile{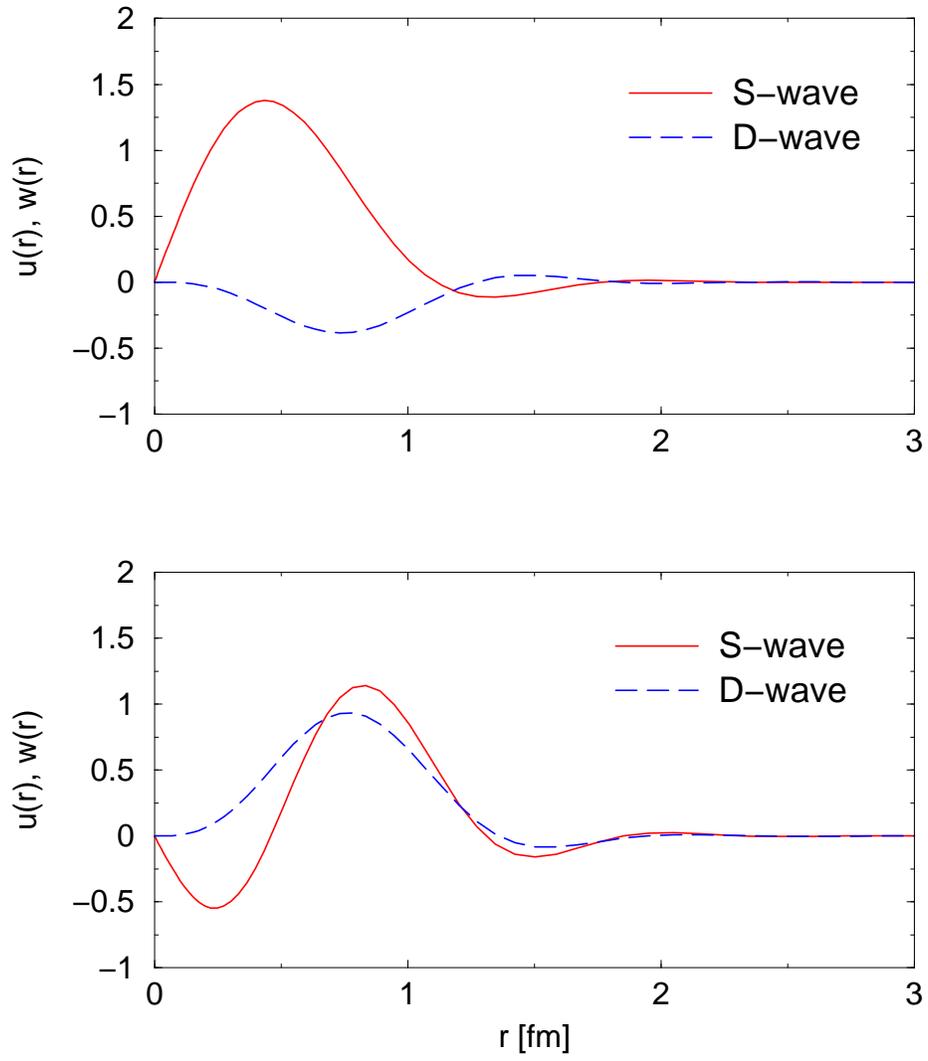}}
   \vspace{2cm}
   \centerline{\parbox{13cm}{\caption{\label{figdeep}
Coordinate space representations of the S-- (red solid line)
and D--wave (blue dashed line) for the two unphysical boundstates.
  }}}
\end{figure}
\pagebreak

\begin{figure}[htb]
   \vspace{2.9cm}
   \epsfysize=14cm
   \centerline{\epsffile{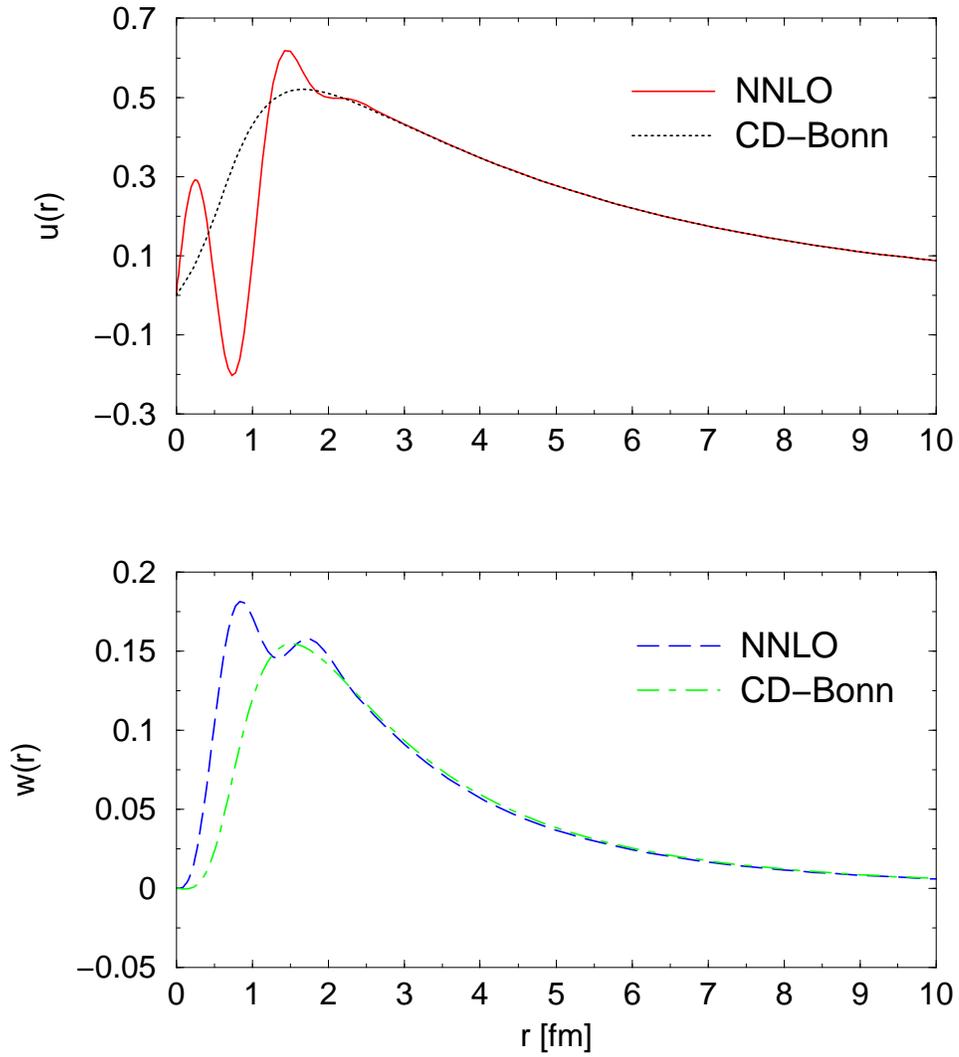}}
   \vspace{2cm}
   \centerline{\parbox{13cm}{\caption{\label{cdbonn}
Coordinate space representations of the S-- (upper panel)
and D--wave (lower panel)  deuteron
wave functions at NNLO compared to the one from the CD-Bonn potential.
  }}}
\end{figure}

\begin{figure}[htb]
\psfrag{1S0}{$^1S_0 (\Delta)$}
\psfrag{3S1}{$^3S_1 (\Delta)$}
\psfrag{E1}{$\epsilon_1 (\Delta)$}
\psfrag{1D2}{$^1D_2 (\Delta)$}
\psfrag{3D3}{$^3D_3 (\Delta)$}
\psfrag{3G5}{$^3G_5 (\Delta)$}%
\vspace{1.5cm}
\parbox{6.5cm}{\centerline{\psfig{file=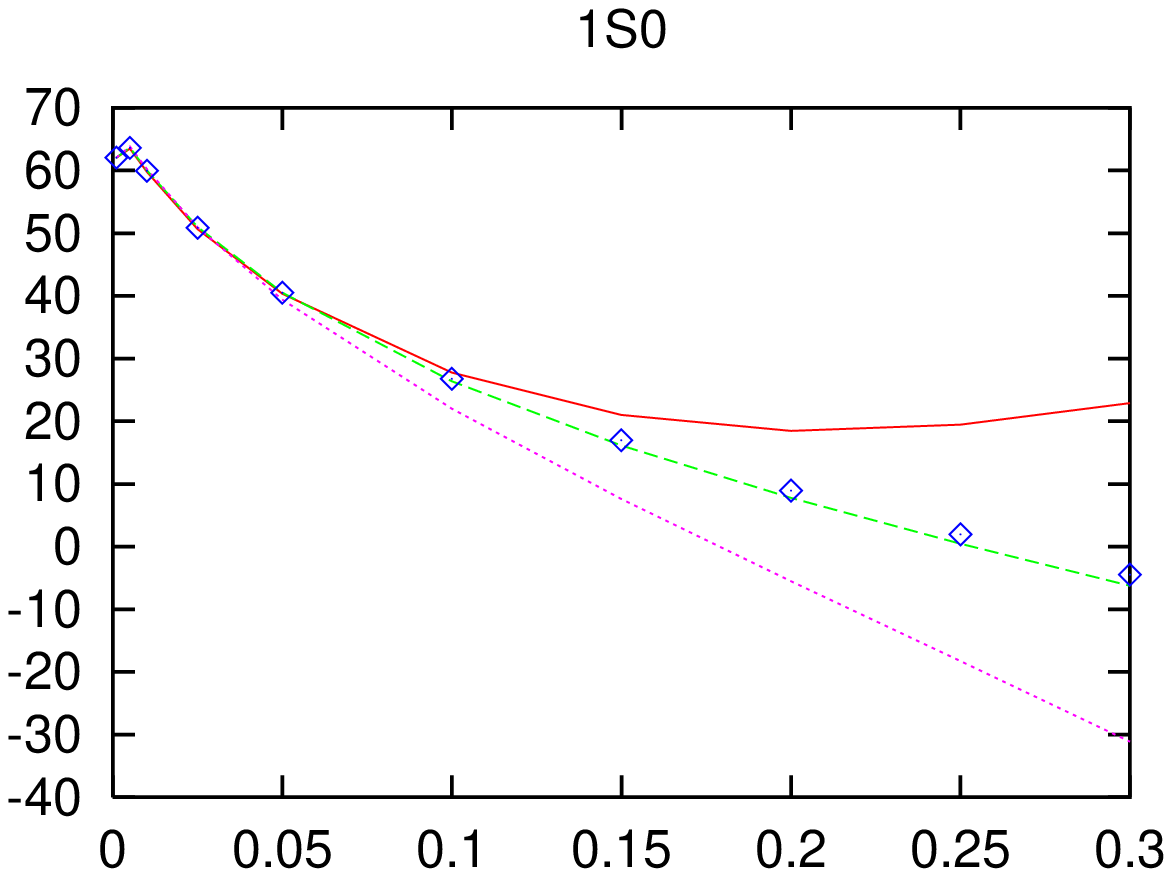,width=2.9in}}}
\hfill
\parbox{6.5cm}{
\centerline{\psfig{file=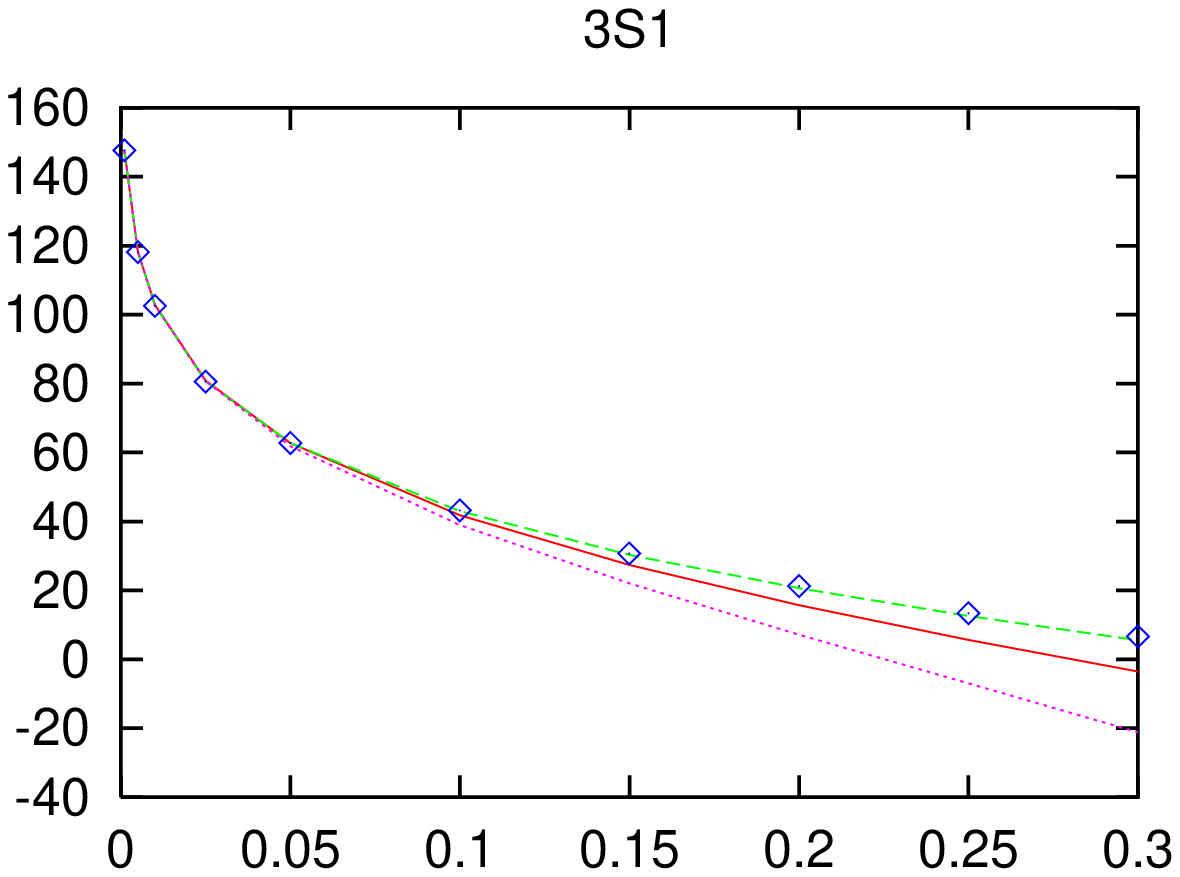,width=2.9in}}}
\parbox{6.5cm}{\centerline{\psfig{file=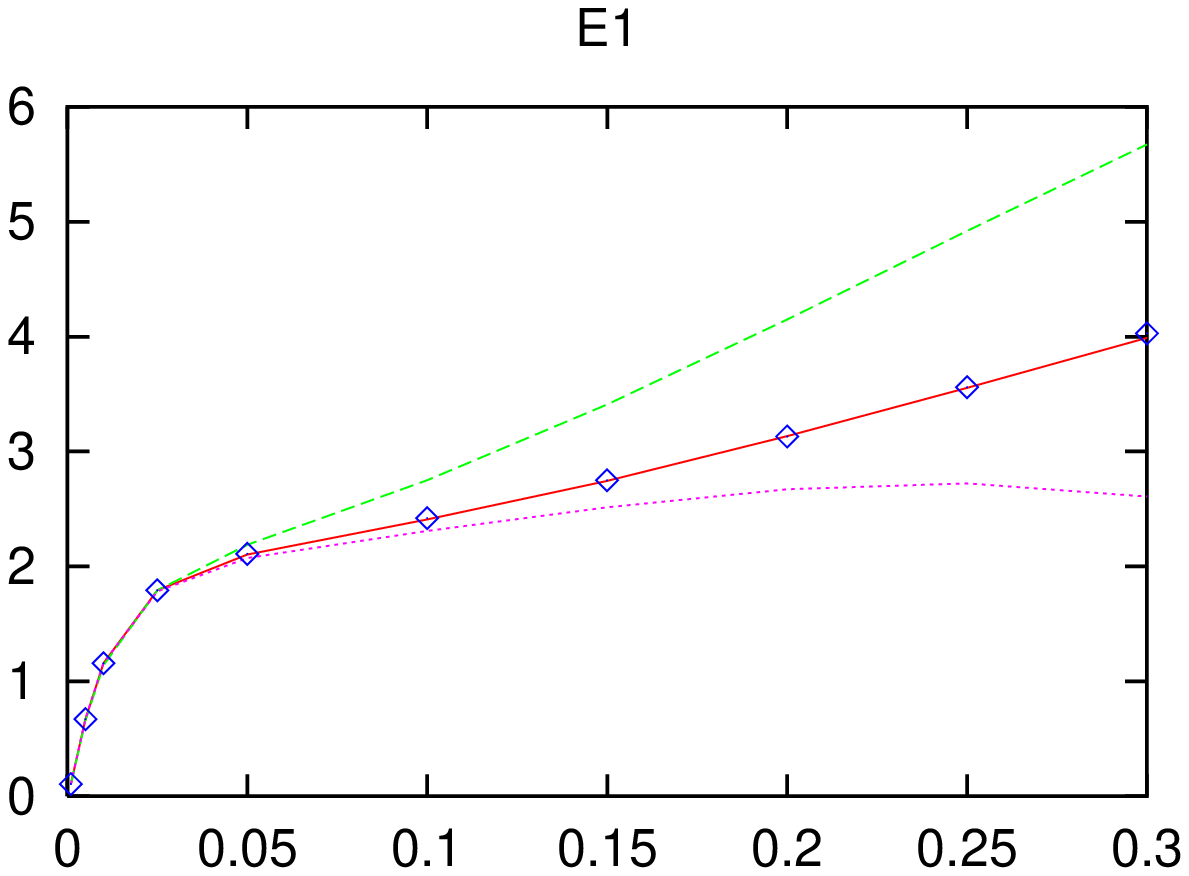,width=2.9in}}}
\hfill
\parbox{6.5cm}{
\centerline{\psfig{file=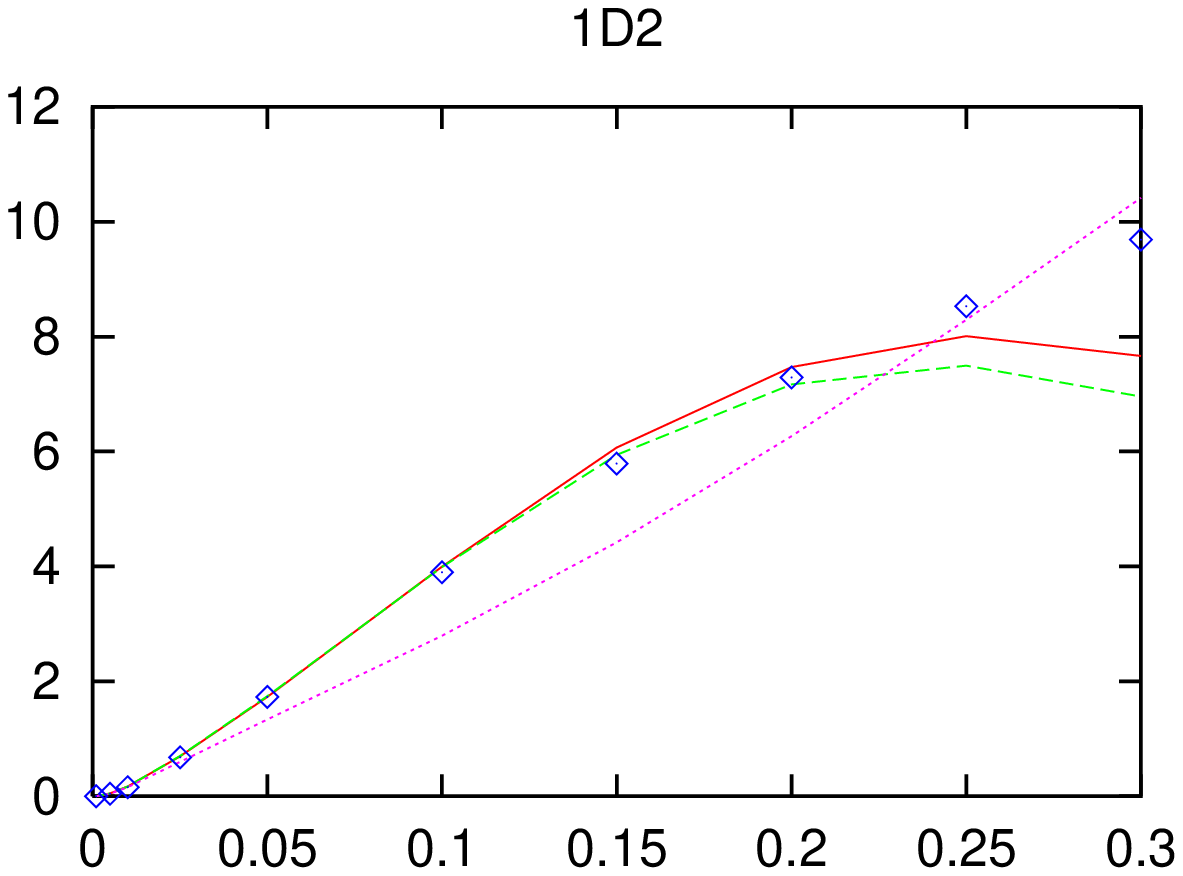,width=2.9in}}}

\vspace{0.2cm}
\parbox{6.5cm}{\centerline{\psfig{file=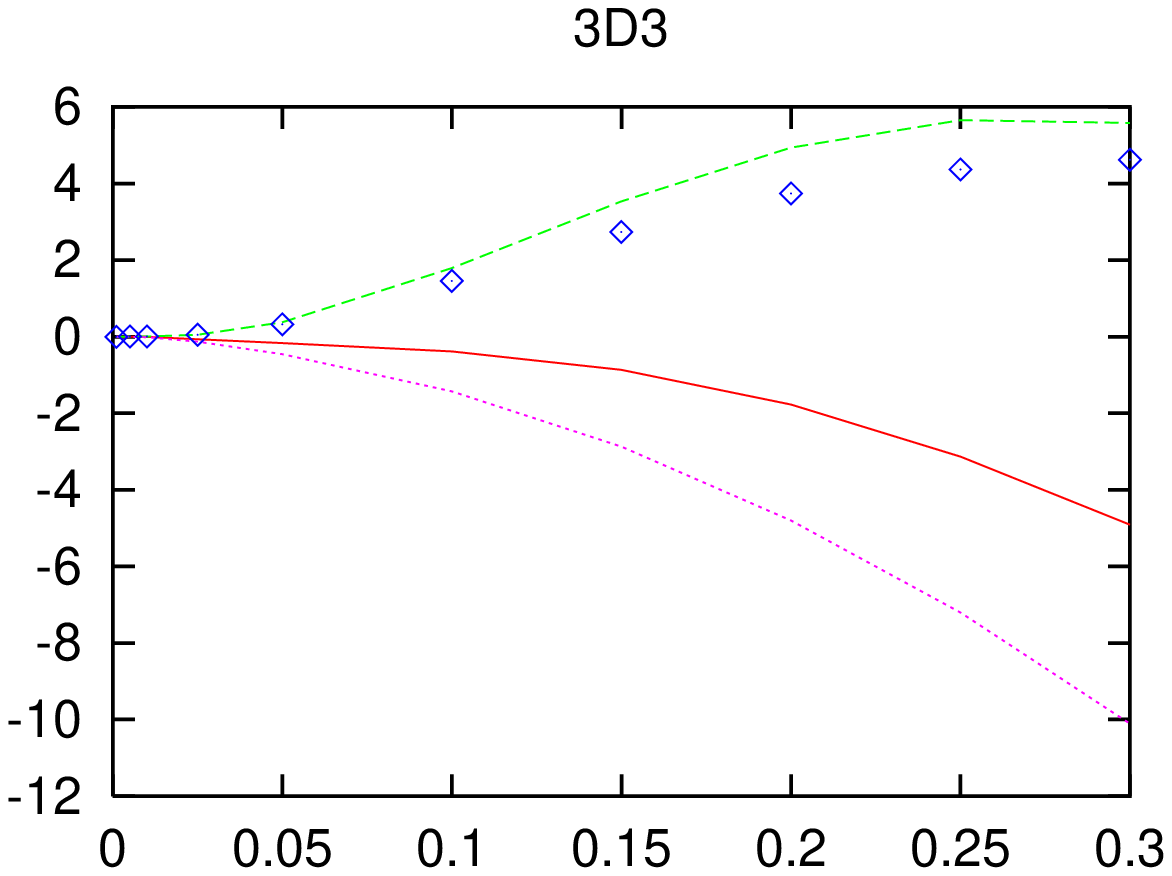,width=2.9in}}}
\hfill
\parbox{6.5cm}{
\centerline{\psfig{file=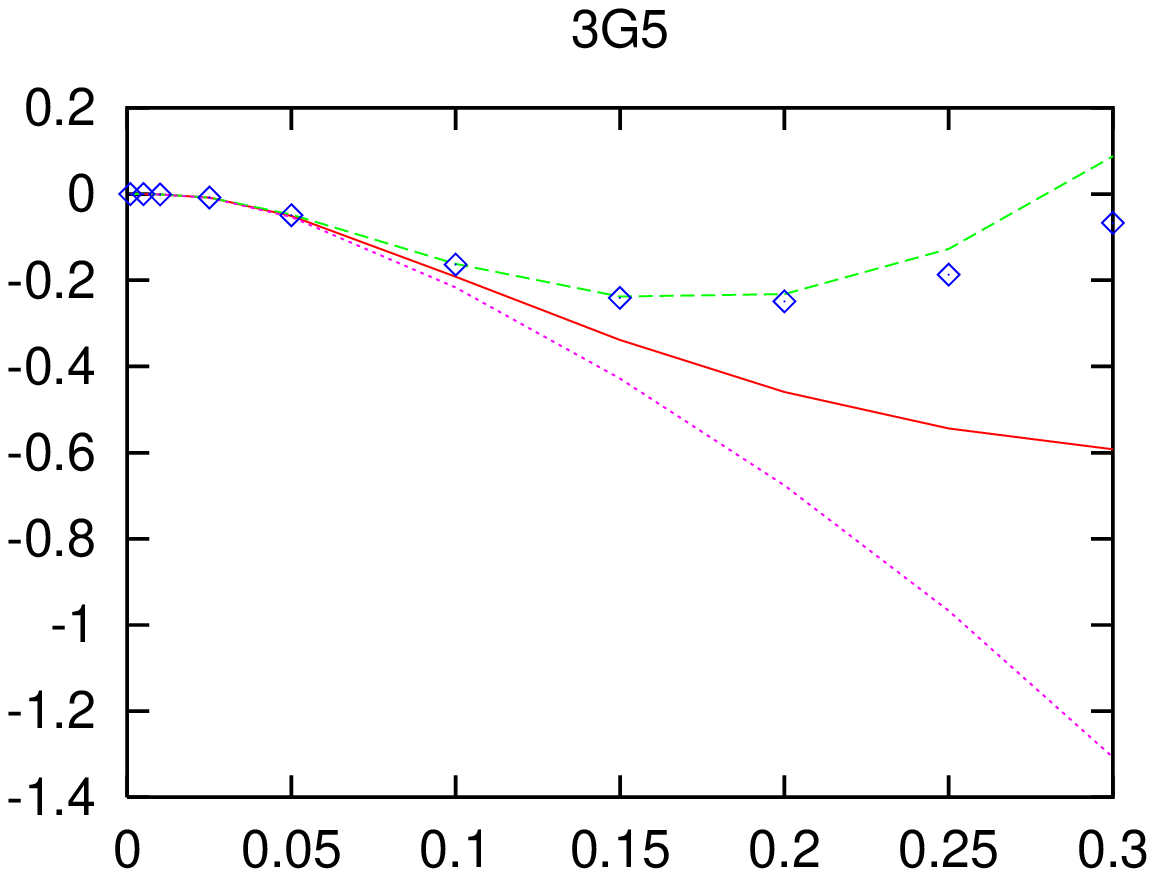,width=2.9in}}}
\vspace{1cm}
\caption{\label{fig:delta}
Phase shifts (in degrees) for the NNLO--$\Delta$ theory (red solid lines)
in comparison to the NNLO (green dashed lines) and NLO (purple
dotted lines) results as a function of the lab energy 
in GeV. For other notations, see fig.\ref{SW}.
}
\end{figure}

\end{document}